\documentclass[aps,prd,reprint,floatfix,amsmath,amssymb,nofootinbib,notitlepage,longbibliography,tabulary,color,usenames,dvipsnames,svgnames,table]{revtex4-1}

\usepackage{amsmath,amssymb,graphics,setspace}
\usepackage{bm,soul,ulem}
\usepackage[pdftex]{graphicx}
\usepackage{multirow,microtype}
\usepackage{color}
\usepackage{courier}
\usepackage{xspace}
\usepackage{tabularx,ragged2e,booktabs}
\usepackage{array}
\usepackage{hhline}

\usepackage{tikz}
\usepackage{mathdots}
\usepackage{yhmath}
\usepackage{cancel}
\usepackage{siunitx}
\usepackage{gensymb}
\usetikzlibrary{fadings}
\usetikzlibrary{patterns}
\usetikzlibrary{shadows.blur}

\usepackage{xcolor}
\usepackage{natbib}
\usepackage{hyperref} 
\hypersetup{colorlinks=true,linkcolor=blue,filecolor=magenta,urlcolor=red,citecolor=blue}

\usepackage{titlesec}

\titlespacing\section{1pt}{12pt plus 4pt minus 2pt}{0pt plus 2pt minus 2pt}
\titlespacing\subsection{0pt}{12pt plus 4pt minus 2pt}{0pt plus 2pt minus 2pt}
\titlespacing\subsubsection{0pt}{12pt plus 4pt minus 2pt}{0pt plus 2pt minus 2pt}

\usepackage{dcolumn}
\usepackage{tabulary}
\usepackage{siunitx}
\usepackage{longtable}
\usepackage{enumitem}
\usepackage{url}

\DeclareFontFamily{OT1}{pzc}{}
\DeclareFontShape{OT1}{pzc}{m}{it}{<-> s * [1.000] pzcmi7t}{}
\DeclareMathAlphabet{\mathpzc}{OT1}{pzc}{m}{it}

\newcommand{\ten}[2]{\ensuremath{{#1}{\times} 10^{#2}}}

\newcommand{\fig}{Fig.}
\newcommand{\figs}{Figs.}

\newcommand{\dskips}{-0.4cm}

\newcommand{\olam}{\Omega_{\Lambda}}
\newcommand{\om}{\Omega_{\rm m}}
\newcommand{\orad}{\Omega_{\rm r}}
\newcommand{\ok}{\Omega_{\rm k}}

\newcommand{\Gev}{{\rm \, GeV}}
\newcommand{\ev}{{\rm \, eV}}

\newcommand{\tnot}{t_0}
\newcommand{\Ho}{H_0}

\newcommand{\Hounits}{km s$^{-1}$Mpc$^{-1}$}

\newcommand{\sourceindex}{{}}

\newcommand{\Yjm}{\ensuremath{Y_{jm}}}
\newcommand{\Yjmminus}{\ensuremath{Y_{j,(-m)}}}
\newcommand{\sYjm}{\ensuremath{{}_{s}\Yjm}}

\newcommand{\nhat}{ {\bm{\hat{n}}} }

\newcommand{\oYjm}{   \ensuremath{ {}_{0}\Yjm }    }
\newcommand{\oYjmminus}{   \ensuremath{ {}_{0}\Yjmminus }    }

\newcommand{\twoYjm}{   \ensuremath{ {}_{\pm2}\Yjm }    }
\newcommand{\twompYjm}{   \ensuremath{ {}_{\mp2}\Yjm }    }
\newcommand{\twoYjmminus}{   \ensuremath{ {}_{\pm2}\Yjmminus }    }
\newcommand{\cdIjm}{\ensuremath{c^{(d)}_{(I)jm}}}
\newcommand{\kdXjm}{\ensuremath{k^{(d)}_{(X)jm}}}
\newcommand{\kdVjm}{\ensuremath{k^{(d)}_{(V)jm}}}
\newcommand{\kdVjmminus}{\ensuremath{k^{(d)}_{(V)j,(-m)}}}

\newcommand{\kdEjm}{\ensuremath{k^{(d)}_{(E)jm}}}
\newcommand{\kdBjm}{\ensuremath{k^{(d)}_{(B)jm}}}

\newcommand{\kdEBjm}{\ensuremath{k^{(d)}_{(E,B)jm}}}
\newcommand{\kdEBjmminus}{\ensuremath{k^{(d)}_{(E,B)j,(-m)}}}

\newcommand{\stwoplus}{\ensuremath{s_{(+2)}}}
\newcommand{\stwominus}{\ensuremath{s_{(-2)}}}
\newcommand{\szero}{\ensuremath{s_{(0)}}}
\newcommand{\szeroz}{\ensuremath{{\szero}_{z}}}
\newcommand{\stwopm}{\ensuremath{s_{(\pm2)}}}

\newcommand{\stwopmz}{\ensuremath{{\stwopm}_{z}}}

\newcommand{\zi}{z}
\newcommand{\mm}{m} 
\newcommand{\oo}{} 

\newcommand{\zz}{z^{\prime}}
\newcommand{\aaa}{a^{\prime}}

\newcommand{\Idz}{I^{(d)}_{\zi}}
\newcommand{\Qdz}{Q^{(d)}_{\zi}}
\newcommand{\Udz}{U^{(d)}_{\zi}}
\newcommand{\Vdz}{V^{(d)}_{\zi}}

\newcommand{\Ldz}{L^{(d)}_{\zi}}
\newcommand{\Pqdz}{q^{(d)}_{\zi}}
\newcommand{\Pudz}{u^{(d)}_{\zi}}
\newcommand{\Pvdz}{v^{(d)}_{\zi}}

\newcommand{\Pqdzprime}{q^{(d)\prime}_{\zi}}
\newcommand{\Pudzprime}{u^{(d)\prime}_{\zi}}

\newcommand{\Pqd}{q^{(d)}} 
\newcommand{\Pud}{u^{(d)}} 
\newcommand{\DPqd}{\Delta \Pqd}
\newcommand{\DPud}{\Delta \Pud}

\newcommand{\Pqdprime}{q^{(d)\prime}} 
\newcommand{\Pudprime}{u^{(d)\prime}} 

\newcommand{\PqdprimeN}{{\mathpzc{q}}^{(d)\prime}} 
\newcommand{\PudprimeN}{{\mathpzc{u}}^{(d)\prime}} 

\newcommand{\DPqdprime}{\Delta \Pqdprime}
\newcommand{\DPudprime}{\Delta \Pudprime}

\newcommand{\Mz}{ {\bm{M}_{\zi}} }
\newcommand{\svec}{ {\bm{s}} }
\newcommand{\svecz}{ {\bm{s}_{\zi}} }
\newcommand{\varsigmavec}{ {\bm{\varsigma}} }

\newcommand{\Muller}{M{\"u}ller\xspace}

\newcommand{\varsigmazero}{ {\varsigma^{(0)}} }
\newcommand{\varsigmaone}{ {\varsigma^{(1)}} }
\newcommand{\varsigmatwo}{ {\varsigma^{(2)}} }
\newcommand{\varsigmathree}{ {\varsigma^{(3)}} }

\newcommand{\varsigmapm}{ \varsigma^{(\pm)} }

\newcommand{\varsigmaplusd}{ {\varsigma^{(+)(d)}} }
\newcommand{\varsigmaminusd}{ {\varsigma^{(-)(d)}} }

\newcommand{\varxi}{ \mathcal{\xi}^{(d)} }
\newcommand{\varxiprime}{ \mathcal{\xi}^{(d)\prime} }

\newcommand{\Sdnhat}{S^{(d)}(\nhat)}

\newcommand{\gammad}{\gamma^{(d)}}
\newcommand{\varthetad}{\vartheta^{(d)}}

\newcommand{\varsigmapmd}{\varsigma^{\pm(d)}}
\newcommand{\varsigmaoned}{\varsigma^{1(d)}}
\newcommand{\varsigmatwod}{\varsigma^{2(d)}}
\newcommand{\varsigmathreed}{\varsigma^{3(d)}}

\newcommand{\phiz}{\Phi^{(d)}_{\zi}}
\newcommand{\cossqphiz}{\cos^2(\phiz)}
\newcommand{\sinsqphiz}{\sin^2(\phiz)}
\newcommand{\costwophiz}{\cos(2\phiz)}
\newcommand{\sintwophiz}{\sin(2\phiz)}
\newcommand{\eivarxip}{e^{i\varxi}}
\newcommand{\eivarxim}{e^{-i\varxi}}
\newcommand{\etwoivarxip}{e^{2i\varxi}}
\newcommand{\etwoivarxim}{e^{-2i\varxi}}

\newcommand{\Psibm}{{\psi}}
\newcommand{\deltabm}{{\delta}}

\newcommand{\psizprime}{ \Psibm^{(d)\prime}_{\zi} }
\newcommand{\psiz}{ \Psibm^{(d)}_{\zi} }
\newcommand{\psio}{ \Psibm^{(d)}_{\oo} }
\newcommand{\psim}{ \Psibm^{(d)}_{\mm} }

\newcommand{\psioprime}{ \Psibm^{(d)\prime}_{\oo} }
\newcommand{\deltapsiz}{\deltabm \psiz}

\newcommand{\deltaphiz}{\deltabm \phiz}

\newcommand{\eitwodeltapsizp}{e^{2 i \deltapsiz}}
\newcommand{\eitwodeltapsizm}{e^{-2 i \deltapsiz}}
\newcommand{\eitwodeltapsizmp}{e^{\mp2 i \deltapsiz}}

\newcommand{\NN}{\mathcal{N}}
\newcommand{\FF}{\mathcal{F}}
\newcommand{\AAA}{\FF}
\newcommand{\BB}{\mathcal{G}}

\newcommand{\GG}{\BB}

\newcommand{\pol}{{\Pi}}

\newcommand{\polhat}{\hat{\Pi}}
\newcommand{\polmm}{\pol_{m}}
\newcommand{\sigmapolm}{\sigma_{\polmm}}
\newcommand{\polmax}{\pol_{\rm max}}
\newcommand{\pmaxd}{\pol^{(d)}_{\rm max}}

\newcommand{\pz}{\pol_{\zi}}
\newcommand{\pzmax}{{\pz}_{\rm max}}

\newcommand{\pzup}{0.7}

\newcommand{\polbar}{\bar{\pol}}
\newcommand{\sigmapolbar}{\bar{\sigma}_{\pol}}

\newcommand{\uzoverpzprime}{\mathcal{u}^{(d)\prime}_{\zi}}

\newcommand{\sigmap}{\sigma_{\pol}}

\newcommand{\orcidwidth}{0.1in}

\newcommand{\andyinadd}[1]{\textcolor{blue}{#1}}

\newcommand{\sign}{{\rm sign}}

\newcommand{\blazars}{Blazars}

\newenvironment{psmallmatrix}
  {\left(\begin{smallmatrix}}
  {\end{smallmatrix}\right)}

\newcommand{\nagn}{1278}  
\newcommand{\nsources}{\nagn}
\newcommand{\nobs}{7554}
\newcommand{\nobsangle}{7376}

\newcommand{\dfourbetterfactor}{\ensuremath{35}}
\newcommand{\dfivebetterfactor}{\ensuremath{10}}
\newcommand{\dfourbettervol}{\ensuremath{15}}
\newcommand{\dfivebettervol}{\ensuremath{16}}
\newcommand{\dfourworsefactorspec}{\ensuremath{2}} 
\newcommand{\dfiveworsefactorspec}{\ensuremath{12}} 

\newcommand{\MM}{N_s}

\renewcommand{\prl}{Phys. Rev. Lett.}
\renewcommand{\rmp}{Rev. Mod. Phys.}
\newcommand{\plb}{Phys. Lett. B}

\newcommand{\mnras}{Mon. Not. R. Astron Soc.}

\newcommand{\aap}{Astron. Astrophys.}
\newcommand{\jcap}{J. Cos. Astropart. Phys.}
\newcommand{\apjl}{Astrophys. J. Lett.}
\newcommand{\apjs}{Astrophys. J. Suppl. Ser.}

\newcommand{\aj}{Astron. J.}

\newcommand{\physrep}{Phys. Rep.}
\newcommand{\app}{Astropart. Phys.}

\newcommand{\aipcs}{Amer. Inst. Phys. Conf. Ser.}

\newcommand{\jpcs}{Jour. Phys. Conf. Ser.}

\newcommand{\mpla}{Mod. Phys. Lett. A}

\newcommand{\cpc}{Chin. Phys. C}

\newcommand{\natp}{Nature Phys.}

\newcommand{\procspie}{Proc. SPIE}

\newcommand{\aapr}{Astron. Astrophys. Rev.}

\newcommand{\actaa}{Acta Astronomica}

\newcommand{\nar}{New Astron. Rev.}

\newcommand{\aaps}{Astron. Astrophys. Suppl. Ser.}

\begin{document}

\title{Improved Constraints on Anisotropic Birefringent Lorentz Invariance and CPT Violation from Broadband Optical Polarimetry of High Redshift Galaxies}

\author{Andrew~S.~Friedman 
\href{https://orcid.org/0000-0003-1334-039X}{\includegraphics[width=\orcidwidth]{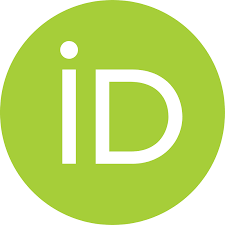}}}
\email{asf@ucsd.edu}
\affiliation{Center for Astrophysics and Space Sciences, University of California, San Diego, La Jolla, California 92093, USA}

\author{Roman~Gerasimov}
\email{romang@ucsd.edu}
\affiliation{Center for Astrophysics and Space Sciences, University of California, San Diego, La Jolla, California 92093, USA}

\author{David~Leon}
\email{dleon@physics.ucsd.edu}
\affiliation{Center for Astrophysics and Space Sciences, University of California, San Diego, La Jolla, California 92093, USA}

\author{Walker~Stevens}
\email{wstevens@physics.ucsd.edu}
\affiliation{Center for Astrophysics and Space Sciences, University of California, San Diego, La Jolla, California 92093, USA}

\author{David~Tytler}
\email{dtytler@physics.ucsd.edu}
\affiliation{Center for Astrophysics and Space Sciences, University of California, San Diego, La Jolla, California 92093, USA}

\author{Brian~G.~Keating
\href{https://orcid.org/0000-0003-3118-5514}{\includegraphics[width=\orcidwidth]{./figs/orcid.png}}}
\email{bkeating@physics.ucsd.edu}
\affiliation{Center for Astrophysics and Space Sciences, University of California, San Diego, La Jolla, California 92093, USA}

\author{Fabian~Kislat
\href{https://orcid.org/0000-0001-7477-0380}{\includegraphics[width=\orcidwidth]{./figs/orcid.png}}}
\email{fabian.kislat@unh.edu}
\affiliation{Department of Physics \& Astronomy and Space Science Center, University of New Hampshire, Durham, NH 03824, USA}

\date{\today}

\begin{abstract}
In the framework of the Standard Model Extension (SME), we present improved constraints on anisotropic Lorentz invariance and Charge-Parity-Time (CPT) violation by searching for astrophysical signals of cosmic vacuum birefringence with broadband optical polarimetry of high redshift astronomical sources, including Active Galactic Nuclei and Gamma-Ray Burst afterglows. We generalize Ref.~\cite{kislat18}, which studied the SME mass dimension $d=4$ case, to arbitrary mass dimension for both the CPT-even and CPT-odd cases. We then present constraints on all 10, 16, and 42 anisotropic birefringent SME coefficients for dimension $d=4$, $d=5$, and $d=6$ models, respectively, using \nobs{} observations for odd $d$ and \nobsangle{} observations for even $d$ of \nsources{} unique sources on the sky, which, to our knowledge, comprises the most complete catalog of optical polarization from extragalactic sources in the literature to date. Compared to the smaller sample of $44$ and $45$ broadband optical polarimetry observations analyzed in Refs.~\cite{kislat18} and \cite{kislat17}, our dimension $d=4$ and $d=5$ average constraints are more sensitive by factors of \dfourbetterfactor{} and \dfivebetterfactor{}, corresponding to a reduction in allowed SME parameter space volume for these studies of \dfourbettervol{} and \dfivebettervol{} orders of magnitude, respectively. Constraints from individual lines of sight can be significantly stronger using spectropolarimetry, due to the steep energy dependence of birefringence effects at increasing mass dimension. Nevertheless, due to the increased number of observations and lines of sight in our catalog, our average $d=4$ and $d=5$ broadband constraints are within factors of \dfourworsefactorspec{} and \dfiveworsefactorspec{} of previous constraints using spectropolarimetry from Refs.~\cite{kislat18} and Ref.~\cite{kislat17}, respectively, using an independent data set and an improved analysis method. By contrast, our anisotropic constraints on all 42 birefringent SME coefficients for $d=6$  are the first to be presented in the literature.
\end{abstract}

\pacs{\bf{Valid PACS appear here}}


\maketitle

\section{Introduction}
\label{sec:intro}

Special relativity and the Standard Model of particle physics obey the symmetries of Lorentz and Charge-Parity-Time (CPT) invariance, which various tests over the past century indicate are obeyed in nature to high precision \cite{kostelecky11}. However, many theoretical approaches seeking to unify quantum theory and general relativity within an underlying theory of quantum gravity predict that Lorentz and CPT invariance may be broken at energies approaching the Planck scale $E_p = \sqrt{c^5 \hbar / G} = \ten{1.22}{19} \Gev$, perhaps due to extra spatial dimensions or the underlying quantized nature of spacetime \cite{myers03,rizzo05,amelino15}. Several well known candidate quantum gravity models including String Theory \cite{kostelecky89}, warped brane worlds \cite{burgess02}, loop quantum gravity \cite{gambini99}, Ho\v{r}ava-Lifshitz gravity \cite{pospelov10}, and Chern-Simons gravity \cite{lim09}, can all lead to Lorentz invariance violation (LIV) or CPT violation (CPTV).

While the Standard Model of particle physics has been remarkably successful, it does not include gravity, dark matter, or dark energy, and thus cannot be the final theory of nature. The failure of the CERN Large Hadron Collider (LHC) to detect evidence of supersymmetry \cite{tanabashi18} --- or any new physics beyond the Standard Model --- has challenged several candidate quantum gravity theories, including String Theory \cite{ishak18}. There is thus a desperate need for experimental input. It has long been known that symmetries such as Lorentz and CPT invariance --- which are taken as axioms in the Standard Model --- may not be true symmetries in nature at a variety of energy scales \cite{mattingly05}. High energy physicists have therefore routinely searched for LIV and CPTV, for example, in Fermilab neutrino experiments and \cite{adamson10,adamson12}, and various LHC tests \cite{aaij16,chanon19}. However, searching for such physics beyond the Standard Model with conventional particle accelerators continues to require progressively larger energy scales that are rapidly becoming unfeasible.

All of this motivates novel astroparticle physics experiments that leverage the vast distances, timescales, and energy scales of the cosmos itself to look for signatures 
of quantum gravity and to constrain, or rule out, alternatives to the Standard Model. Using the universe as a laboratory ultimately enables searches for exotic physical effects which would likely be impossible to detect with experiments on Earth. Since such approaches are far less explored than terrestrial tests, this represents a huge untapped opportunity.

Since the relevant energies are not accessible to any foreseeable Earth-bound tests, most astrophysical tests use observations of extragalactic sources to exploit small effects that may accumulate to detectable levels over cosmological distances and timescales \cite{kostelecky09,kislat17,friedman19b}. Still, since no strong evidence yet exists for LIV or CPTV in nature, some models have already been effectively ruled out \cite{kostelecky11}. However, since the full parameter space is largely unconstrained, astrophysical observations of cosmological sources at broader wavelength ranges, higher redshifts and energies, and varied positions on the sky, represent ideal data to constraint LIV/CPTV effects in our universe.

The Standard Model Extension (SME) is an exhaustive and general effective field theory framework for constraining new physics beyond the Standard Model, including LIV and CPTV effects (See \cite{kostelecky09} for a review). While others have considered LIV and CPTV tests in the SME (and other frameworks) for massive particles like cosmic rays \cite{scully09,stecker10,bietenholz11,cowsik12,lang17} and neutrinos \cite{jacob07,jacob08,chakraborty13,stecker14,amelino15}, in this work, we consider only LIV and CPTV in the photon sector. In addition, this paper focuses exclusively on astrophysical SME tests, although see \cite{kostelecky11} for a review of SME constraints from various laboratory and other tests. 

SME models are typically ordered and labelled by the mass dimension $d \geq 3$ of the relevant operator in the expansion of terms that modify the Standard Model Lagrangian to incorporate Lorentz invariance and/or CPT violation \cite{kostelecky09}. Nonzero coefficients in the SME expansion can yield a modified vacuum dispersion relation for photons and ``vacuum birefringence". A modified vacuum dispersion relation would mean that the speed of light became energy dependent, which would cause a time delay (or early arrival) for promptly emitted photons of different energies \cite{jacob08,kostelecky08}. Vacuum birefringence for $d > 3$ refers to an energy dependent rotation of the plane of linear polarization for photons emitted promptly with the same initial polarization angle. We do not consider circular polarization in this work.

Constraints on models with vacuum dispersion from LIV can be obtained from astronomical observations of time delays from astronomical sources at higher redshifts and energies \cite{jacob07,jacob08,kostelecky08,kostelecky09}. However, since optical time delay constraints on vacuum dispersion SME models are not competitive with high time resolution $\gamma$-ray observations of GRBs \cite{amelino98,boggs04,ellis06,rodriguez06,kahniashvili06,biesiada07,xiao09,laurent11,stecker10,stecker11,toma12,kostelecky13,vasileiou13,pan15,zhangs15,kislat15,chang16,lin16,wei17} or TeV flares from \blazars{}  \cite{biller99,albert08,aharonian08,kostelecky08,shao10,tavecchio16}, this work does not employ time delay studies. 

Rather, we focus on constraining vacuum birefringent SME models, which can be tested with much higher sensitivity using broadband polarimetry \cite{kostelecky09}. We further focus only on linear polarization, since the observed circular polarization is often consistent with zero for the high redshift sources of interest (e.g.~\cite{hutsemekers10,matsumiya03,sagiv04,toma08}) and there is insufficient circular polarization data in the literature to meaningfully constrain any circular polarization induced by vacuum birefringence. 

The tests we perform in this work do not seek to directly detect positive evidence of Lorentz or CPT violation in the universe. Rather, we assume the null hypothesis that the Standard Model is correct, and we seek to constrain how large any LIV or CPTV effects could be, in the framework of the SME, given the observed data. Our constraints are therefore presented as upper bounds on the relevant SME coefficients. As such, while this approach is explicitly designed to progressively rule out increasingly larger sectors of the SME parameter space, different approaches would be required if the aim was instead to potentially detect non-zero signals of Lorentz invariance and/or CPT violation with astrophysical observations.

To date, astrophysical observations have primarily been used to constrain models using measurements along individual lines-of-sight, including ``vacuum isotropic" models with a single SME coefficient over the whole sky, and linear combinations of anisotropic SME coefficients \cite{kostelecky01,kostelecky02,kostelecky06,kostelecky09,kostelecky13,friedman19b}. However, the most general SME models are anisotropic, where LIV and CPTV effects can vary with direction on the sky. As such, these models require astrophysical observations along many independent lines-of-sight to fully constrain all the parameters for a given SME model \cite{kislat15,kislat17,kislat18}. 

Ultimately, astronomical polarimetry can constrain birefringent SME effects which would increasingly suppress the observed polarization of intrinsically more highly linearly polarized cosmological sources via an energy-dependent drift in polarization angle. 
In this work, we present new and more sensitive SME constraints on anisotropic Lorentz invariance violation and CPT violation than those found using only the sample of broadband optical polarimetry of high redshift sources, including Active Galactic Nuclei (AGN) and the optical afterglows of Gamma-Ray Bursts (GRBs), that were analyzed in previous work \cite{kislat17,kislat18}.

The recent work in Refs.~\cite{kislat15,kislat17,kislat18} was the first to constrain all SME coefficients for various anisotropic models. While Ref.~\cite{kislat15} was the first to constrain all 25 non-birefringent $d=6$ SME coefficients using $\gamma$-ray time delay studies of AGN observed by Fermi-LAT, in this work, we restrict our analysis to constraining birefringent SME coefficients.  Subsequently, Refs.~\cite{kislat17,kislat18},
were the first studies to constrain all 16 (10) birefringent SME coefficients for $d=5$ ($d=4$) SME models using a small sample of archival optical polarimetry and spectropolarimetry.

\begin{figure*}
\centering
\begin{tabular}{@{}c@{}c@{}}

\includegraphics[width=4.0in]{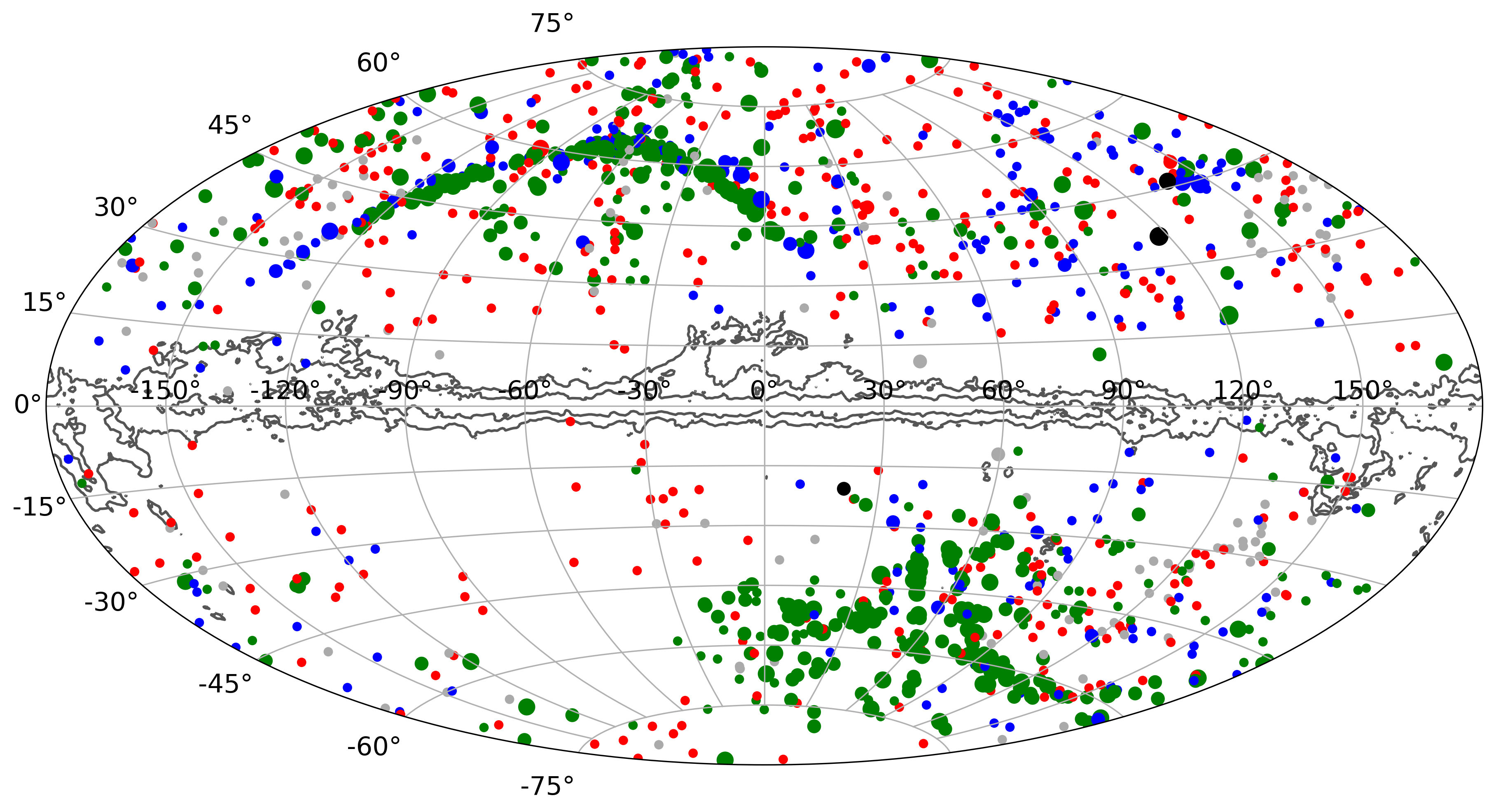} &
\includegraphics[width=3.0in]{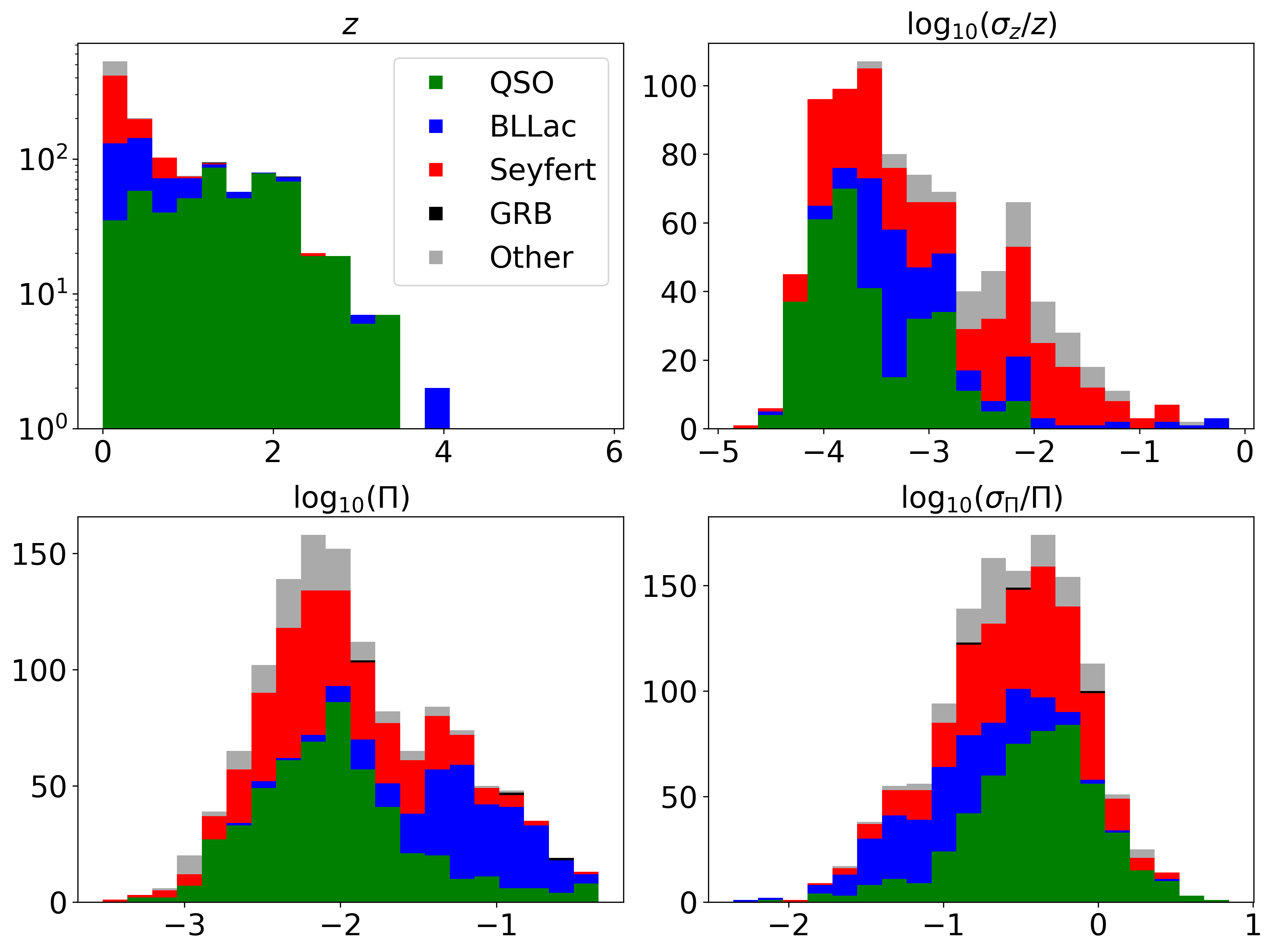} \\
\vspace{-1.0cm}
\end{tabular}
\caption{
\baselineskip 11pt (\textit{Left}) Sky catalog Aitoff projection in galactic coordinates of \nagn{} AGN and GRBs with broadband optical polarimetry
\cite{Steele_2017,Hovatta_2016,Pavlidou_2014,Heidt_2011,Angelakis_2018,Kumar_2018,Borguet_2008,Smith_2002,Tadhunter_2002,Jones_2012,Almeida_2016,Gorosabel_2014,Brindle_1986,Brindle_1990a,Brindle_1990b,Brindle_1991,Martin_1983,Cimatti_1993,Angelakis_2016,Itoh_2016,Sluse_2005,Wills_2011,Hutsemekers_2017}. The Milky Way is shown with gray contours of optical color excess $E(B-V)=0.7$ and $2.0$ from the Ref.~\cite{schlegel98} galactic reddening map (\url{https://lambda.gsfc.nasa.gov/product/foreground/fg_sfd_get.cfm}). Plot symbol size increases with redshift. Plot colors indicate object type from the \texttt{Simbad} database: quasars=QSO (green), BL Lac (blue), Seyfert (red), GRB optical afterglows (black), and Other/Unknown (gray). (\textit{Right}) For these \nagn{} objects, we show histograms, with same color coding by object type, of the key inputs to test
anisotropic birefringent SME models with broadband optical polarimetry: redshift $z$ (\textit{upper left}), and the $\log_{10}$ of: the fractional redshift error $(\sigma_z/z)$ (\textit{upper right}), the maximum linear polarization fraction $\pol$ (\textit{lower left}), and the mean fractional polarization error $(\sigmap/\pol)$ (\textit{lower right}). The observed polarization angle, as shown for our catalog in \fig{}~\ref{fig:angles_catalog}, is also needed for the CPT-even case, but not for the CPT-odd case.
\vspace{\dskips}
}
\label{fig:catalog}
\end{figure*}

While Refs.~\cite{kislat17} (\cite{kislat18}) analyzed a preliminary set of less than 100 AGN and GRB afterglows, thousands of AGN have broadband optical polarization data 
in the literature 
(e.g.~\cite{sluse05,smith07,wills11,goyal12,itoh16,jermak16b,angelakis16,hutsemekers17,marscher17}), and hundreds have published spectropolarimetry (e.g. \cite{schmidt92,sluse05,smith09}). See Fig.~\ref{fig:catalog} for sky coverage and histograms of a broadband polarimetry database that we have compiled of \nsources{} highly polarized AGN and GRB afterglows with linear polarization fraction $p \gtrsim 2\%$ and redshift $z < 3.5$. This work thus aims to significantly improve upon the broadband only analyses in Refs.~\cite{kislat17,kislat18} by analyzing more than an order of magnitude more individual sources and over two orders of magnitude more individual observations, and by also including multiple observations of each source, where available, to improve our constraints.  

\begin{figure}[h!]
    \centering
    \includegraphics[width=0.99\columnwidth]{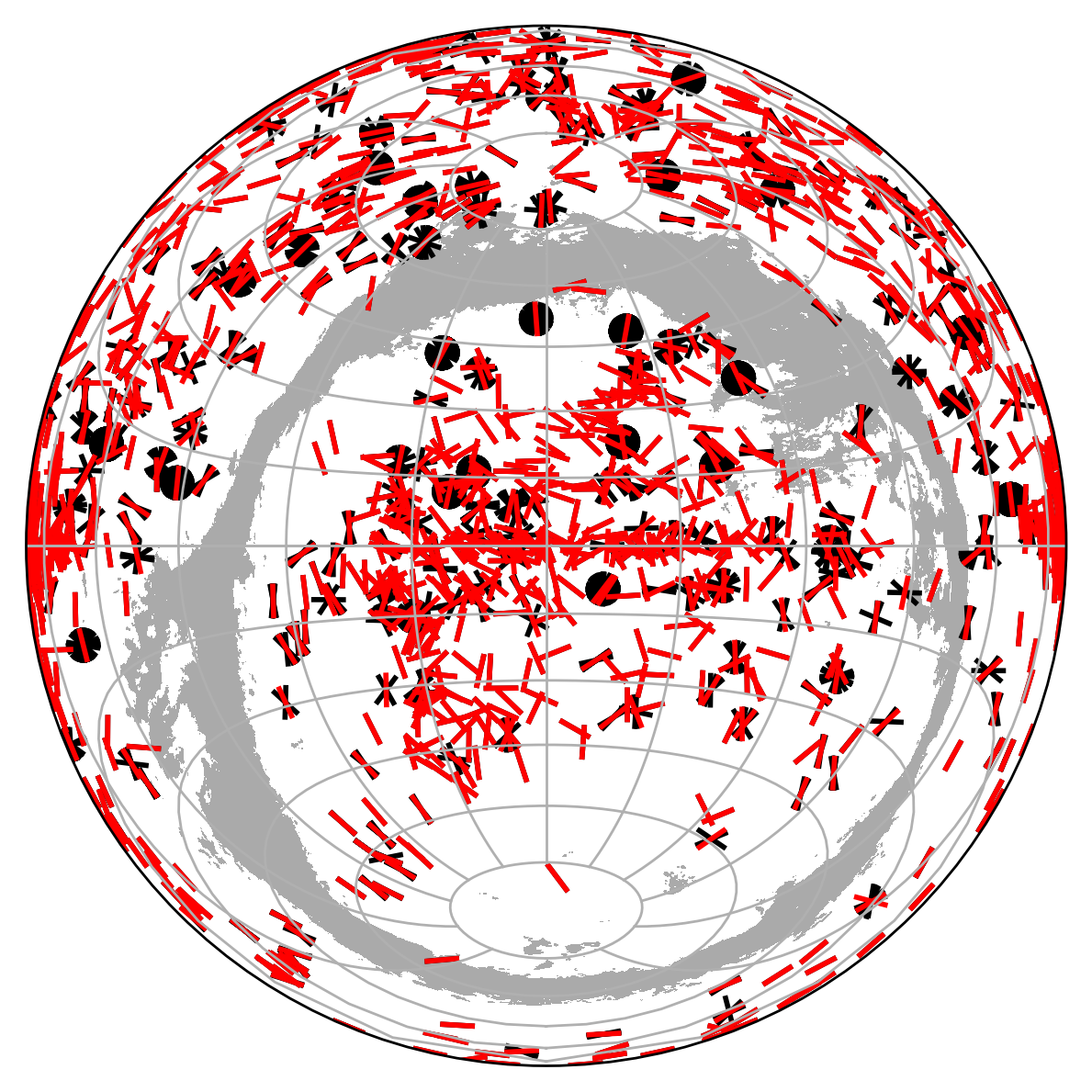}
    \vspace{\dskips}
    \vspace{\dskips}
    \caption{Polarization angle measurements from the compiled catalog of extragalactic sources in ICRS J2000 equatorial coordinates using a Lambert azimuthal projection centered at the vernal equinox. Black strokes represent all available polarization angles, including cases of multiple measurements per line of sight. Red strokes are averages for each unique line of sight. Note that while some sources appear stable, others undergo rapid rotation spanning the entire range of possible angles (black circles). Polarization angles serve as a probe of the direction of the birefringence axis and, therefore, must be measured to constrain CPT-even SME cases, where said direction is not known \textit{a priori}. The apparent gap in the data encircling the center of the projection is due to the galactic equator, where foregrounds render extra-galactic observations extremely challenging. Parts of the sky with $E(B-V)>0.5$ are shaded in gray to display the band of the Milky Way, based on the same Ref.~\cite{schlegel98} reddening map used in \fig{}~\ref{fig:catalog}. At the center of the projection, North is up and East is right.
    }
    \label{fig:angles_catalog}
\end{figure}

While Refs.~\cite{kislat15,kislat17} used a linear least squares approach to upper bound the relevant SME coefficients, Ref.~\cite{kislat18}, which focused on the CPT-even $d=4$ birefringent case, developed a more principled approach that uses Markov Chain Monte Carlo (MCMC) methods to compute the posterior probability distribution of birefringent SME coefficients, given the observed data. In this work, we extend and refine the Ref.~\cite{kislat18} analysis method to arbitrary mass dimension $d$, including both the CPT-even and CPT-odd cases, and present constraints, using only broadband optical polarimetry, which significantly improve upon the broadband-only constraints in Refs.~\cite{kislat17,kislat18}.

This paper is organized as follows. In \S\ref{sec:back}-\ref{sec:cosmo}, we provide the relevant theoretical background for photon sector tests in the SME, including cosmological effects. Secs.~\S\ref{sec:stokessme}-\ref{sec:stokes} describe how SME polarization angle drift from LIV or CPTV induced birefringence corresponds to changes in Stokes parameters from the source to the observer. In \S\ref{sec:broad}, we detail our method for constraining LIV and CPTV effects using broadband polarimetry, while \S\ref{sec:constrain} outlines the assumptions underlying our MCMC analysis of SME parameters. Sec.~\S\ref{sec:catalog} describes the archival catalog of broadband optical AGN polarimetry analyzed in this paper, with further details in Appendices~\ref{sec:catalog1}, and~\ref{sec:catalog2}. In \S\ref{sec:constraints}, we present our constraints on all 16, 10, and 42 birefringent SME coefficients for mass dimensions $d=4$, $5$, and $6$, respectively. Sec.~\S\ref{sec:sys} addresses systematic uncertainties. Further discussion and conclusions are presented in \S\ref{sec:disc}.

\section{Background: Cosmic Birefringence in the SME}
\label{sec:back}

In natural units with $c=\hbar=1$, the photon vacuum dispersion relation in the SME is given by \cite{kostelecky09}
\begin{equation}
E \simeq p \Big(1 - \varsigmazero \pm \sqrt{(\varsigmaone)^2 + (\varsigmatwo)^2 + (\varsigmathree)^2} \Big)\, ,
\label{eq:SMEphotdisp}    
\end{equation}
\noindent where $E$ is the energy, $p$ is the momentum, and the various $\varsigma^{(x)}$ represent the new terms in the SME expansion, which vanish identically in the Standard Model. Following the notation and phase conventions in Ref.~\cite{kostelecky09}, using an expansion of spin-weighted spherical harmonics $\sYjm$ and operator mass dimension $d$, 
\begin{eqnarray}
\varsigmazero     & = & \sum_{\substack{djm \\ \ d \ {\rm even}}} E^{d-4} \oYjm(\nhat) \cdIjm \, , \label{eq:varsigmas0}    \\
\varsigmapm & = & \varsigmaone \mp i \varsigmatwo \nonumber \\
 & = & \sum_{\substack{djm, \\ d \ {\rm even}}} E^{d-4} \twoYjm(\nhat)   \Big(\kdEjm \pm i \kdBjm\Big)\, , \label{eq:varsigmaspm}   \\
\varsigmathree     & = & \sum_{\substack{djm, \\ d \ {\rm odd}}} E^{d-4} \oYjm(\nhat) \kdVjm\, , \label{eq:varsigmas3}   
\end{eqnarray}
where $\nhat = (\textrm{RA},\textrm{Dec})$
are the ICRS J2000 spherical polar coordinates in the direction of the astrophysical source.\footnote{In this work, we use $\textrm{RA}$ and $\textrm{Dec}$ to refer to the ICRS J2000 right ascension and declination; however, any consistent spherical polar coordinate system may be adopted for this purpose.} In the CPT-odd case (odd $d$), there are $(d-1)^2$ vacuum birefringent SME coefficients $\kdVjm$. For the CPT-even case (even $d$), there are $(d-1)^2$ non-birefringent SME  coefficients $\cdIjm$ that are uniquely constrained by time-delay measurements, and $(d-1)^2-4$ birefringent SME coefficients for each of $\kdEjm$ and $\kdBjm$. Overall, the CPT-even vacuum birefringent SME parameters $\kdEjm$, $\kdBjm$, and vacuum dispersion parameters $\cdIjm$ characterize CPT-preserving LIV, while the vacuum birefringent CPT-odd parameters $\kdVjm$ also lead to CPTV \cite{kostelecky09}. 

For all SME models, the sum in Eqs.~\eqref{eq:varsigmas0}-\eqref{eq:varsigmas3} runs over mass dimension $d$ from $d=3$ or $d=4$ to $\infty$, (with $d$ even or odd as indicated) accounting for all possible LIV or CPTV contributions in the SME framework. However, in this work, we will only consider the case of arbitrary fixed values of mass dimension,
e.g. $d=4$, $d=5$, or $d=6$. For any model that could produce operators with multiple values of $d$, the dominant contribution would be predicted to come from the leading order term, so it is reasonable to consider only fixed values of $d$ for this work. For any mass dimension $d$, the spherical harmonic indices $j$ and $m$ run over the following ranges
\begin{eqnarray}
-j \leq m \leq j \text{ , }
\begin{cases}
    j \in 0,1,\ldots,d-2, & \text{ odd } d\, , \\  
    j \in 2,3,\ldots,d-2, & \text{ even } d\, .  
\end{cases}
\label{eq:jm}
\end{eqnarray}
Eq.~\eqref{eq:jm} shows that vacuum isotropic $j=m=0$ models containing a single SME coefficient over the whole sky exist only in the CPT-odd case. As such, CPT-even models are of particular interest because they are, by definition, \textit{anisotropic}.

At fixed mass dimension $d$, the birefringent SME coefficients can be written 
\begin{alignat}{3}
    \varsigmathreed  &    = E^{d-4}   \sum_{jm} \oYjm(\nhat) \kdVjm\,, && \text{ odd } d\, , \label{eq:varsigmas3d}  \\
     \varsigmapmd 
     & =  E^{d-4}  \sum_{jm} \twoYjm(\nhat)   \Big(\kdEjm \pm i \kdBjm\Big),                 && \text{ even } d\, , \label{eq:varsigmaspmd} 
\end{alignat}
where $\varsigmapmd = \varsigmaoned \mp i \varsigmatwod$. The convenience of converting the SME coefficients into this complex spin-weighted basis will become apparent shortly.
Using the following parity relations (where $^*$ denotes complex conjugation) for the spherical harmonics 
\begin{alignat}{3} 
    \oYjmminus   & =  (-1)^m (\oYjm)^* && \text{ odd } d\, ,
    \label{eq:parityodd} \\
    \twoYjmminus  & =  (-1)^{m \mp 2} (\twompYjm)^* && \text{ even } d\, ,    
    \label{eq:parityeven}  
\end{alignat}
and birefringent SME coefficients
\begin{alignat}{3} 
    \kdVjmminus  & = (-1)^m \Big(\kdVjm\Big)^*,   && \text{ odd } d\, , \label{eq:parityoddk} \\
    \kdEBjmminus  & = (-1)^{m} \Big(\kdEBjm\Big)^*,  && \text{ even } d\, , \label{eq:parityevenk}  
\end{alignat}
yields $N(d)$ unique real components for each mass dimension $d$ given by
\begin{eqnarray}
N(d)=
\begin{cases}
    (d-1)^2\, ,       & \text{ odd } d\, , \\ 
    2(d-1)^2 - 8\, ,  & \text{ even } d\, . 
\end{cases}
\label{eq:Nd}
\end{eqnarray}
Therefore, there are a total of $N(d)=4,10,16,42,36,90,64,154,\ldots$ birefringent SME coefficients for $d=3,4,5,6,7,8,9,10,\ldots$. If the number of sources $\MM < N(d)$, one can only constrain linear combinations of the relevant SME coefficients. To constrain all $N(d)$ parameters for a given $d$, astrophysical studies must therefore observe $\MM > N(d)$ sources along different lines-of-sight. This work compiles and analyzes the largest such database to date, including $\MM=\nsources{}$ sources, with $\MM \gg N(d)$ for all mass dimensions $d=4$, $5$, and $6$ considered here. Vacuum birefringence for $d=3$ in the SME is energy-independent and cannot be studied using our approach. However, see \cite{kostelecky08} for $d=3$ constraints using observations of the Cosmic Microwave Background (also see Refs.~\cite{kostelecky07,kostelecky08,kostelecky09,komatsu09,gubitosi09,kahniashvili09,kaufman16a,leon17}).

\section{Cosmology}
\label{sec:cosmo}

For a fixed mass dimension, the effective comoving distance $\Ldz$ traveled by the photons over cosmological distances is 
\begin{eqnarray}
\Ldz = \int_{0}^{z_{\sourceindex}} \frac{ (1+\zz)^{d-4} }{ H(\zz) }d\zz =  \int_{a_{\sourceindex}}^{1} \frac{d\aaa}{ (\aaa)^{d-2} H(\aaa)} \, ,
\label{eq:Lz}
\end{eqnarray}
which includes the relevant cosmological effects in an expanding universe \cite{jacob08,kostelecky09}. Setting $d=4$ in Eq.~\eqref{eq:Lz} recovers the usual expression for comoving distance. In Eq.~(\ref{eq:Lz}), $H(z)=H(a)$ is the Hubble expansion rate at a redshift $z_{\sourceindex}$ with scale factor $a_{\sourceindex}^{-1}=1+z_{\sourceindex}$ (with the usual normalization $a(\tnot)=1$ at the present cosmic time $t=\tnot$ at $z=0$) given by
\begin{eqnarray}
H(a) = \Ho \Big[\orad a^{-4} + \om a^{-3} + \ok a^{-2} + \olam \Big]^{1/2}\, ,
\label{eq:Hz}
\end{eqnarray}
in terms of the present day Hubble constant, which we fix to $\Ho=67.66$ \Hounits{} and best fit cosmological parameters for matter $\om = 0.3111$, radiation 
$\orad = \om/(1+z_{\rm eq}) =\ten{9.182}{-5}$ (with the matter-radiation equality redshift $z_{\rm eq}=3387$), vacuum energy $\olam = 0.6889$, and curvature $\ok = 1 - \orad - \om - \olam \approx 0$ using the \textit{Planck} satellite 2018 data release \cite{planck18}.\footnote{
We use cosmological parameters reported in Table 2 column 7 of \cite{planck18} and assume zero uncertainties. These are the joint cosmological constraints (TT,TE,EE+lowE+lensing+BAO 68\% limits). However, based on recent tension between the Hubble constant $\Ho$ determined using CMB data and distance ladder measurements from Type Ia supernovae (SN Ia), we note that even if we used the SN Ia Hubble constant 
$\Ho=73.48$ \Hounits \cite{riess18b} rather than $\Ho=67.66$ \Hounits from Table 2 column 7 of Planck 2018 \cite{planck18}, and include $2$-$\sigma$ uncertainties on the cosmological parameters, it would have a negligible effect on the final numerical values of our SME coefficient constraints.}

\section{Stokes Parameters in the SME}
\label{sec:stokessme}

The Stokes parameters $I,Q,U$, and $V$ completely describe the general elliptical polarization of light, where $I$ is the intensity, $Q$ and $U$ describe linear polarization (with relative angle $45^{\circ}$), and $V$ describes circular polarization. Since circular polarization is generally measured to be small, and is expected to be intrinsically small for the cosmological sources of interest at optical wavelengths, including AGN (e.g. \cite{hutsemekers10}) and GRBs (e.g. \cite{matsumiya03,sagiv04,toma08}; although see \cite{wiersema14}), we assume $\Vdz=0$ at the source at redshift $z$ for a dimension $d$ SME model throughout the remainder of this work. Furthermore, due to the scarcity of extragalactic circular polarization measurements in the literature, we only search for SME effects in linear polarization measurements and neglect any non-zero observed values of $V$ that may have been induced by vacuum birefringence.

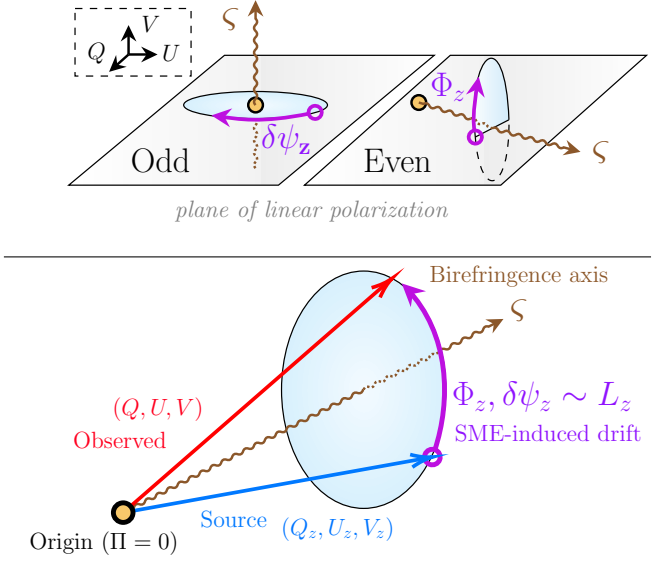
\begin{figure}[ht]
\centering
\resizebox{\columnwidth}{!}{

  
\tikzset {_xmffkuwrh/.code = {\pgfsetadditionalshadetransform{ \pgftransformshift{\pgfpoint{0 bp } { 0 bp }  }  \pgftransformrotate{0 }  \pgftransformscale{2 }  }}}
\pgfdeclarehorizontalshading{_dkvfxudt3}{150bp}{rgb(0bp)=(1,1,1);
rgb(37.5bp)=(1,1,1);
rgb(50bp)=(0.95,0.95,0.95);
rgb(50.25bp)=(0.93,0.93,0.93);
rgb(62.5bp)=(1,1,1);
rgb(100bp)=(1,1,1)}

  
\tikzset {_sl7oj9ter/.code = {\pgfsetadditionalshadetransform{ \pgftransformshift{\pgfpoint{0 bp } { 0 bp }  }  \pgftransformrotate{0 }  \pgftransformscale{2 }  }}}
\pgfdeclarehorizontalshading{_d9wk1dihd}{150bp}{rgb(0bp)=(1,1,1);
rgb(37.5bp)=(1,1,1);
rgb(50bp)=(0.95,0.95,0.95);
rgb(50.25bp)=(0.93,0.93,0.93);
rgb(62.5bp)=(1,1,1);
rgb(100bp)=(1,1,1)}

  
\tikzset {_njlisp1py/.code = {\pgfsetadditionalshadetransform{ \pgftransformshift{\pgfpoint{0 bp } { 0 bp }  }  \pgftransformrotate{0 }  \pgftransformscale{2 }  }}}
\pgfdeclarehorizontalshading{_625afopd7}{150bp}{rgb(0bp)=(1,1,1);
rgb(37.5bp)=(1,1,1);
rgb(62.5bp)=(0.82,0.92,0.98);
rgb(100bp)=(0.82,0.92,0.98)}

  
\tikzset {_e775ey2ur/.code = {\pgfsetadditionalshadetransform{ \pgftransformshift{\pgfpoint{0 bp } { 0 bp }  }  \pgftransformrotate{0 }  \pgftransformscale{2 }  }}}
\pgfdeclarehorizontalshading{_ya3s2n0u1}{150bp}{rgb(0bp)=(1,1,1);
rgb(37.5bp)=(1,1,1);
rgb(62.5bp)=(0.82,0.92,0.98);
rgb(100bp)=(0.82,0.92,0.98)}

  
\tikzset {_zg7970hrg/.code = {\pgfsetadditionalshadetransform{ \pgftransformshift{\pgfpoint{0 bp } { 0 bp }  }  \pgftransformscale{1 }  }}}
\pgfdeclareradialshading{_n8u2qydpa}{\pgfpoint{0bp}{0bp}}{rgb(0bp)=(0.95,0.98,1);
rgb(0bp)=(0.95,0.98,1);
rgb(25bp)=(0.84,0.94,0.99);
rgb(400bp)=(0.84,0.94,0.99)}
\tikzset{every picture/.style={line width=0.75pt}} 

\begin{tikzpicture}[x=0.75pt,y=0.75pt,yscale=-1,xscale=1]

\path  [shading=_dkvfxudt3,_xmffkuwrh] (402.74,559.39) -- (561.74,559.39) -- (443.65,664.66) -- (284.65,664.66) -- cycle ; 
 \draw   (402.74,559.39) -- (561.74,559.39) -- (443.65,664.66) -- (284.65,664.66) -- cycle ; 

\path  [shading=_d9wk1dihd,_sl7oj9ter] (219.83,559.39) -- (389.74,559.39) -- (263.55,664.66) -- (93.65,664.66) -- cycle ; 
 \draw   (219.83,559.39) -- (389.74,559.39) -- (263.55,664.66) -- (93.65,664.66) -- cycle ; 

\draw [color={rgb, 255:red, 139; green, 87; blue, 42 }  ,draw opacity=1 ][line width=1.5]  [dash pattern={on 1.5pt off 1.5pt}]  (245,650.84) .. controls (243.33,649.17) and (243.33,647.51) .. (245,645.84) .. controls (246.67,644.17) and (246.67,642.51) .. (245,640.84) .. controls (243.33,639.17) and (243.33,637.51) .. (245,635.84) .. controls (246.67,634.17) and (246.67,632.51) .. (245,630.84) .. controls (243.33,629.17) and (243.33,627.51) .. (245,625.84) .. controls (246.67,624.17) and (246.67,622.51) .. (245,620.84) .. controls (243.33,619.17) and (243.33,617.51) .. (245,615.84) .. controls (246.67,614.17) and (246.67,612.51) .. (245,610.84) .. controls (243.33,609.17) and (243.33,607.51) .. (245,605.84) .. controls (246.67,604.17) and (246.67,602.51) .. (245,600.84) -- (245,599.69) -- (245,599.69) ;

\draw [shading=_625afopd7,_njlisp1py]   (423,625.65) .. controls (416.25,557.66) and (447.25,530.66) .. (450.25,611.66) ;

\draw [shading=_ya3s2n0u1,_e775ey2ur]   (423,625.65) -- (450.25,611.66) ;

\draw  [dash pattern={on 4.5pt off 4.5pt}]  (423,625.65) .. controls (427.25,678.66) and (451.25,670.66) .. (450.25,611.66) ;

\draw [color={rgb, 255:red, 139; green, 87; blue, 42 }  ,draw opacity=1 ][line width=1.5]    (436.62,618.66) .. controls (438.66,617.47) and (440.27,617.9) .. (441.46,619.94) .. controls (442.64,621.98) and (444.25,622.41) .. (446.29,621.22) .. controls (448.33,620.03) and (449.94,620.46) .. (451.12,622.5) .. controls (452.3,624.54) and (453.91,624.97) .. (455.95,623.79) .. controls (457.99,622.6) and (459.6,623.03) .. (460.79,625.07) .. controls (461.97,627.11) and (463.58,627.54) .. (465.62,626.35) .. controls (467.66,625.16) and (469.27,625.59) .. (470.45,627.63) .. controls (471.64,629.67) and (473.25,630.1) .. (475.29,628.91) .. controls (477.33,627.73) and (478.94,628.16) .. (480.12,630.2) .. controls (481.3,632.24) and (482.91,632.67) .. (484.95,631.48) .. controls (486.99,630.29) and (488.6,630.72) .. (489.78,632.76) .. controls (490.97,634.8) and (492.58,635.23) .. (494.62,634.04) -- (496.65,634.58) -- (504.38,636.63) ;
\draw [shift={(508.25,637.66)}, rotate = 194.86] [fill={rgb, 255:red, 139; green, 87; blue, 42 }  ,fill opacity=1 ][line width=0.08]  [draw opacity=0] (13.4,-6.43) -- (0,0) -- (13.4,6.44) -- (8.9,0) -- cycle    ;

\draw [color={rgb, 255:red, 139; green, 87; blue, 42 }  ,draw opacity=1 ][line width=1.5]  [dash pattern={on 1.5pt off 1.5pt}]  (419.25,610.66) .. controls (421.46,609.84) and (422.97,610.54) .. (423.79,612.75) .. controls (424.6,614.96) and (426.12,615.66) .. (428.33,614.84) .. controls (430.54,614.02) and (432.06,614.72) .. (432.87,616.93) -- (436.62,618.66) -- (436.62,618.66) ;

\draw [line width=1.5]    (129.31,570.15) -- (142.11,558.71) ;
\draw [line width=1.5]    (161.63,558.71) -- (142.11,558.71) ;
\draw [line width=1.5]    (142.11,538.87) -- (142.11,558.71) ;
\draw [shift={(142.11,534.87)}, rotate = 90] [fill={rgb, 255:red, 0; green, 0; blue, 0 }  ][line width=0.08]  [draw opacity=0] (13.4,-6.43) -- (0,0) -- (13.4,6.44) -- (8.9,0) -- cycle    ;
\draw [shift={(126.32,572.82)}, rotate = 318.21] [fill={rgb, 255:red, 0; green, 0; blue, 0 }  ][line width=0.08]  [draw opacity=0] (13.4,-6.43) -- (0,0) -- (13.4,6.44) -- (8.9,0) -- cycle    ;
\draw [shift={(165.63,558.71)}, rotate = 180] [fill={rgb, 255:red, 0; green, 0; blue, 0 }  ][line width=0.08]  [draw opacity=0] (13.4,-6.43) -- (0,0) -- (13.4,6.44) -- (8.9,0) -- cycle    ;
\draw  [dash pattern={on 4.5pt off 4.5pt}] (98.74,521.44) -- (192.74,521.44) -- (192.74,576.2) -- (98.74,576.2) -- cycle ;
\draw (160.25,533.81) node  [font=\Large]  {$V$};
\draw (116.34,557.53) node  [font=\Large]  {$Q$};
\draw (176.65,557.53) node  [font=\Large]  {$U$};

\draw [color={rgb, 255:red, 189; green, 16; blue, 224 }  ,draw opacity=1 ][line width=2.25]    (422.39,621.3) .. controls (421.37,614.09) and (420.74,607.05) .. (424.95,581.34) ;
\draw [shift={(425.74,576.66)}, rotate = 459.73] [fill={rgb, 255:red, 189; green, 16; blue, 224 }  ,fill opacity=1 ][line width=0.08]  [draw opacity=0] (16.07,-7.72) -- (0,0) -- (16.07,7.72) -- (10.67,0) -- cycle    ;
\draw [shift={(423,625.65)}, rotate = 262.79] [color={rgb, 255:red, 189; green, 16; blue, 224 }  ,draw opacity=1 ][line width=2.25]      (0, 0) circle [x radius= 5.36, y radius= 5.36]   ;
\path  [shading=_n8u2qydpa,_zg7970hrg] (186.75,599.69) .. controls (186.75,593.63) and (212.83,588.72) .. (245,588.72) .. controls (277.17,588.72) and (303.25,593.63) .. (303.25,599.69) .. controls (303.25,605.75) and (277.17,610.66) .. (245,610.66) .. controls (212.83,610.66) and (186.75,605.75) .. (186.75,599.69) -- cycle ; 
 \draw   (186.75,599.69) .. controls (186.75,593.63) and (212.83,588.72) .. (245,588.72) .. controls (277.17,588.72) and (303.25,593.63) .. (303.25,599.69) .. controls (303.25,605.75) and (277.17,610.66) .. (245,610.66) .. controls (212.83,610.66) and (186.75,605.75) .. (186.75,599.69) -- cycle ; 

\draw [color={rgb, 255:red, 189; green, 16; blue, 224 }  ,draw opacity=1 ][line width=2.25]    (214.21,608.76) .. controls (232.44,611.29) and (248.99,614.65) .. (289.43,606.45) ;
\draw [shift={(293.25,605.66)}, rotate = 347.99] [color={rgb, 255:red, 189; green, 16; blue, 224 }  ,draw opacity=1 ][line width=2.25]      (0, 0) circle [x radius= 5.36, y radius= 5.36]   ;
\draw [shift={(209.25,608.11)}, rotate = 6.91] [fill={rgb, 255:red, 189; green, 16; blue, 224 }  ,fill opacity=1 ][line width=0.08]  [draw opacity=0] (16.07,-7.72) -- (0,0) -- (16.07,7.72) -- (10.67,0) -- cycle    ;
\draw  [color={rgb, 255:red, 0; green, 0; blue, 0 }  ,draw opacity=1 ][fill={rgb, 255:red, 248; green, 200; blue, 107 }  ,fill opacity=1 ][line width=1.5]  (371.3,597.66) .. controls (371.3,594.38) and (373.96,591.71) .. (377.25,591.71) .. controls (380.53,591.71) and (383.19,594.38) .. (383.19,597.66) .. controls (383.19,600.94) and (380.53,603.6) .. (377.25,603.6) .. controls (373.96,603.6) and (371.3,600.94) .. (371.3,597.66) -- cycle ;
\draw  [color={rgb, 255:red, 0; green, 0; blue, 0 }  ,draw opacity=1 ][fill={rgb, 255:red, 248; green, 200; blue, 107 }  ,fill opacity=1 ][line width=1.5]  (239.05,599.69) .. controls (239.05,596.41) and (241.71,593.75) .. (245,593.75) .. controls (248.28,593.75) and (250.94,596.41) .. (250.94,599.69) .. controls (250.94,602.97) and (248.28,605.64) .. (245,605.64) .. controls (241.71,605.64) and (239.05,602.97) .. (239.05,599.69) -- cycle ;
\draw [color={rgb, 255:red, 139; green, 87; blue, 42 }  ,draw opacity=1 ][line width=1.5]    (245,599.69) .. controls (243.33,598.02) and (243.33,596.36) .. (245,594.69) .. controls (246.67,593.02) and (246.67,591.36) .. (245,589.69) .. controls (243.33,588.02) and (243.33,586.36) .. (245,584.69) .. controls (246.67,583.02) and (246.67,581.36) .. (245,579.69) .. controls (243.33,578.02) and (243.33,576.36) .. (245,574.69) .. controls (246.67,573.02) and (246.67,571.36) .. (245,569.69) .. controls (243.33,568.02) and (243.33,566.36) .. (245,564.69) .. controls (246.67,563.02) and (246.67,561.36) .. (245,559.69) .. controls (243.33,558.02) and (243.33,556.36) .. (245,554.69) .. controls (246.67,553.02) and (246.67,551.36) .. (245,549.69) .. controls (243.33,548.02) and (243.33,546.36) .. (245,544.69) .. controls (246.67,543.02) and (246.67,541.36) .. (245,539.69) .. controls (243.33,538.02) and (243.33,536.36) .. (245,534.69) .. controls (246.67,533.02) and (246.67,531.36) .. (245,529.69) -- (245,526.66) -- (245,518.66) ;
\draw [shift={(245,514.66)}, rotate = 450] [fill={rgb, 255:red, 139; green, 87; blue, 42 }  ,fill opacity=1 ][line width=0.08]  [draw opacity=0] (13.4,-6.43) -- (0,0) -- (13.4,6.44) -- (8.9,0) -- cycle    ;

\draw [color={rgb, 255:red, 139; green, 87; blue, 42 }  ,draw opacity=1 ][line width=1.5]    (419.25,610.66) .. controls (417.16,611.76) and (415.57,611.27) .. (414.47,609.18) .. controls (413.37,607.09) and (411.78,606.6) .. (409.69,607.7) .. controls (407.6,608.8) and (406.01,608.31) .. (404.92,606.22) .. controls (403.81,604.14) and (402.22,603.65) .. (400.14,604.75) .. controls (398.05,605.85) and (396.46,605.36) .. (395.37,603.27) .. controls (394.27,601.18) and (392.68,600.69) .. (390.59,601.79) .. controls (388.5,602.89) and (386.91,602.4) .. (385.81,600.31) .. controls (384.72,598.22) and (383.13,597.73) .. (381.04,598.83) -- (377.25,597.66) -- (377.25,597.66) ;

\draw (168,644.65) node  [font=\huge] [align=left] {Odd};
\draw (360,644.65) node  [font=\huge] [align=left] {Even};
\draw (401,585) node  [font=\huge,color={rgb, 255:red, 144; green, 19; blue, 254 }  ,opacity=1 ]  {$\Phi _{z}$};
\draw (270,628) node  [font=\huge,color={rgb, 255:red, 144; green, 19; blue, 254 }  ,opacity=1 ]  {$\mathbf{\delta \psi _{z}}$};
\draw (290,685.65) node  [font=\Large] [align=left] {\textit{\textcolor[rgb]{0.5,0.5,0.5}{plane of linear polarization}}};
\draw (525,638) node  [font=\Huge]  {$\textcolor[rgb]{0.55,0.34,0.16}{{\textstyle \mathbf{\varsigma }}}$};
\draw (225,530) node  [font=\Huge]  {$\textcolor[rgb]{0.55,0.34,0.16}{{\textstyle \mathbf{\varsigma }}}$};

\end{tikzpicture}

}
\noindent\rule{0.48\textwidth}{0.4pt}
\resizebox{\columnwidth}{!}{
  
\tikzset {_k3os2pq9s/.code = {\pgfsetadditionalshadetransform{ \pgftransformshift{\pgfpoint{0 bp } { 0 bp }  }  \pgftransformscale{1 }  }}}
\pgfdeclareradialshading{_22g8c6urd}{\pgfpoint{0bp}{0bp}}{rgb(0bp)=(0.95,0.98,1);
rgb(0bp)=(0.95,0.98,1);
rgb(25bp)=(0.84,0.94,0.99);
rgb(400bp)=(0.84,0.94,0.99)}
\tikzset{every picture/.style={line width=0.75pt}} 

\begin{tikzpicture}[x=0.75pt,y=0.75pt,yscale=-1,xscale=1]

\path  [shading=_22g8c6urd,_k3os2pq9s] (254,176.06) .. controls (254,122.45) and (284.05,79) .. (321.12,79) .. controls (358.19,79) and (388.25,122.45) .. (388.25,176.06) .. controls (388.25,229.66) and (358.19,273.11) .. (321.12,273.11) .. controls (284.05,273.11) and (254,229.66) .. (254,176.06) -- cycle ; 
 \draw   (254,176.06) .. controls (254,122.45) and (284.05,79) .. (321.12,79) .. controls (358.19,79) and (388.25,122.45) .. (388.25,176.06) .. controls (388.25,229.66) and (358.19,273.11) .. (321.12,273.11) .. controls (284.05,273.11) and (254,229.66) .. (254,176.06) -- cycle ; 

\draw [color={rgb, 255:red, 139; green, 87; blue, 42 }  ,draw opacity=1 ][line width=1.5]    (124.62,276.56) .. controls (125.35,274.31) and (126.83,273.55) .. (129.08,274.28) .. controls (131.32,275) and (132.8,274.24) .. (133.53,272) .. controls (134.26,269.76) and (135.74,269) .. (137.98,269.73) .. controls (140.22,270.45) and (141.7,269.69) .. (142.43,267.45) .. controls (143.16,265.21) and (144.64,264.45) .. (146.88,265.17) .. controls (149.12,265.9) and (150.6,265.14) .. (151.33,262.9) .. controls (152.06,260.66) and (153.54,259.9) .. (155.78,260.62) .. controls (158.03,261.35) and (159.51,260.59) .. (160.24,258.34) .. controls (160.97,256.1) and (162.45,255.34) .. (164.69,256.07) .. controls (166.93,256.79) and (168.41,256.03) .. (169.14,253.79) .. controls (169.87,251.55) and (171.35,250.79) .. (173.59,251.51) .. controls (175.83,252.24) and (177.31,251.48) .. (178.04,249.24) .. controls (178.77,247) and (180.25,246.24) .. (182.49,246.96) .. controls (184.74,247.69) and (186.22,246.93) .. (186.95,244.68) .. controls (187.68,242.44) and (189.16,241.68) .. (191.4,242.41) .. controls (193.64,243.13) and (195.12,242.37) .. (195.85,240.13) .. controls (196.58,237.89) and (198.06,237.13) .. (200.3,237.85) .. controls (202.54,238.58) and (204.02,237.82) .. (204.75,235.58) .. controls (205.48,233.34) and (206.96,232.58) .. (209.2,233.3) .. controls (211.44,234.02) and (212.92,233.26) .. (213.65,231.02) .. controls (214.38,228.78) and (215.87,228.02) .. (218.11,228.75) .. controls (220.35,229.47) and (221.83,228.71) .. (222.56,226.47) .. controls (223.29,224.23) and (224.77,223.47) .. (227.01,224.19) .. controls (229.25,224.91) and (230.73,224.15) .. (231.46,221.91) .. controls (232.19,219.67) and (233.67,218.91) .. (235.91,219.64) .. controls (238.15,220.36) and (239.63,219.6) .. (240.36,217.36) .. controls (241.09,215.11) and (242.57,214.35) .. (244.82,215.08) .. controls (247.06,215.81) and (248.54,215.05) .. (249.27,212.81) .. controls (250,210.57) and (251.48,209.81) .. (253.72,210.53) .. controls (255.96,211.25) and (257.44,210.49) .. (258.17,208.25) .. controls (258.9,206.01) and (260.38,205.25) .. (262.62,205.98) .. controls (264.86,206.7) and (266.34,205.94) .. (267.07,203.7) .. controls (267.8,201.45) and (269.28,200.69) .. (271.53,201.42) .. controls (273.77,202.15) and (275.25,201.39) .. (275.98,199.15) .. controls (276.71,196.91) and (278.19,196.15) .. (280.43,196.87) .. controls (282.67,197.59) and (284.15,196.83) .. (284.88,194.59) .. controls (285.61,192.35) and (287.09,191.59) .. (289.33,192.32) .. controls (291.57,193.04) and (293.05,192.28) .. (293.78,190.04) .. controls (294.51,187.8) and (295.99,187.04) .. (298.23,187.76) .. controls (300.47,188.49) and (301.96,187.73) .. (302.69,185.49) .. controls (303.42,183.25) and (304.9,182.49) .. (307.14,183.21) .. controls (309.38,183.93) and (310.86,183.17) .. (311.59,180.93) .. controls (312.32,178.69) and (313.8,177.93) .. (316.04,178.66) .. controls (318.28,179.38) and (319.76,178.62) .. (320.49,176.38) -- (321.12,176.06) -- (321.12,176.06) ;

\draw [color={rgb, 255:red, 139; green, 87; blue, 42 }  ,draw opacity=1 ][line width=1.5]  [dash pattern={on 1.5pt off 1.5pt}]  (321.12,176.06) .. controls (321.87,173.83) and (323.37,173.08) .. (325.6,173.83) .. controls (327.83,174.58) and (329.32,173.83) .. (330.07,171.6) .. controls (330.82,169.37) and (332.32,168.62) .. (334.55,169.37) .. controls (336.78,170.12) and (338.28,169.37) .. (339.03,167.14) .. controls (339.78,164.91) and (341.27,164.16) .. (343.5,164.91) .. controls (345.73,165.66) and (347.23,164.91) .. (347.98,162.68) .. controls (348.73,160.45) and (350.22,159.7) .. (352.45,160.45) .. controls (354.68,161.2) and (356.18,160.45) .. (356.93,158.22) .. controls (357.68,155.99) and (359.17,155.24) .. (361.4,155.99) .. controls (363.63,156.74) and (365.13,155.99) .. (365.88,153.76) .. controls (366.63,151.53) and (368.12,150.79) .. (370.35,151.54) .. controls (372.58,152.29) and (374.08,151.54) .. (374.83,149.31) .. controls (375.58,147.08) and (377.08,146.33) .. (379.31,147.08) -- (383.25,145.11) -- (383.25,145.11) ;

\draw [color={rgb, 255:red, 255; green, 0; blue, 0 }  ,draw opacity=1 ][line width=2.25]    (124.62,276.56) -- (341.23,87.74) ;
\draw [shift={(344.25,85.11)}, rotate = 498.92] [color={rgb, 255:red, 255; green, 0; blue, 0 }  ,draw opacity=1 ][line width=2.25]    (17.49,-5.26) .. controls (11.12,-2.23) and (5.29,-0.48) .. (0,0) .. controls (5.29,0.48) and (11.12,2.23) .. (17.49,5.26)   ;

\draw [color={rgb, 255:red, 139; green, 87; blue, 42 }  ,draw opacity=1 ][line width=1.5]    (383.25,145.11) .. controls (383.96,142.87) and (385.44,142.11) .. (387.69,142.82) .. controls (389.94,143.53) and (391.42,142.77) .. (392.13,140.52) .. controls (392.84,138.27) and (394.32,137.51) .. (396.57,138.23) .. controls (398.82,138.94) and (400.3,138.18) .. (401.01,135.93) .. controls (401.73,133.68) and (403.21,132.92) .. (405.46,133.63) .. controls (407.71,134.35) and (409.19,133.59) .. (409.9,131.34) .. controls (410.61,129.09) and (412.09,128.33) .. (414.34,129.04) .. controls (416.59,129.76) and (418.07,129) .. (418.78,126.75) .. controls (419.49,124.5) and (420.97,123.74) .. (423.22,124.45) -- (424.71,123.68) -- (431.82,120.01) ;
\draw [shift={(435.37,118.17)}, rotate = 512.6700000000001] [fill={rgb, 255:red, 139; green, 87; blue, 42 }  ,fill opacity=1 ][line width=0.08]  [draw opacity=0] (13.4,-6.43) -- (0,0) -- (13.4,6.44) -- (8.9,0) -- cycle    ;

\draw [color={rgb, 255:red, 189; green, 16; blue, 224 }  ,draw opacity=1 ][line width=3]    (379.09,225.87) .. controls (392.69,184.84) and (391.7,130.51) .. (356.24,93.11) ;
\draw [shift={(352.25,89.11)}, rotate = 403.57] [fill={rgb, 255:red, 189; green, 16; blue, 224 }  ,fill opacity=1 ][line width=0.08]  [draw opacity=0] (18.75,-9.01) -- (0,0) -- (18.75,9.01) -- (12.45,0) -- cycle    ;
\draw [shift={(377.25,231.11)}, rotate = 290.41] [color={rgb, 255:red, 189; green, 16; blue, 224 }  ,draw opacity=1 ][line width=3]      (0, 0) circle [x radius= 6.37, y radius= 6.37]   ;
\draw [color={rgb, 255:red, 0; green, 121; blue, 255 }  ,draw opacity=1 ][line width=2.25]    (124.62,276.56) -- (373.31,231.82) ;
\draw [shift={(377.25,231.11)}, rotate = 529.8] [color={rgb, 255:red, 0; green, 121; blue, 255 }  ,draw opacity=1 ][line width=2.25]    (17.49,-5.26) .. controls (11.12,-2.23) and (5.29,-0.48) .. (0,0) .. controls (5.29,0.48) and (11.12,2.23) .. (17.49,5.26)   ;

\draw  [color={rgb, 255:red, 0; green, 0; blue, 0 }  ,draw opacity=1 ][fill={rgb, 255:red, 248; green, 200; blue, 107 }  ,fill opacity=1 ][line width=3]  (116.62,276.56) .. controls (116.62,272.14) and (120.2,268.55) .. (124.62,268.55) .. controls (129.04,268.55) and (132.63,272.14) .. (132.63,276.56) .. controls (132.63,280.98) and (129.04,284.56) .. (124.62,284.56) .. controls (120.2,284.56) and (116.62,280.98) .. (116.62,276.56) -- cycle ;

\draw (448,83) node   [font=\Large,align=left] {\textcolor[rgb]{0.55,0.34,0.16}{Birefringence axis}};
\draw (448,111) node  [font=\Huge]  {$\textcolor[rgb]{0.55,0.34,0.16}{{\textstyle \mathbf{\varsigma }}}$};
\draw (215,282) node  [font=\Large,color={rgb, 255:red, 0; green, 121; blue, 255 }  ,opacity=1 ] [align=left] {Source};
\draw (300,290) node  [font=\Large,color={rgb, 255:red, 0; green, 117; blue, 255 }  ,opacity=1 ]  {$( Q_{z} ,U_{z} ,V_{z})$};
\draw (122,217) node  [font=\Large,color={rgb, 255:red, 255; green, 0; blue, 32 }  ,opacity=1 ] [align=left] {Observed};
\draw (153,192) node  [font=\Large,color={rgb, 255:red, 255; green, 0; blue, 32 }  ,opacity=1 ]  {$( Q,U,V)$};
\draw (467,182) node  [font=\huge,color={rgb, 255:red, 144; green, 19; blue, 254 }  ,opacity=1 ]  {$\Phi _{z} ,\delta \psi _{z} \sim L_{z}$};
\draw (472,211) node  [font=\Large,color={rgb, 255:red, 144; green, 19; blue, 254 }  ,opacity=1 ] [align=left] {SME-induced drift};
\draw (109,301.65) node [font=\Large]  [align=left] {Origin ($\displaystyle \Pi =0$)};

\end{tikzpicture}
}
\vspace{\dskips}
\caption{Depiction of the SME-induced polarization drift. Here, each point in space represents a polarization state given by three coordinates corresponding to the $Q$, $U$ and $V$ Stokes parameters, with the origin at the yellow circle. SME effects cause the state of the photon to precess around the birefringence axis, $\varsigmavec$, by the angle of $\deltapsiz$ (if $d$ is odd) or $\phiz$ (if $d$ is even), which increases with the comoving distance to the source ($\Ldz$). The $^{(d)}$ superscripts are omitted in the figure. (\textit{Top panel}) For odd $d$, the precession occurs in the plane of linear polarization ($V=0$). By contrast, in the even $d$ case, the plane of precession is perpendicular to the $V$-axis and confined to the $Q-U$ plane. (\textit{Bottom panel}) enlarged representation, detailing the labeling in use. 
}
\label{fig:stokes_rotation}
\end{figure}

In the SME, photons emitted with energy $E$ will have their polarization change as they propagate over cosmological distances due to vacuum birefringence via:
\begin{equation}
    \frac{d\svec}{dt} = 2 E \varsigmavec \times \svec \, ,
    \label{eq:stokesdt}
\end{equation}
where $\svec=(Q,U,V)^T$ is the Stokes vector in the Cartesian basis, describing the polarization state of the photons, and $\varsigmavec=(\varsigmaoned,\varsigmatwod,\varsigmathreed)^T$ is the so-called \textit{birefringence axis}. Since $Q,U$, and $V$ are all real valued, one can draw Stokes space diagrams such as \fig{}~\ref{fig:stokes_rotation} that illustrate how a photon's polarization would be rotated due to vacuum birefringence. However, by noting that Eq.~\eqref{eq:stokesdt} describes a rotation of $\svec$ around the axis $\varsigmavec$ and by using the rotational properties of Stokes parameters \cite{hovenier_domke_mee_2004}, further convenience may be gained by switching into the spin-weighted basis, where the Stokes $Q$ and $U$ parameters are combined in a single complex number $Q\mp iU$ and a Cartesian rotation through the angle $\deltapsiz$ amounts to a multiplication by $\eitwodeltapsizmp$. In this basis, we can write
\begin{equation}
    \svec =(\stwoplus,\szero,\stwominus)^T = (Q - iU, V, Q + iU)^T \, ,
    \label{eq:spinstokes}
\end{equation}
where $\szero = V$, $\stwopm = Q \mp i U$, with a SME birefringence axis in this basis given by $\varsigmavec = (\varsigmaplusd,\varsigmathreed,\varsigmaminusd)^T$ \cite{kostelecky09}. 

Following \cite{kostelecky09,kislat18}, the observed Stokes vector $\svec$ can be computed from the Stokes vector $\svecz$ emitted at the source using the \Muller matrix $\Mz$, via
\begin{equation}
    \svec = \Mz \cdot \svecz\, ,
    \label{eq:stokesmuller}
\end{equation}
where $\Mz$ is given by 
\begin{alignat}{2}
\footnotesize
\setlength{\arraycolsep}{2.5pt}
\medmuskip = 1mu 
\Mz & =  \nonumber \\
& \begin{cases}
\begin{psmallmatrix}
   \eitwodeltapsizm & 0 & 0  \\
   0 & 1 & 0  \\
   0 & 0 & \eitwodeltapsizp  
   \end{psmallmatrix}, \\
   \text{ odd } d\, ,  \\
\begin{psmallmatrix}
   \cossqphiz & -i \sintwophiz \eivarxim & \sinsqphiz \etwoivarxim\\
   -\frac{i}{2}\sintwophiz \eivarxip & \costwophiz & \frac{i}{2}\sintwophiz \eivarxim  \\
   \sinsqphiz \etwoivarxip & i \sintwophiz \eivarxip & \cossqphiz
   \end{psmallmatrix},  \\
   \text{ even } d\, ,  \\
\end{cases}
\label{eq:mullers}
\end{alignat}
with Eqs.~\eqref{eq:psizdelta}-\eqref{eq:varxi} defining $\deltapsiz$, $\phiz$, and $\varxi$. 

In the CPT-odd case, having $\varsigmavec$ aligned with the $V$-axis (see \fig{}~\ref{fig:stokes_rotation}), Eq.~\eqref{eq:mullers} yields a particularly simple form, diagonal in the spin-weighted Stokes basis, given by
\begin{alignat}{2}
\stwopm = e^{\mp i 2 \deltapsiz} \stwopmz\, , &  \ \ \ \ \ \ \ \ & \szero = \szeroz .
\label{eq:Seven}
\end{alignat}
In this case, both Stokes $V$ and the linear polarization fraction remain constant as the photon travels to the observer. The theoretically predicted linear polarization angle $\psio$ in the SME is related to the intrinsic polarization angle for the source at redshift $z$ via
\begin{equation}
\psio = \psiz + \deltapsiz\, ,
\label{eq:psizdelta}
\end{equation}
with an SME induced polarization angle change of
\begin{alignat}{2}
& \deltapsiz & =  E^{d-3} \Ldz \sum_{jm} \oYjm(\nhat) \kdVjm = E \Ldz \varsigmathreed\, . 
\label{eq:deltaphiz1}
\end{alignat}
Note that the parity relationships of the spherical harmonics in Eq.~\eqref{eq:parityodd} ensure that $\deltapsiz$ is real valued in the CPT-odd case.

By contrast, in the CPT-even case, $\varsigmavec$ lies in the plane of linear polarization, implying that (1) Stokes $V$ polarization may be induced in-flight (although we ignore it in this analysis) and (2) the drift in polarization angle can no longer be described with a single phase as a simple rotation around the $V$-axis through the origin (see \fig{}~\ref{fig:stokes_rotation}). Mathematically, the additional complexity can be modelled by allowing the CPT-even equivalent of $\deltapsiz$, which we will call $\deltaphiz$, to be a complex number, composed of magnitude $\phiz$ and argument $\varxi$, which are each real. As such, the complex quantity $\deltaphiz$ is given by 
\begin{eqnarray}
    \deltaphiz = \phiz e^{\mp i \varxi} & = & E \Ldz \varsigmapmd \, ,
    \label{eq:phiz1a}
\end{eqnarray}
with the real-valued angle $\phiz$ in Eq.~\eqref{eq:mullers} given by
\begin{alignat}{2}
& \phiz & =  E^{d-3} \Ldz \Big| \sum_{jm} \twoYjm(\nhat)   \Big(\kdEjm \pm i \kdBjm\Big) \Big| \, , 
\label{eq:phiz1}
\end{alignat}
where the phase angle $\varxi$ for the CPT-even case in Eq.~\eqref{eq:mullers} is given by
\begin{eqnarray}
\varxi = \mp \arg\Big(\Sdnhat\Big) \, , 
\label{eq:varxi}
\end{eqnarray}
where we also define the abbreviation $\Sdnhat$ for the complex linear combination of SME coefficients, given by
\begin{alignat}{3}
\Sdnhat \equiv 
\begin{cases}
      \sum_{jm} \oYjm(\nhat) \kdVjm\,, & \text{ odd } d\, , \\
      \sum_{jm} \twoYjm(\nhat)   \Big(\kdEjm \pm i \kdBjm\Big),                 & \text{ even } d\, .
\end{cases}
\label{eq:S}
\end{alignat}
We further define the abbreviation
\begin{alignat}{3}
\gammad(\nhat) \equiv
\begin{cases}
    \Sdnhat\,.             & \text{ odd } d\, , \\
    \Big|\Sdnhat\Big|\, ,  & \text{ even } d\, .
\end{cases}
\label{eq:gamma}
\end{alignat}
This allows us to write Eqs.~\eqref{eq:deltaphiz1} and \eqref{eq:phiz1} for both the CPT-odd and CPT-even cases as
\begin{alignat}{3}
E^{d-3} \varthetad(\nhat) \equiv
\begin{cases}
    \deltapsiz\,,             & \text{ odd } d\, , \\
    \phiz\, ,  & \text{ even } d\, ,
\end{cases}
\label{eq:phiza}
\end{alignat}
with 
\begin{equation}
    \varthetad(\nhat) \equiv \Ldz \gammad(\nhat)\, .
    \label{eq:varthetad}
\end{equation}

\section{Stokes Parameters and Polarization Angles}
\label{sec:stokes}

Since the measured optical circular polarization is generally small for the extragalactic sources of interest, and since there are relatively few such measurements in the literature (e.g.~\cite{hutsemekers18,matsumiya03,sagiv04,toma08}), we ignore circular polarization in this work and write and write the intensity normalized Stokes parameters at the source at redshift $z$ as
\begin{alignat}{3}
    \Pqdz & = \frac{\Qdz}{\Idz} = \pz \cos\left(2 \psiz\right)\, , \nonumber \\[5pt]
    \Pudz & = \frac{\Udz}{\Idz} = \pz\sin\left(2 \psiz\right)\, , \nonumber \\[5pt]
    \Pvdz & = \frac{\Vdz}{\Idz} \ \approx \ 0 \, ,
    \label{eq:stokes}
\end{alignat}
where $\pz$ and $\psiz$ are the intrinsic linear polarization fraction and polarization angle at the source, respectively. We conservatively assume both $\pz$ and $\psiz$ (and thus the source frame Stokes parameters) to be independent of wavelength. 
Previous analyses \cite{kislat17,kislat18} have assumed a 100\% intrinsic linear polarization fraction at all wavelengths such that $\pz=1$, which leads to the most conservative possible SME constraints. However, we will relax this assumption in this work based on more realistic AGN source models with conservative upper limits $\pz < \pzmax=\pzup$ at optical wavelengths for even the most highly intrinsically polarized AGN subclass of BL Lac objects \cite{kartje95,goosmann07,zhangh13,marscher14,lobos18,zhangh19}.\footnote{For consistency, if we assume $\pzmax < 1$, we would need to exclude all data from the analysis with observed polarization $\pol > \pzmax$. However, the maximum polarization value in our catalog is $0.45$, which does not violate $\pol > \pzmax=\pzup$, so this does not affect the inclusion of any data.} More detailed and realistic source models where $\pz$ and $\psiz$ (and thus $\Pqdz$ and $\Pudz$), depended on wavelength --- with smaller maximum values for different AGN sub-classes other than BL Lac objects --- would yield even stronger SME constraints, so our assumptions are still reasonably conservative.

Using Eqs.~\eqref{eq:spinstokes}-\eqref{eq:stokes}, the observer frame Stokes parameters $\Pqd$ and $\Pud$ can be written as
\begin{alignat}{3}
\Pqd = \pz
\begin{cases}
     \cos \left(2 \left(\deltapsiz + \psiz\right) \right)   \, ,             & \text{ odd } d\, , \\
     \Big[ \cos \left(2 \psiz\right) \cos^{2}\left(\phiz\right)  \\
     \ +\cos \left(2\left(\varxi-\psiz\right)\right) \sin^{2}\left(\phiz\right) \Big]\, ,  & \text{ even } d\, .
\end{cases}
\label{eq:q}
\end{alignat}
and
\begin{alignat}{3}
\Pud = \pz
\begin{cases}
      \sin \left(2 \left(\deltapsiz + \psiz\right) \right)  \, ,             & \text{ odd } d\, , \\
     \Big[ \sin \left(2\left(\varxi-\psiz\right)\right) \sin ^{2}\left(\phiz)\right) \\
     \ +\sin \left(2 \psiz\right) \cos ^{2}\left(\phiz\right) \Big]\, ,  & \text{ even } d\, .
\end{cases}
\label{eq:u}
\end{alignat}
The changes in Stokes parameters from the observed frame to the source frame are then given by
\begin{alignat}{2}
\label{eq:deltaq}
& \DPqd  =\Pqd-\Pqdz \\
& = -2 \pz
\begin{cases}
       \sin\left(\deltapsiz\right) \sin\left(\deltapsiz + 2 \psiz \right) \, ,  
       & 
       \text{ odd } d\, , \nonumber  \\
     \sin ^{2}\left(\phiz\right) \sin \left(\varxi\right) \sin \left(\varxi-2 \psiz\right)\, ,  
     & 
     \text{ even } d\, ,
\end{cases}
\end{alignat}
and
\begin{alignat}{2}
\label{eq:deltau}
& \DPud  = \Pud-\Pudz  \\
& = 2 \pz
\begin{cases}
      \sin\left(\deltapsiz\right) \cos\left(\deltapsiz + 2 \psiz \right)  \, , 
      & 
      \text{ odd } d\, , \nonumber \\
     \sin^{2}\left(\phiz\right) \cos \left(\varxi\right) \sin \left(\varxi-2 \psiz\right)\, ,  
     & 
     \text{ even } d\, .
\end{cases}
\end{alignat}

\begin{figure*}[t]
    \centering
    \includegraphics[width=0.9\textwidth]{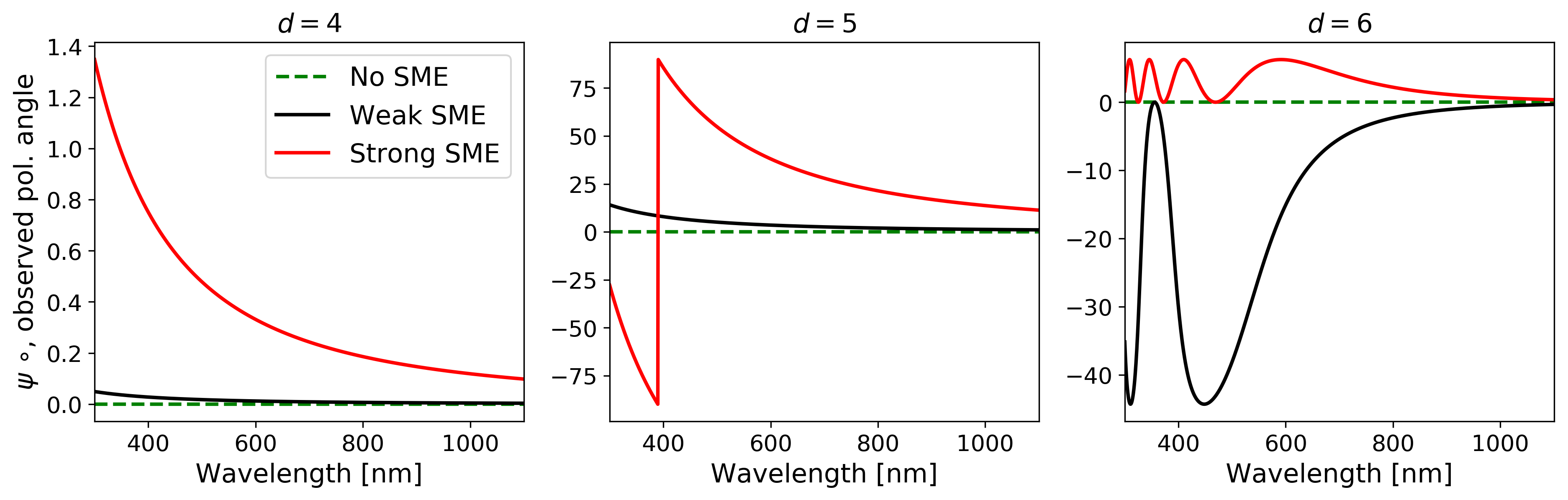}
    \vspace{\dskips}
    \caption{Expected polarization angle spectra of a cosmologically distant source after 
    LIV and/or CPTV induced birefringence. In this demonstration, the source is placed at RA $=2\mathrm{h}$, Dec $=-60\degree$ and $z=3$. The emitted (pre-birefringence) spectrum is assumed to be flat with a polarization angle of $0$ at all wavelengths. The \textit{No SME} case has all SME coefficients set to $0$, yielding an observed spectrum identical to the emitted spectrum. For the \textit{Weak SME} case, all real components of the SME coefficients are set to $10^{-35}\ \ev^{4-d}$ except $\mathrm{Re}[k^{(4)}_{{(B)2,1}}]$, $\mathrm{Im}[k^{(5)}_{{(V)2,1}}]$ and $\mathrm{Re}[k^{(6)}_{{(E)3,1}}]$ which are set to $-10^{-35}\ \ev^{4-d}$. Finally, the \textit{Strong SME} case fixes all real components at $5\times 10^{-35}\ \ev^{4-d}$ except $\mathrm{Im}[k^{(4)}_{{(E)2,1}}]$, $k^{(5)}_{{(V)1,0}}$, $\mathrm{Re}[k^{(5)}_{{(V)1,1}}]$ and $\mathrm{Im}[k^{(6)}_{{(E)2,1}}]$, each kept at $-5\times 10^{-35}\ \ev^{4-d}$. The choices are arbitrary and only intended to demonstrate typical behaviors. In a CPT-odd universe, the birefringent drift spans all angles in the range [-90$^{\circ}$,+90$^{\circ}$], while CPT-even universes are often restricted to oscillations between two bounds, one of which corresponds to the emitted polarization angle. This result follows directly from the Stokes space geometry illustrated in \fig{}~\ref{fig:stokes_rotation}. Larger SME coefficients tend to accelerate the rate of drift with wavelength. Note that in CPT-even cases, the magnitude of the SME coefficients sets the rate of polarization angle drift but not its amplitude, which is instead determined by the distance between the initial polarization and the birefringence axis. Therefore, even with large SME coefficients, certain sources may display very little birefringence, further justifying our use of an extensive catalog of measurements. The $d=6$ panel of this figure illustrates this peculiar property.
    }
    \label{fig:psi_vs_wl}
\end{figure*}

Ref.~\cite{kislat18} noted that, for the CPT-even case, Eqs.~\eqref{eq:deltaq}-\eqref{eq:deltau} can be simplified by choosing the reference direction for the polarization angle, by transforming to a primed coordinate frame 
\begin{equation}
    \psizprime = \psiz - \varxi/2\, ,
    \label{eq:primeeven}
\end{equation}
and choosing a reference angle such that $\varxiprime=0$. For the CPT-odd case,
such a transformation is not possible since the birefringence axis $\varsigmavec$ is along the Stokes V axis and $\varxi$ can not be defined, but we will apply Eq.~\eqref{eq:primeeven} for even $d$ and label the coordinate systems as primed for both even and odd $d$ from now on for convenience.

It will now be useful to present Eqs.~\eqref{eq:deltaq}-\eqref{eq:deltau} in terms of the source frame Stokes parameters $\Pqdzprime$ and $\Pudzprime$ as
\begin{alignat}{2}
\label{eq:deltaqprime}
& \DPqdprime = \Pqdprime-\Pqdzprime  \\
& =
\begin{cases}
    - 2\sin^2\left(\deltapsiz\right) \Pqdzprime -\sin\left(2\deltapsiz\right) \Pudzprime \, , 
    & \text{ odd } d\, , \nonumber \\
    0 \, ,  
    & \text{ even } d\, , 
\end{cases}
\end{alignat}
and
\begin{alignat}{2}
\label{eq:deltauprime}
& \DPudprime = \Pudprime-\Pudzprime  \\
& =
\begin{cases}
       - 2\sin^2\left(\deltapsiz\right) \Pudzprime+ \sin\left(2\deltapsiz\right) \Pqdzprime   \, , 
       & \text{ odd } d\, , \nonumber \\
      -2\sin^{2}\left(\phiz\right) \Pudzprime \, ,  
      & \text{ even } d\, ,
\end{cases}
\end{alignat}
where in the first lines of Eqs.~\eqref{eq:deltaqprime}-\eqref{eq:deltauprime}, we used trigonometric identities, along with the definitions $\Pqdzprime = \pz\cos(2 \psizprime)$ and $\Pudzprime = \pz\sin(2 \psizprime)$ from Eq.~\eqref{eq:stokes}, and we note that these are \textit{different} primed coordinate systems for the CPT-odd and CPT-even cases.

\begin{figure}[h!]
    \centering
    \includegraphics[width=0.99\columnwidth]{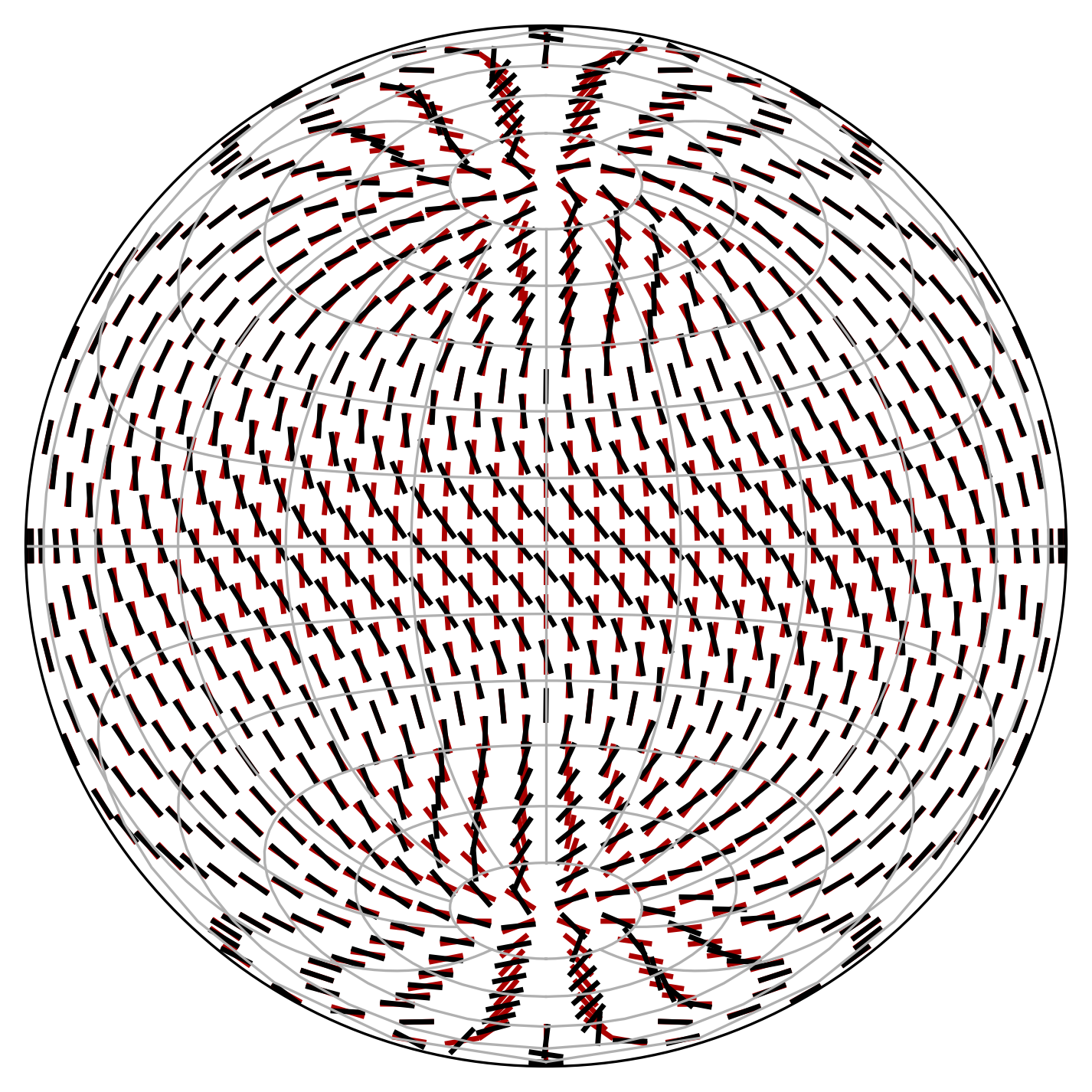}
    \vspace{\dskips}
    \caption{Expected polarization angles after SME birefringence at different positions on the sky, assuming a universe where the only non-zero SME coefficient is $k^{(5)}_{{(V)2,0}}=10^{-33}\ \mathrm{eV}^{-1}$ (CPT-odd). Black strokes represents the observed polarization angle of a $1\ \mathrm{eV}$ photon from a test source placed at the location of the stroke and $z=3$. In each case, a polarization angle of $0$ (North) (shown with red strokes) is assumed at emission ($\psiz$). The projection is identical to that in \fig{}~\ref{fig:angles_catalog}.
    }
    \label{fig:sme_map}
\end{figure}

Using Eqs.~\eqref{eq:deltaqprime}-\eqref{eq:deltauprime}, and the definitions of $\deltapsiz$ and $\phiz$ in Eqs.~\eqref{eq:phiza}-\eqref{eq:varthetad}, we can write
$\Pqdzprime$ and $\Pudzprime$ in terms of $\left(E,\varthetad,\Pqdprime,\Pudprime\right)$ as
\begin{alignat}{2}
& \Pqdzprime = 
\begin{cases}
      \Pqdprime \cos\left(2 E^{d-3}\varthetad\right)  \\
      \  + \Pudprime \sin\left(2 E^{d-3}\varthetad\right)\, , 
      & 
      \text{ odd } d\, , \\
      \Pqdprime \, ,  
      & 
      \text{ even } d\, ,
\end{cases}
\label{eq:qzprime}
\end{alignat}
and
\begin{alignat}{2}
& \Pudzprime = 
\begin{cases}
        \Pudprime \cos\left(2 E^{d-3}\varthetad\right)  \\
       \   - \Pqdprime \sin\left(2 E^{d-3}\varthetad\right) \, , 
       & 
       \text{ odd } d\, , \\
      \Pudprime \sec\left(2 E^{d-3}\varthetad\right)  \, ,  
      & 
      \text{ even } d\, ,
\end{cases}
\label{eq:uzprime}
\end{alignat}
so that in each case
\begin{equation}
\psizprime = \frac{1}{2} \arctan\Bigg(\frac{\Pudzprime}{\Pqdzprime}\Bigg)\, .
    \label{eq:psizprimearctan}
\end{equation}
The dependence of the observed polarization angle after the SME-induced drift in various mass dimensions is illustrated in \fig{}~\ref{fig:psi_vs_wl} for arbitrarily selected SME coefficients and a test source with a flat, pre-birefringence, polarization angle spectrum of $\psiz(E)=0$ at all energies, where zero degrees polarization is defined with the polarization vector pointing North. \fig{}~\ref{fig:sme_map} shows a Lambert all-sky projection of the polarization vectors for a universe where one of the CPT-odd coefficients has a non-zero value.

\section{Broadband Polarimetry}
\label{sec:broad}

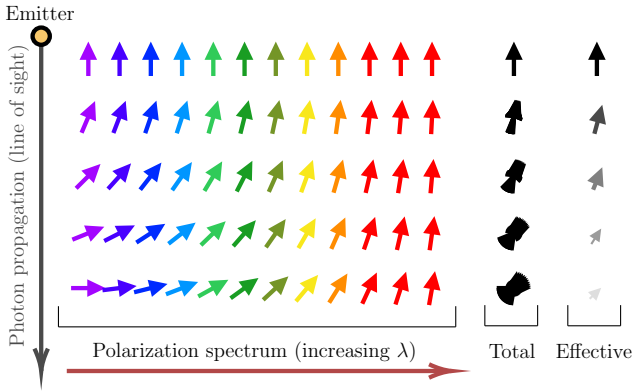
\begin{figure}[h!]
\centering
\resizebox{\columnwidth}{!}{

\tikzset{every picture/.style={line width=0.75pt}} 

\begin{tikzpicture}[x=0.75pt,y=0.75pt,yscale=-1,xscale=1]

\draw [color={rgb, 255:red, 163; green, 5; blue, 255 }  ,draw opacity=1 ][fill={rgb, 255:red, 200; green, 14; blue, 253 }  ,fill opacity=1 ][line width=3]    (90.74,818.11) -- (90.74,792.11) ;
\draw [shift={(90.74,786.11)}, rotate = 450] [fill={rgb, 255:red, 163; green, 5; blue, 255 }  ,fill opacity=1 ][line width=0.08]  [draw opacity=0] (16.97,-8.15) -- (0,0) -- (16.97,8.15) -- cycle    ;

\draw [color={rgb, 255:red, 72; green, 0; blue, 255 }  ,draw opacity=1 ][fill={rgb, 255:red, 99; green, 0; blue, 255 }  ,fill opacity=1 ][line width=3]    (119.74,818.11) -- (119.74,792.11) ;
\draw [shift={(119.74,786.11)}, rotate = 450] [fill={rgb, 255:red, 72; green, 0; blue, 255 }  ,fill opacity=1 ][line width=0.08]  [draw opacity=0] (16.97,-8.15) -- (0,0) -- (16.97,8.15) -- cycle    ;

\draw [color={rgb, 255:red, 0; green, 43; blue, 255 }  ,draw opacity=1 ][fill={rgb, 255:red, 10; green, 0; blue, 255 }  ,fill opacity=1 ][line width=3]    (148.74,818.11) -- (148.74,792.11) ;
\draw [shift={(148.74,786.11)}, rotate = 450] [fill={rgb, 255:red, 0; green, 43; blue, 255 }  ,fill opacity=1 ][line width=0.08]  [draw opacity=0] (16.97,-8.15) -- (0,0) -- (16.97,8.15) -- cycle    ;

\draw [color={rgb, 255:red, 0; green, 150; blue, 255 }  ,draw opacity=1 ][line width=3]    (177.74,818.11) -- (177.74,792.11) ;
\draw [shift={(177.74,786.11)}, rotate = 450] [fill={rgb, 255:red, 0; green, 150; blue, 255 }  ,fill opacity=1 ][line width=0.08]  [draw opacity=0] (16.97,-8.15) -- (0,0) -- (16.97,8.15) -- cycle    ;

\draw [color={rgb, 255:red, 48; green, 203; blue, 89 }  ,draw opacity=1 ][line width=3]    (206.74,818.11) -- (206.74,792.11) ;
\draw [shift={(206.74,786.11)}, rotate = 450] [fill={rgb, 255:red, 48; green, 203; blue, 89 }  ,fill opacity=1 ][line width=0.08]  [draw opacity=0] (16.97,-8.15) -- (0,0) -- (16.97,8.15) -- cycle    ;

\draw [color={rgb, 255:red, 25; green, 157; blue, 35 }  ,draw opacity=1 ][line width=3]    (235.74,818.11) -- (235.74,792.11) ;
\draw [shift={(235.74,786.11)}, rotate = 450] [fill={rgb, 255:red, 25; green, 157; blue, 35 }  ,fill opacity=1 ][line width=0.08]  [draw opacity=0] (16.97,-8.15) -- (0,0) -- (16.97,8.15) -- cycle    ;

\draw [color={rgb, 255:red, 116; green, 149; blue, 39 }  ,draw opacity=1 ][line width=3]    (264.74,818.11) -- (264.74,792.11) ;
\draw [shift={(264.74,786.11)}, rotate = 450] [fill={rgb, 255:red, 116; green, 149; blue, 39 }  ,fill opacity=1 ][line width=0.08]  [draw opacity=0] (16.97,-8.15) -- (0,0) -- (16.97,8.15) -- cycle    ;

\draw [color={rgb, 255:red, 248; green, 231; blue, 28 }  ,draw opacity=1 ][line width=3]    (293.74,818.11) -- (293.74,792.11) ;
\draw [shift={(293.74,786.11)}, rotate = 450] [fill={rgb, 255:red, 248; green, 231; blue, 28 }  ,fill opacity=1 ][line width=0.08]  [draw opacity=0] (16.97,-8.15) -- (0,0) -- (16.97,8.15) -- cycle    ;

\draw [color={rgb, 255:red, 255; green, 140; blue, 0 }  ,draw opacity=1 ][line width=3]    (322.74,818.11) -- (322.74,792.11) ;
\draw [shift={(322.74,786.11)}, rotate = 450] [fill={rgb, 255:red, 255; green, 140; blue, 0 }  ,fill opacity=1 ][line width=0.08]  [draw opacity=0] (16.97,-8.15) -- (0,0) -- (16.97,8.15) -- cycle    ;

\draw [color={rgb, 255:red, 255; green, 0; blue, 0 }  ,draw opacity=1 ][fill={rgb, 255:red, 255; green, 0; blue, 0 }  ,fill opacity=1 ][line width=3]    (351.74,818.11) -- (351.74,792.11) ;
\draw [shift={(351.74,786.11)}, rotate = 450] [fill={rgb, 255:red, 255; green, 0; blue, 0 }  ,fill opacity=1 ][line width=0.08]  [draw opacity=0] (16.97,-8.15) -- (0,0) -- (16.97,8.15) -- cycle    ;

\draw [color={rgb, 255:red, 255; green, 0; blue, 0 }  ,draw opacity=1 ][fill={rgb, 255:red, 255; green, 0; blue, 0 }  ,fill opacity=1 ][line width=3]    (380.74,818.11) -- (380.74,792.11) ;
\draw [shift={(380.74,786.11)}, rotate = 450] [fill={rgb, 255:red, 255; green, 0; blue, 0 }  ,fill opacity=1 ][line width=0.08]  [draw opacity=0] (16.97,-8.15) -- (0,0) -- (16.97,8.15) -- cycle    ;

\draw [color={rgb, 255:red, 255; green, 0; blue, 0 }  ,draw opacity=1 ][line width=3]    (409.74,818.11) -- (409.74,792.11) ;
\draw [shift={(409.74,786.11)}, rotate = 450] [fill={rgb, 255:red, 255; green, 0; blue, 0 }  ,fill opacity=1 ][line width=0.08]  [draw opacity=0] (16.97,-8.15) -- (0,0) -- (16.97,8.15) -- cycle    ;

\draw [color={rgb, 255:red, 163; green, 5; blue, 255 }  ,draw opacity=1 ][fill={rgb, 255:red, 200; green, 14; blue, 253 }  ,fill opacity=1 ][line width=3]    (84.48,870.84) -- (94.64,846.91) ;
\draw [shift={(96.99,841.39)}, rotate = 473] [fill={rgb, 255:red, 163; green, 5; blue, 255 }  ,fill opacity=1 ][line width=0.08]  [draw opacity=0] (16.97,-8.15) -- (0,0) -- (16.97,8.15) -- cycle    ;

\draw [color={rgb, 255:red, 72; green, 0; blue, 255 }  ,draw opacity=1 ][fill={rgb, 255:red, 99; green, 0; blue, 255 }  ,fill opacity=1 ][line width=3]    (114,871.05) -- (123.32,846.78) ;
\draw [shift={(125.47,841.18)}, rotate = 471] [fill={rgb, 255:red, 72; green, 0; blue, 255 }  ,fill opacity=1 ][line width=0.08]  [draw opacity=0] (16.97,-8.15) -- (0,0) -- (16.97,8.15) -- cycle    ;

\draw [color={rgb, 255:red, 0; green, 43; blue, 255 }  ,draw opacity=1 ][fill={rgb, 255:red, 10; green, 0; blue, 255 }  ,fill opacity=1 ][line width=3]    (143.53,871.24) -- (151.99,846.66) ;
\draw [shift={(153.94,840.99)}, rotate = 469] [fill={rgb, 255:red, 0; green, 43; blue, 255 }  ,fill opacity=1 ][line width=0.08]  [draw opacity=0] (16.97,-8.15) -- (0,0) -- (16.97,8.15) -- cycle    ;

\draw [color={rgb, 255:red, 0; green, 150; blue, 255 }  ,draw opacity=1 ][line width=3]    (172.79,871.33) -- (180.83,846.6) ;
\draw [shift={(182.68,840.9)}, rotate = 468] [fill={rgb, 255:red, 0; green, 150; blue, 255 }  ,fill opacity=1 ][line width=0.08]  [draw opacity=0] (16.97,-8.15) -- (0,0) -- (16.97,8.15) -- cycle    ;

\draw [color={rgb, 255:red, 48; green, 203; blue, 89 }  ,draw opacity=1 ][line width=3]    (202.33,871.49) -- (209.49,846.5) ;
\draw [shift={(211.15,840.73)}, rotate = 466] [fill={rgb, 255:red, 48; green, 203; blue, 89 }  ,fill opacity=1 ][line width=0.08]  [draw opacity=0] (16.97,-8.15) -- (0,0) -- (16.97,8.15) -- cycle    ;

\draw [color={rgb, 255:red, 25; green, 157; blue, 35 }  ,draw opacity=1 ][line width=3]    (231.87,871.64) -- (238.16,846.41) ;
\draw [shift={(239.61,840.59)}, rotate = 464] [fill={rgb, 255:red, 25; green, 157; blue, 35 }  ,fill opacity=1 ][line width=0.08]  [draw opacity=0] (16.97,-8.15) -- (0,0) -- (16.97,8.15) -- cycle    ;

\draw [color={rgb, 255:red, 116; green, 149; blue, 39 }  ,draw opacity=1 ][line width=3]    (261.41,871.76) -- (266.81,846.33) ;
\draw [shift={(268.06,840.46)}, rotate = 462] [fill={rgb, 255:red, 116; green, 149; blue, 39 }  ,fill opacity=1 ][line width=0.08]  [draw opacity=0] (16.97,-8.15) -- (0,0) -- (16.97,8.15) -- cycle    ;

\draw [color={rgb, 255:red, 248; green, 231; blue, 28 }  ,draw opacity=1 ][line width=3]    (290.96,871.87) -- (295.47,846.27) ;
\draw [shift={(296.51,840.36)}, rotate = 460] [fill={rgb, 255:red, 248; green, 231; blue, 28 }  ,fill opacity=1 ][line width=0.08]  [draw opacity=0] (16.97,-8.15) -- (0,0) -- (16.97,8.15) -- cycle    ;

\draw [color={rgb, 255:red, 255; green, 140; blue, 0 }  ,draw opacity=1 ][line width=3]    (320.51,871.96) -- (324.13,846.21) ;
\draw [shift={(324.96,840.27)}, rotate = 458] [fill={rgb, 255:red, 255; green, 140; blue, 0 }  ,fill opacity=1 ][line width=0.08]  [draw opacity=0] (16.97,-8.15) -- (0,0) -- (16.97,8.15) -- cycle    ;

\draw [color={rgb, 255:red, 255; green, 0; blue, 0 }  ,draw opacity=1 ][fill={rgb, 255:red, 255; green, 0; blue, 0 }  ,fill opacity=1 ][line width=3]    (349.79,871.99) -- (352.95,846.19) ;
\draw [shift={(353.69,840.23)}, rotate = 457] [fill={rgb, 255:red, 255; green, 0; blue, 0 }  ,fill opacity=1 ][line width=0.08]  [draw opacity=0] (16.97,-8.15) -- (0,0) -- (16.97,8.15) -- cycle    ;

\draw [color={rgb, 255:red, 255; green, 0; blue, 0 }  ,draw opacity=1 ][fill={rgb, 255:red, 255; green, 0; blue, 0 }  ,fill opacity=1 ][line width=3]    (379.34,872.05) -- (381.61,846.15) ;
\draw [shift={(382.13,840.17)}, rotate = 455] [fill={rgb, 255:red, 255; green, 0; blue, 0 }  ,fill opacity=1 ][line width=0.08]  [draw opacity=0] (16.97,-8.15) -- (0,0) -- (16.97,8.15) -- cycle    ;

\draw [color={rgb, 255:red, 255; green, 0; blue, 0 }  ,draw opacity=1 ][line width=3]    (408.9,872.09) -- (410.26,846.13) ;
\draw [shift={(410.57,840.14)}, rotate = 453] [fill={rgb, 255:red, 255; green, 0; blue, 0 }  ,fill opacity=1 ][line width=0.08]  [draw opacity=0] (16.97,-8.15) -- (0,0) -- (16.97,8.15) -- cycle    ;

\draw [color={rgb, 255:red, 163; green, 5; blue, 255 }  ,draw opacity=1 ][fill={rgb, 255:red, 200; green, 14; blue, 253 }  ,fill opacity=1 ][line width=3]    (79.42,922.43) -- (97.81,904.04) ;
\draw [shift={(102.05,899.8)}, rotate = 495] [fill={rgb, 255:red, 163; green, 5; blue, 255 }  ,fill opacity=1 ][line width=0.08]  [draw opacity=0] (16.97,-8.15) -- (0,0) -- (16.97,8.15) -- cycle    ;

\draw [color={rgb, 255:red, 72; green, 0; blue, 255 }  ,draw opacity=1 ][fill={rgb, 255:red, 99; green, 0; blue, 255 }  ,fill opacity=1 ][line width=3]    (109.03,923) -- (126.43,903.68) ;
\draw [shift={(130.44,899.22)}, rotate = 492] [fill={rgb, 255:red, 72; green, 0; blue, 255 }  ,fill opacity=1 ][line width=0.08]  [draw opacity=0] (16.97,-8.15) -- (0,0) -- (16.97,8.15) -- cycle    ;

\draw [color={rgb, 255:red, 0; green, 43; blue, 255 }  ,draw opacity=1 ][fill={rgb, 255:red, 10; green, 0; blue, 255 }  ,fill opacity=1 ][line width=3]    (138.89,923.72) -- (154.89,903.23) ;
\draw [shift={(158.59,898.51)}, rotate = 488] [fill={rgb, 255:red, 0; green, 43; blue, 255 }  ,fill opacity=1 ][line width=0.08]  [draw opacity=0] (16.97,-8.15) -- (0,0) -- (16.97,8.15) -- cycle    ;

\draw [color={rgb, 255:red, 0; green, 150; blue, 255 }  ,draw opacity=1 ][line width=3]    (168.56,924.22) -- (183.47,902.92) ;
\draw [shift={(186.91,898.01)}, rotate = 485] [fill={rgb, 255:red, 0; green, 150; blue, 255 }  ,fill opacity=1 ][line width=0.08]  [draw opacity=0] (16.97,-8.15) -- (0,0) -- (16.97,8.15) -- cycle    ;

\draw [color={rgb, 255:red, 48; green, 203; blue, 89 }  ,draw opacity=1 ][line width=3]    (198.5,924.83) -- (211.89,902.54) ;
\draw [shift={(214.98,897.4)}, rotate = 481] [fill={rgb, 255:red, 48; green, 203; blue, 89 }  ,fill opacity=1 ][line width=0.08]  [draw opacity=0] (16.97,-8.15) -- (0,0) -- (16.97,8.15) -- cycle    ;

\draw [color={rgb, 255:red, 25; green, 157; blue, 35 }  ,draw opacity=1 ][line width=3]    (228.47,925.37) -- (240.28,902.2) ;
\draw [shift={(243,896.86)}, rotate = 477] [fill={rgb, 255:red, 25; green, 157; blue, 35 }  ,fill opacity=1 ][line width=0.08]  [draw opacity=0] (16.97,-8.15) -- (0,0) -- (16.97,8.15) -- cycle    ;

\draw [color={rgb, 255:red, 116; green, 149; blue, 39 }  ,draw opacity=1 ][line width=3]    (258.23,925.73) -- (268.8,901.98) ;
\draw [shift={(271.24,896.5)}, rotate = 474] [fill={rgb, 255:red, 116; green, 149; blue, 39 }  ,fill opacity=1 ][line width=0.08]  [draw opacity=0] (16.97,-8.15) -- (0,0) -- (16.97,8.15) -- cycle    ;

\draw [color={rgb, 255:red, 248; green, 231; blue, 28 }  ,draw opacity=1 ][line width=3]    (288.26,926.15) -- (297.16,901.72) ;
\draw [shift={(299.21,896.08)}, rotate = 470] [fill={rgb, 255:red, 248; green, 231; blue, 28 }  ,fill opacity=1 ][line width=0.08]  [draw opacity=0] (16.97,-8.15) -- (0,0) -- (16.97,8.15) -- cycle    ;

\draw [color={rgb, 255:red, 255; green, 140; blue, 0 }  ,draw opacity=1 ][line width=3]    (318.33,926.49) -- (325.49,901.5) ;
\draw [shift={(327.15,895.73)}, rotate = 466] [fill={rgb, 255:red, 255; green, 140; blue, 0 }  ,fill opacity=1 ][line width=0.08]  [draw opacity=0] (16.97,-8.15) -- (0,0) -- (16.97,8.15) -- cycle    ;

\draw [color={rgb, 255:red, 255; green, 0; blue, 0 }  ,draw opacity=1 ][fill={rgb, 255:red, 255; green, 0; blue, 0 }  ,fill opacity=1 ][line width=3]    (348.14,926.7) -- (353.99,901.37) ;
\draw [shift={(355.34,895.52)}, rotate = 463] [fill={rgb, 255:red, 255; green, 0; blue, 0 }  ,fill opacity=1 ][line width=0.08]  [draw opacity=0] (16.97,-8.15) -- (0,0) -- (16.97,8.15) -- cycle    ;

\draw [color={rgb, 255:red, 255; green, 0; blue, 0 }  ,draw opacity=1 ][fill={rgb, 255:red, 255; green, 0; blue, 0 }  ,fill opacity=1 ][line width=3]    (378.23,926.92) -- (382.3,901.24) ;
\draw [shift={(383.24,895.31)}, rotate = 459] [fill={rgb, 255:red, 255; green, 0; blue, 0 }  ,fill opacity=1 ][line width=0.08]  [draw opacity=0] (16.97,-8.15) -- (0,0) -- (16.97,8.15) -- cycle    ;

\draw [color={rgb, 255:red, 255; green, 0; blue, 0 }  ,draw opacity=1 ][line width=3]    (408.34,927.05) -- (410.61,901.15) ;
\draw [shift={(411.13,895.17)}, rotate = 455] [fill={rgb, 255:red, 255; green, 0; blue, 0 }  ,fill opacity=1 ][line width=0.08]  [draw opacity=0] (16.97,-8.15) -- (0,0) -- (16.97,8.15) -- cycle    ;

\draw [color={rgb, 255:red, 163; green, 5; blue, 255 }  ,draw opacity=1 ][fill={rgb, 255:red, 200; green, 14; blue, 253 }  ,fill opacity=1 ][line width=3]    (75.9,970.11) -- (100.01,960.37) ;
\draw [shift={(105.57,958.12)}, rotate = 518] [fill={rgb, 255:red, 163; green, 5; blue, 255 }  ,fill opacity=1 ][line width=0.08]  [draw opacity=0] (16.97,-8.15) -- (0,0) -- (16.97,8.15) -- cycle    ;

\draw [color={rgb, 255:red, 72; green, 0; blue, 255 }  ,draw opacity=1 ][fill={rgb, 255:red, 99; green, 0; blue, 255 }  ,fill opacity=1 ][line width=3]    (105.48,971.38) -- (128.65,959.57) ;
\draw [shift={(133.99,956.85)}, rotate = 513] [fill={rgb, 255:red, 72; green, 0; blue, 255 }  ,fill opacity=1 ][line width=0.08]  [draw opacity=0] (16.97,-8.15) -- (0,0) -- (16.97,8.15) -- cycle    ;

\draw [color={rgb, 255:red, 0; green, 43; blue, 255 }  ,draw opacity=1 ][fill={rgb, 255:red, 10; green, 0; blue, 255 }  ,fill opacity=1 ][line width=3]    (135.32,972.83) -- (157.12,958.67) ;
\draw [shift={(162.15,955.4)}, rotate = 507] [fill={rgb, 255:red, 0; green, 43; blue, 255 }  ,fill opacity=1 ][line width=0.08]  [draw opacity=0] (16.97,-8.15) -- (0,0) -- (16.97,8.15) -- cycle    ;

\draw [color={rgb, 255:red, 0; green, 150; blue, 255 }  ,draw opacity=1 ][line width=3]    (165.13,973.96) -- (185.62,957.96) ;
\draw [shift={(190.34,954.26)}, rotate = 502] [fill={rgb, 255:red, 0; green, 150; blue, 255 }  ,fill opacity=1 ][line width=0.08]  [draw opacity=0] (16.97,-8.15) -- (0,0) -- (16.97,8.15) -- cycle    ;

\draw [color={rgb, 255:red, 48; green, 203; blue, 89 }  ,draw opacity=1 ][line width=3]    (195.23,975.23) -- (213.93,957.17) ;
\draw [shift={(218.25,953)}, rotate = 496] [fill={rgb, 255:red, 48; green, 203; blue, 89 }  ,fill opacity=1 ][line width=0.08]  [draw opacity=0] (16.97,-8.15) -- (0,0) -- (16.97,8.15) -- cycle    ;

\draw [color={rgb, 255:red, 25; green, 157; blue, 35 }  ,draw opacity=1 ][line width=3]    (225.24,976.19) -- (242.3,956.57) ;
\draw [shift={(246.23,952.04)}, rotate = 491] [fill={rgb, 255:red, 25; green, 157; blue, 35 }  ,fill opacity=1 ][line width=0.08]  [draw opacity=0] (16.97,-8.15) -- (0,0) -- (16.97,8.15) -- cycle    ;

\draw [color={rgb, 255:red, 116; green, 149; blue, 39 }  ,draw opacity=1 ][line width=3]    (255.56,977.22) -- (270.47,955.92) ;
\draw [shift={(273.91,951.01)}, rotate = 485] [fill={rgb, 255:red, 116; green, 149; blue, 39 }  ,fill opacity=1 ][line width=0.08]  [draw opacity=0] (16.97,-8.15) -- (0,0) -- (16.97,8.15) -- cycle    ;

\draw [color={rgb, 255:red, 248; green, 231; blue, 28 }  ,draw opacity=1 ][line width=3]    (285.74,977.97) -- (298.74,955.45) ;
\draw [shift={(301.74,950.26)}, rotate = 480] [fill={rgb, 255:red, 248; green, 231; blue, 28 }  ,fill opacity=1 ][line width=0.08]  [draw opacity=0] (16.97,-8.15) -- (0,0) -- (16.97,8.15) -- cycle    ;

\draw [color={rgb, 255:red, 255; green, 140; blue, 0 }  ,draw opacity=1 ][line width=3]    (316.23,978.73) -- (326.8,954.98) ;
\draw [shift={(329.24,949.5)}, rotate = 474] [fill={rgb, 255:red, 255; green, 140; blue, 0 }  ,fill opacity=1 ][line width=0.08]  [draw opacity=0] (16.97,-8.15) -- (0,0) -- (16.97,8.15) -- cycle    ;

\draw [color={rgb, 255:red, 255; green, 0; blue, 0 }  ,draw opacity=1 ][fill={rgb, 255:red, 255; green, 0; blue, 0 }  ,fill opacity=1 ][line width=3]    (346.53,979.24) -- (354.99,954.66) ;
\draw [shift={(356.94,948.99)}, rotate = 469] [fill={rgb, 255:red, 255; green, 0; blue, 0 }  ,fill opacity=1 ][line width=0.08]  [draw opacity=0] (16.97,-8.15) -- (0,0) -- (16.97,8.15) -- cycle    ;

\draw [color={rgb, 255:red, 255; green, 0; blue, 0 }  ,draw opacity=1 ][fill={rgb, 255:red, 255; green, 0; blue, 0 }  ,fill opacity=1 ][line width=3]    (377.14,979.7) -- (382.99,954.37) ;
\draw [shift={(384.34,948.52)}, rotate = 463] [fill={rgb, 255:red, 255; green, 0; blue, 0 }  ,fill opacity=1 ][line width=0.08]  [draw opacity=0] (16.97,-8.15) -- (0,0) -- (16.97,8.15) -- cycle    ;

\draw [color={rgb, 255:red, 255; green, 0; blue, 0 }  ,draw opacity=1 ][line width=3]    (407.51,979.96) -- (411.13,954.21) ;
\draw [shift={(411.96,948.27)}, rotate = 458] [fill={rgb, 255:red, 255; green, 0; blue, 0 }  ,fill opacity=1 ][line width=0.08]  [draw opacity=0] (16.97,-8.15) -- (0,0) -- (16.97,8.15) -- cycle    ;

\draw [color={rgb, 255:red, 163; green, 5; blue, 255 }  ,draw opacity=1 ][fill={rgb, 255:red, 200; green, 14; blue, 253 }  ,fill opacity=1 ][line width=3]    (74.74,1015.11) -- (100.74,1015.11) ;
\draw [shift={(106.74,1015.11)}, rotate = 180] [fill={rgb, 255:red, 163; green, 5; blue, 255 }  ,fill opacity=1 ][line width=0.08]  [draw opacity=0] (16.97,-8.15) -- (0,0) -- (16.97,8.15) -- cycle    ;

\draw [color={rgb, 255:red, 72; green, 0; blue, 255 }  ,draw opacity=1 ][fill={rgb, 255:red, 99; green, 0; blue, 255 }  ,fill opacity=1 ][line width=3]    (103.86,1017.06) -- (129.66,1013.89) ;
\draw [shift={(135.62,1013.16)}, rotate = 533] [fill={rgb, 255:red, 72; green, 0; blue, 255 }  ,fill opacity=1 ][line width=0.08]  [draw opacity=0] (16.97,-8.15) -- (0,0) -- (16.97,8.15) -- cycle    ;

\draw [color={rgb, 255:red, 0; green, 43; blue, 255 }  ,draw opacity=1 ][fill={rgb, 255:red, 10; green, 0; blue, 255 }  ,fill opacity=1 ][line width=3]    (133.21,1018.98) -- (158.44,1012.69) ;
\draw [shift={(164.26,1011.24)}, rotate = 526] [fill={rgb, 255:red, 0; green, 43; blue, 255 }  ,fill opacity=1 ][line width=0.08]  [draw opacity=0] (16.97,-8.15) -- (0,0) -- (16.97,8.15) -- cycle    ;

\draw [color={rgb, 255:red, 0; green, 150; blue, 255 }  ,draw opacity=1 ][line width=3]    (162.8,1020.85) -- (187.07,1011.53) ;
\draw [shift={(192.67,1009.38)}, rotate = 519] [fill={rgb, 255:red, 0; green, 150; blue, 255 }  ,fill opacity=1 ][line width=0.08]  [draw opacity=0] (16.97,-8.15) -- (0,0) -- (16.97,8.15) -- cycle    ;

\draw [color={rgb, 255:red, 48; green, 203; blue, 89 }  ,draw opacity=1 ][line width=3]    (192.74,1022.87) -- (215.48,1010.27) ;
\draw [shift={(220.73,1007.36)}, rotate = 511] [fill={rgb, 255:red, 48; green, 203; blue, 89 }  ,fill opacity=1 ][line width=0.08]  [draw opacity=0] (16.97,-8.15) -- (0,0) -- (16.97,8.15) -- cycle    ;

\draw [color={rgb, 255:red, 25; green, 157; blue, 35 }  ,draw opacity=1 ][line width=3]    (222.79,1024.52) -- (243.83,1009.24) ;
\draw [shift={(248.68,1005.71)}, rotate = 504] [fill={rgb, 255:red, 25; green, 157; blue, 35 }  ,fill opacity=1 ][line width=0.08]  [draw opacity=0] (16.97,-8.15) -- (0,0) -- (16.97,8.15) -- cycle    ;

\draw [color={rgb, 255:red, 116; green, 149; blue, 39 }  ,draw opacity=1 ][line width=3]    (253.03,1026.03) -- (272.05,1008.29) ;
\draw [shift={(276.44,1004.2)}, rotate = 497] [fill={rgb, 255:red, 116; green, 149; blue, 39 }  ,fill opacity=1 ][line width=0.08]  [draw opacity=0] (16.97,-8.15) -- (0,0) -- (16.97,8.15) -- cycle    ;

\draw [color={rgb, 255:red, 248; green, 231; blue, 28 }  ,draw opacity=1 ][line width=3]    (283.45,1027.37) -- (300.16,1007.45) ;
\draw [shift={(304.02,1002.86)}, rotate = 490] [fill={rgb, 255:red, 248; green, 231; blue, 28 }  ,fill opacity=1 ][line width=0.08]  [draw opacity=0] (16.97,-8.15) -- (0,0) -- (16.97,8.15) -- cycle    ;

\draw [color={rgb, 255:red, 255; green, 140; blue, 0 }  ,draw opacity=1 ][line width=3]    (314.26,1028.68) -- (328.04,1006.63) ;
\draw [shift={(331.21,1001.54)}, rotate = 482] [fill={rgb, 255:red, 255; green, 140; blue, 0 }  ,fill opacity=1 ][line width=0.08]  [draw opacity=0] (16.97,-8.15) -- (0,0) -- (16.97,8.15) -- cycle    ;

\draw [color={rgb, 255:red, 255; green, 0; blue, 0 }  ,draw opacity=1 ][fill={rgb, 255:red, 255; green, 0; blue, 0 }  ,fill opacity=1 ][line width=3]    (344.97,1029.61) -- (355.96,1006.05) ;
\draw [shift={(358.5,1000.61)}, rotate = 475] [fill={rgb, 255:red, 255; green, 0; blue, 0 }  ,fill opacity=1 ][line width=0.08]  [draw opacity=0] (16.97,-8.15) -- (0,0) -- (16.97,8.15) -- cycle    ;

\draw [color={rgb, 255:red, 255; green, 0; blue, 0 }  ,draw opacity=1 ][fill={rgb, 255:red, 255; green, 0; blue, 0 }  ,fill opacity=1 ][line width=3]    (375.79,1030.33) -- (383.83,1005.6) ;
\draw [shift={(385.68,999.9)}, rotate = 468] [fill={rgb, 255:red, 255; green, 0; blue, 0 }  ,fill opacity=1 ][line width=0.08]  [draw opacity=0] (16.97,-8.15) -- (0,0) -- (16.97,8.15) -- cycle    ;

\draw [color={rgb, 255:red, 255; green, 0; blue, 0 }  ,draw opacity=1 ][line width=3]    (406.96,1030.87) -- (411.47,1005.27) ;
\draw [shift={(412.51,999.36)}, rotate = 460] [fill={rgb, 255:red, 255; green, 0; blue, 0 }  ,fill opacity=1 ][line width=0.08]  [draw opacity=0] (16.97,-8.15) -- (0,0) -- (16.97,8.15) -- cycle    ;

\draw [color={rgb, 255:red, 0; green, 0; blue, 0 }  ,draw opacity=1 ][fill={rgb, 255:red, 200; green, 14; blue, 253 }  ,fill opacity=1 ][line width=3]    (485.74,818.11) -- (485.74,792.11) ;
\draw [shift={(485.74,786.11)}, rotate = 450] [fill={rgb, 255:red, 0; green, 0; blue, 0 }  ,fill opacity=1 ][line width=0.08]  [draw opacity=0] (16.97,-8.15) -- (0,0) -- (16.97,8.15) -- cycle    ;

\draw [line width=3]    (479.48,870.84) -- (489.64,846.91) ;
\draw [shift={(491.99,841.39)}, rotate = 473] [fill={rgb, 255:red, 0; green, 0; blue, 0 }  ][line width=0.08]  [draw opacity=0] (16.97,-8.15) -- (0,0) -- (16.97,8.15) -- cycle    ;

\draw [line width=3]    (480,871.05) -- (489.32,846.78) ;
\draw [shift={(491.47,841.18)}, rotate = 471] [fill={rgb, 255:red, 0; green, 0; blue, 0 }  ][line width=0.08]  [draw opacity=0] (16.97,-8.15) -- (0,0) -- (16.97,8.15) -- cycle    ;

\draw [line width=3]    (480.53,871.24) -- (488.99,846.66) ;
\draw [shift={(490.94,840.99)}, rotate = 469] [fill={rgb, 255:red, 0; green, 0; blue, 0 }  ][line width=0.08]  [draw opacity=0] (16.97,-8.15) -- (0,0) -- (16.97,8.15) -- cycle    ;

\draw [line width=3]    (480.79,871.33) -- (488.83,846.6) ;
\draw [shift={(490.68,840.9)}, rotate = 468] [fill={rgb, 255:red, 0; green, 0; blue, 0 }  ][line width=0.08]  [draw opacity=0] (16.97,-8.15) -- (0,0) -- (16.97,8.15) -- cycle    ;

\draw [line width=3]    (481.33,871.49) -- (488.49,846.5) ;
\draw [shift={(490.15,840.73)}, rotate = 466] [fill={rgb, 255:red, 0; green, 0; blue, 0 }  ][line width=0.08]  [draw opacity=0] (16.97,-8.15) -- (0,0) -- (16.97,8.15) -- cycle    ;

\draw [line width=3]    (481.87,871.64) -- (488.16,846.41) ;
\draw [shift={(489.61,840.59)}, rotate = 464] [fill={rgb, 255:red, 0; green, 0; blue, 0 }  ][line width=0.08]  [draw opacity=0] (16.97,-8.15) -- (0,0) -- (16.97,8.15) -- cycle    ;

\draw [line width=3]    (482.41,871.76) -- (487.81,846.33) ;
\draw [shift={(489.06,840.46)}, rotate = 462] [fill={rgb, 255:red, 0; green, 0; blue, 0 }  ][line width=0.08]  [draw opacity=0] (16.97,-8.15) -- (0,0) -- (16.97,8.15) -- cycle    ;

\draw [line width=3]    (482.96,871.87) -- (487.47,846.27) ;
\draw [shift={(488.51,840.36)}, rotate = 460] [fill={rgb, 255:red, 0; green, 0; blue, 0 }  ][line width=0.08]  [draw opacity=0] (16.97,-8.15) -- (0,0) -- (16.97,8.15) -- cycle    ;

\draw [line width=3]    (483.51,871.96) -- (487.13,846.21) ;
\draw [shift={(487.96,840.27)}, rotate = 458] [fill={rgb, 255:red, 0; green, 0; blue, 0 }  ][line width=0.08]  [draw opacity=0] (16.97,-8.15) -- (0,0) -- (16.97,8.15) -- cycle    ;

\draw [line width=3]    (483.79,871.99) -- (486.95,846.19) ;
\draw [shift={(487.69,840.23)}, rotate = 457] [fill={rgb, 255:red, 0; green, 0; blue, 0 }  ][line width=0.08]  [draw opacity=0] (16.97,-8.15) -- (0,0) -- (16.97,8.15) -- cycle    ;

\draw [line width=3]    (484.34,872.05) -- (486.61,846.15) ;
\draw [shift={(487.13,840.17)}, rotate = 455] [fill={rgb, 255:red, 0; green, 0; blue, 0 }  ][line width=0.08]  [draw opacity=0] (16.97,-8.15) -- (0,0) -- (16.97,8.15) -- cycle    ;

\draw [line width=3]    (484.9,872.09) -- (486.26,846.13) ;
\draw [shift={(486.57,840.14)}, rotate = 453] [fill={rgb, 255:red, 0; green, 0; blue, 0 }  ][line width=0.08]  [draw opacity=0] (16.97,-8.15) -- (0,0) -- (16.97,8.15) -- cycle    ;

\draw [line width=3]    (474.42,922.43) -- (492.81,904.04) ;
\draw [shift={(497.05,899.8)}, rotate = 495] [fill={rgb, 255:red, 0; green, 0; blue, 0 }  ][line width=0.08]  [draw opacity=0] (16.97,-8.15) -- (0,0) -- (16.97,8.15) -- cycle    ;

\draw [line width=3]    (475.03,923) -- (492.43,903.68) ;
\draw [shift={(496.44,899.22)}, rotate = 492] [fill={rgb, 255:red, 0; green, 0; blue, 0 }  ][line width=0.08]  [draw opacity=0] (16.97,-8.15) -- (0,0) -- (16.97,8.15) -- cycle    ;

\draw [line width=3]    (475.89,923.72) -- (491.89,903.23) ;
\draw [shift={(495.59,898.51)}, rotate = 488] [fill={rgb, 255:red, 0; green, 0; blue, 0 }  ][line width=0.08]  [draw opacity=0] (16.97,-8.15) -- (0,0) -- (16.97,8.15) -- cycle    ;

\draw [line width=3]    (476.56,924.22) -- (491.47,902.92) ;
\draw [shift={(494.91,898.01)}, rotate = 485] [fill={rgb, 255:red, 0; green, 0; blue, 0 }  ][line width=0.08]  [draw opacity=0] (16.97,-8.15) -- (0,0) -- (16.97,8.15) -- cycle    ;

\draw [line width=3]    (477.5,924.83) -- (490.89,902.54) ;
\draw [shift={(493.98,897.4)}, rotate = 481] [fill={rgb, 255:red, 0; green, 0; blue, 0 }  ][line width=0.08]  [draw opacity=0] (16.97,-8.15) -- (0,0) -- (16.97,8.15) -- cycle    ;

\draw [line width=3]    (478.47,925.37) -- (490.28,902.2) ;
\draw [shift={(493,896.86)}, rotate = 477] [fill={rgb, 255:red, 0; green, 0; blue, 0 }  ][line width=0.08]  [draw opacity=0] (16.97,-8.15) -- (0,0) -- (16.97,8.15) -- cycle    ;

\draw [line width=3]    (479.23,925.73) -- (489.8,901.98) ;
\draw [shift={(492.24,896.5)}, rotate = 474] [fill={rgb, 255:red, 0; green, 0; blue, 0 }  ][line width=0.08]  [draw opacity=0] (16.97,-8.15) -- (0,0) -- (16.97,8.15) -- cycle    ;

\draw [line width=3]    (480.26,926.15) -- (489.16,901.72) ;
\draw [shift={(491.21,896.08)}, rotate = 470] [fill={rgb, 255:red, 0; green, 0; blue, 0 }  ][line width=0.08]  [draw opacity=0] (16.97,-8.15) -- (0,0) -- (16.97,8.15) -- cycle    ;

\draw [line width=3]    (481.33,926.49) -- (488.49,901.5) ;
\draw [shift={(490.15,895.73)}, rotate = 466] [fill={rgb, 255:red, 0; green, 0; blue, 0 }  ][line width=0.08]  [draw opacity=0] (16.97,-8.15) -- (0,0) -- (16.97,8.15) -- cycle    ;

\draw [line width=3]    (482.14,926.7) -- (487.99,901.37) ;
\draw [shift={(489.34,895.52)}, rotate = 463] [fill={rgb, 255:red, 0; green, 0; blue, 0 }  ][line width=0.08]  [draw opacity=0] (16.97,-8.15) -- (0,0) -- (16.97,8.15) -- cycle    ;

\draw [line width=3]    (483.23,926.92) -- (487.3,901.24) ;
\draw [shift={(488.24,895.31)}, rotate = 459] [fill={rgb, 255:red, 0; green, 0; blue, 0 }  ][line width=0.08]  [draw opacity=0] (16.97,-8.15) -- (0,0) -- (16.97,8.15) -- cycle    ;

\draw [line width=3]    (484.34,927.05) -- (486.61,901.15) ;
\draw [shift={(487.13,895.17)}, rotate = 455] [fill={rgb, 255:red, 0; green, 0; blue, 0 }  ][line width=0.08]  [draw opacity=0] (16.97,-8.15) -- (0,0) -- (16.97,8.15) -- cycle    ;

\draw [line width=3]    (470.9,970.11) -- (495.01,960.37) ;
\draw [shift={(500.57,958.12)}, rotate = 518] [fill={rgb, 255:red, 0; green, 0; blue, 0 }  ][line width=0.08]  [draw opacity=0] (16.97,-8.15) -- (0,0) -- (16.97,8.15) -- cycle    ;

\draw [line width=3]    (471.48,971.38) -- (494.65,959.57) ;
\draw [shift={(499.99,956.85)}, rotate = 513] [fill={rgb, 255:red, 0; green, 0; blue, 0 }  ][line width=0.08]  [draw opacity=0] (16.97,-8.15) -- (0,0) -- (16.97,8.15) -- cycle    ;

\draw [line width=3]    (472.32,972.83) -- (494.12,958.67) ;
\draw [shift={(499.15,955.4)}, rotate = 507] [fill={rgb, 255:red, 0; green, 0; blue, 0 }  ][line width=0.08]  [draw opacity=0] (16.97,-8.15) -- (0,0) -- (16.97,8.15) -- cycle    ;

\draw [line width=3]    (473.13,973.96) -- (493.62,957.96) ;
\draw [shift={(498.34,954.26)}, rotate = 502] [fill={rgb, 255:red, 0; green, 0; blue, 0 }  ][line width=0.08]  [draw opacity=0] (16.97,-8.15) -- (0,0) -- (16.97,8.15) -- cycle    ;

\draw [line width=3]    (474.23,975.23) -- (492.93,957.17) ;
\draw [shift={(497.25,953)}, rotate = 496] [fill={rgb, 255:red, 0; green, 0; blue, 0 }  ][line width=0.08]  [draw opacity=0] (16.97,-8.15) -- (0,0) -- (16.97,8.15) -- cycle    ;

\draw [line width=3]    (475.24,976.19) -- (492.3,956.57) ;
\draw [shift={(496.23,952.04)}, rotate = 491] [fill={rgb, 255:red, 0; green, 0; blue, 0 }  ][line width=0.08]  [draw opacity=0] (16.97,-8.15) -- (0,0) -- (16.97,8.15) -- cycle    ;

\draw [line width=3]    (476.56,977.22) -- (491.47,955.92) ;
\draw [shift={(494.91,951.01)}, rotate = 485] [fill={rgb, 255:red, 0; green, 0; blue, 0 }  ][line width=0.08]  [draw opacity=0] (16.97,-8.15) -- (0,0) -- (16.97,8.15) -- cycle    ;

\draw [line width=3]    (477.74,977.97) -- (490.74,955.45) ;
\draw [shift={(493.74,950.26)}, rotate = 480] [fill={rgb, 255:red, 0; green, 0; blue, 0 }  ][line width=0.08]  [draw opacity=0] (16.97,-8.15) -- (0,0) -- (16.97,8.15) -- cycle    ;

\draw [line width=3]    (479.23,978.73) -- (489.8,954.98) ;
\draw [shift={(492.24,949.5)}, rotate = 474] [fill={rgb, 255:red, 0; green, 0; blue, 0 }  ][line width=0.08]  [draw opacity=0] (16.97,-8.15) -- (0,0) -- (16.97,8.15) -- cycle    ;

\draw [line width=3]    (480.53,979.24) -- (488.99,954.66) ;
\draw [shift={(490.94,948.99)}, rotate = 469] [fill={rgb, 255:red, 0; green, 0; blue, 0 }  ][line width=0.08]  [draw opacity=0] (16.97,-8.15) -- (0,0) -- (16.97,8.15) -- cycle    ;

\draw [line width=3]    (482.14,979.7) -- (487.99,954.37) ;
\draw [shift={(489.34,948.52)}, rotate = 463] [fill={rgb, 255:red, 0; green, 0; blue, 0 }  ][line width=0.08]  [draw opacity=0] (16.97,-8.15) -- (0,0) -- (16.97,8.15) -- cycle    ;

\draw [line width=3]    (483.51,979.96) -- (487.13,954.21) ;
\draw [shift={(487.96,948.27)}, rotate = 458] [fill={rgb, 255:red, 0; green, 0; blue, 0 }  ][line width=0.08]  [draw opacity=0] (16.97,-8.15) -- (0,0) -- (16.97,8.15) -- cycle    ;

\draw [line width=3]    (469.74,1015.11) -- (495.74,1015.11) ;
\draw [shift={(501.74,1015.11)}, rotate = 180] [fill={rgb, 255:red, 0; green, 0; blue, 0 }  ][line width=0.08]  [draw opacity=0] (16.97,-8.15) -- (0,0) -- (16.97,8.15) -- cycle    ;

\draw [line width=3]    (469.86,1017.06) -- (495.66,1013.89) ;
\draw [shift={(501.62,1013.16)}, rotate = 533] [fill={rgb, 255:red, 0; green, 0; blue, 0 }  ][line width=0.08]  [draw opacity=0] (16.97,-8.15) -- (0,0) -- (16.97,8.15) -- cycle    ;

\draw [line width=3]    (470.21,1018.98) -- (495.44,1012.69) ;
\draw [shift={(501.26,1011.24)}, rotate = 526] [fill={rgb, 255:red, 0; green, 0; blue, 0 }  ][line width=0.08]  [draw opacity=0] (16.97,-8.15) -- (0,0) -- (16.97,8.15) -- cycle    ;

\draw [line width=3]    (470.8,1020.85) -- (495.07,1011.53) ;
\draw [shift={(500.67,1009.38)}, rotate = 519] [fill={rgb, 255:red, 0; green, 0; blue, 0 }  ][line width=0.08]  [draw opacity=0] (16.97,-8.15) -- (0,0) -- (16.97,8.15) -- cycle    ;

\draw [line width=3]    (471.74,1022.87) -- (494.48,1010.27) ;
\draw [shift={(499.73,1007.36)}, rotate = 511] [fill={rgb, 255:red, 0; green, 0; blue, 0 }  ][line width=0.08]  [draw opacity=0] (16.97,-8.15) -- (0,0) -- (16.97,8.15) -- cycle    ;

\draw [line width=3]    (472.79,1024.52) -- (493.83,1009.24) ;
\draw [shift={(498.68,1005.71)}, rotate = 504] [fill={rgb, 255:red, 0; green, 0; blue, 0 }  ][line width=0.08]  [draw opacity=0] (16.97,-8.15) -- (0,0) -- (16.97,8.15) -- cycle    ;

\draw [line width=3]    (474.03,1026.03) -- (493.05,1008.29) ;
\draw [shift={(497.44,1004.2)}, rotate = 497] [fill={rgb, 255:red, 0; green, 0; blue, 0 }  ][line width=0.08]  [draw opacity=0] (16.97,-8.15) -- (0,0) -- (16.97,8.15) -- cycle    ;

\draw [line width=3]    (475.45,1027.37) -- (492.16,1007.45) ;
\draw [shift={(496.02,1002.86)}, rotate = 490] [fill={rgb, 255:red, 0; green, 0; blue, 0 }  ][line width=0.08]  [draw opacity=0] (16.97,-8.15) -- (0,0) -- (16.97,8.15) -- cycle    ;

\draw [line width=3]    (477.26,1028.68) -- (491.04,1006.63) ;
\draw [shift={(494.21,1001.54)}, rotate = 482] [fill={rgb, 255:red, 0; green, 0; blue, 0 }  ][line width=0.08]  [draw opacity=0] (16.97,-8.15) -- (0,0) -- (16.97,8.15) -- cycle    ;

\draw [line width=3]    (478.97,1029.61) -- (489.96,1006.05) ;
\draw [shift={(492.5,1000.61)}, rotate = 475] [fill={rgb, 255:red, 0; green, 0; blue, 0 }  ][line width=0.08]  [draw opacity=0] (16.97,-8.15) -- (0,0) -- (16.97,8.15) -- cycle    ;

\draw [line width=3]    (480.79,1030.33) -- (488.83,1005.6) ;
\draw [shift={(490.68,999.9)}, rotate = 468] [fill={rgb, 255:red, 0; green, 0; blue, 0 }  ][line width=0.08]  [draw opacity=0] (16.97,-8.15) -- (0,0) -- (16.97,8.15) -- cycle    ;

\draw [line width=3]    (482.96,1030.87) -- (487.47,1005.27) ;
\draw [shift={(488.51,999.36)}, rotate = 460] [fill={rgb, 255:red, 0; green, 0; blue, 0 }  ][line width=0.08]  [draw opacity=0] (16.97,-8.15) -- (0,0) -- (16.97,8.15) -- cycle    ;

\draw [color={rgb, 255:red, 0; green, 0; blue, 0 }  ,draw opacity=1 ][fill={rgb, 255:red, 200; green, 14; blue, 253 }  ,fill opacity=1 ][line width=3]    (562.74,818.11) -- (562.74,792.11) ;
\draw [shift={(562.74,786.11)}, rotate = 450] [fill={rgb, 255:red, 0; green, 0; blue, 0 }  ,fill opacity=1 ][line width=0.08]  [draw opacity=0] (16.97,-8.15) -- (0,0) -- (16.97,8.15) -- cycle    ;

\draw [color={rgb, 255:red, 74; green, 74; blue, 74 }  ,draw opacity=1 ][line width=3]    (560.41,871.76) -- (566.16,850.63) ;
\draw [shift={(567.74,844.84)}, rotate = 465.22] [fill={rgb, 255:red, 74; green, 74; blue, 74 }  ,fill opacity=1 ][line width=0.08]  [draw opacity=0] (16.97,-8.15) -- (0,0) -- (16.97,8.15) -- cycle    ;

\draw [color={rgb, 255:red, 128; green, 128; blue, 128 }  ,draw opacity=1 ][line width=3]    (559.23,923.73) -- (565.47,908.4) ;
\draw [shift={(567.74,902.84)}, rotate = 472.16] [fill={rgb, 255:red, 128; green, 128; blue, 128 }  ,fill opacity=1 ][line width=0.08]  [draw opacity=0] (16.97,-8.15) -- (0,0) -- (16.97,8.15) -- cycle    ;

\draw [color={rgb, 255:red, 155; green, 155; blue, 155 }  ,draw opacity=1 ][line width=1.5]    (557.56,974.22) -- (564.58,963.21) ;
\draw [shift={(566.74,959.84)}, rotate = 482.55] [fill={rgb, 255:red, 155; green, 155; blue, 155 }  ,fill opacity=1 ][line width=0.08]  [draw opacity=0] (11.61,-5.58) -- (0,0) -- (11.61,5.58) -- cycle    ;

\draw [color={rgb, 255:red, 222; green, 222; blue, 222 }  ,draw opacity=1 ][line width=1.5]    (556.03,1025.03) -- (563.84,1017.6) ;
\draw [shift={(566.74,1014.84)}, rotate = 496.42] [fill={rgb, 255:red, 222; green, 222; blue, 222 }  ,fill opacity=1 ][line width=0.08]  [draw opacity=0] (11.61,-5.58) -- (0,0) -- (11.61,5.58) -- cycle    ;

\draw    (63,1031.93) -- (63,1051.11) ;

\draw    (63,1051.11) -- (431.74,1051.11) ;

\draw    (431.74,1031.93) -- (431.74,1051.11) ;

\draw    (458.74,1030.93) -- (458.74,1050.11) ;

\draw    (458.74,1050.11) -- (507.74,1050.11) ;

\draw    (507.74,1030.93) -- (507.74,1050.11) ;

\draw    (535.74,1030.93) -- (535.74,1050.11) ;

\draw    (535.74,1050.11) -- (584.74,1050.11) ;

\draw    (584.74,1030.93) -- (584.74,1050.11) ;

\draw [color={rgb, 255:red, 74; green, 74; blue, 74 }  ,draw opacity=1 ][line width=3]    (47,781.11) -- (47,1087.11) ;
\draw [shift={(47,1092.11)}, rotate = 270] [color={rgb, 255:red, 74; green, 74; blue, 74 }  ,draw opacity=1 ][line width=3]    (20.77,-6.25) .. controls (13.2,-2.65) and (6.28,-0.57) .. (0,0) .. controls (6.28,0.57) and (13.2,2.66) .. (20.77,6.25)   ;

\draw  [color={rgb, 255:red, 0; green, 0; blue, 0 }  ,draw opacity=1 ][fill={rgb, 255:red, 248; green, 200; blue, 107 }  ,fill opacity=1 ][line width=3]  (39,781.11) .. controls (39,776.69) and (42.58,773.11) .. (47,773.11) .. controls (51.42,773.11) and (55,776.69) .. (55,781.11) .. controls (55,785.53) and (51.42,789.11) .. (47,789.11) .. controls (42.58,789.11) and (39,785.53) .. (39,781.11) -- cycle ;

\draw (245,1074.11) node [font=\Large]  [align=left] {{Polarization spectrum (increasing $\lambda$)}};
\draw (484,1073.11) node  [font=\Large] [align=left] {Total};
\draw (560,1073.11) node  [font=\Large] [align=left] {Effective};
\draw (26,929.11) node  [font=\Large,color={rgb, 255:red, 74; green, 74; blue, 74 }  ,opacity=1 ,rotate=-270] [align=left] {Photon propagation (line of sight)};
\draw (46,759.74) node  [font=\Large] [align=left] {Emitter};

\draw [color={rgb, 255:red, 178; green, 72; blue, 72 }  ,draw opacity=1 ][line width=3]    (70.74,1092.11) -- (422.74,1092.11) ;
\draw [shift={(427.74,1092.11)}, rotate = 180] [color={rgb, 255:red, 178; green, 72; blue, 72 }  ,draw opacity=1 ][line width=3]    (20.77,-6.25) .. controls (13.2,-2.65) and (6.28,-0.57) .. (0,0) .. controls (6.28,0.57) and (13.2,2.66) .. (20.77,6.25)   ;

\end{tikzpicture}

}
\vspace{\dskips}
\caption{Schematic depiction of the SME polarization angle drift in a spectrum of photons. Each arrow represents a polarization state with a given direction. Under the most conservative assumption, the initial spectrum (top row) is uniform, with the same polarization angle at all wavelengths. In-flight, the increasing influence of SME effects with wavelength and distance from the source causes the initially identical polarization angles to diverge, resulting in a smaller ``Effective" polarization if measured across the entire band, as illustrated by the ``Effective" arrow, which averages over the superposition of colored arrows in each row, as shown in the ``Total" column.
}
\label{fig:broadband_rotation}
\end{figure}

In general, the initial polarization state of an individual photon at the source (before any birefringence) is unknown, making it challenging to infer its in-flight drift due to potential Lorentz or CPT violation. However, the energy-dependence of the drift shown in \fig{}~\ref{fig:psi_vs_wl} implies that the polarization states of \textit{multiple} photons of different energies will gradually diverge, thereby reducing the overall linear polarization fraction measured across a broad range of energies. This reasoning is schematically illustrated in \fig{}~\ref{fig:broadband_rotation}. For both the CPT-odd and CPT-even cases, SME effects will tend to depolarize light coming from sources at cosmological distances, so to test the SME using broadband polarimetry, we must derive the largest theoretically possible linear polarization fraction $\pmaxd$, observable through a bandpass with energy transmission profile $T(E)$, for a given set of SME coefficients. 

\begin{figure}[h!]
    \centering
    \includegraphics[width=0.9\columnwidth]{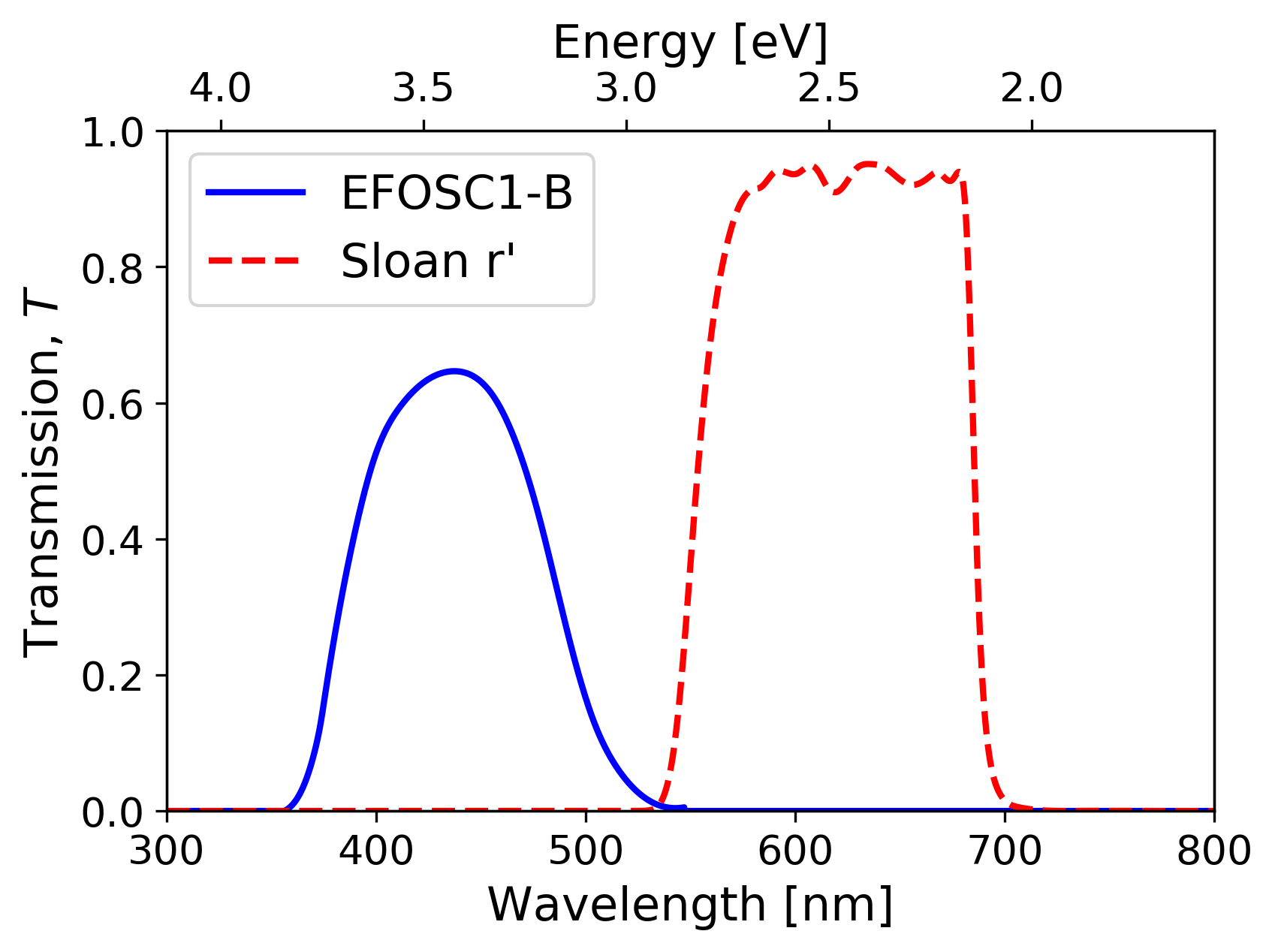}
    \vspace{\dskips}
    \caption{Transmission profiles of two arbitrarily selected bands from the compiled catalog of polarization measurements: the Bessel $V$-band on the ESO Faint Object Spectrograph and Camera \cite{EFOSC1} and the standard $r^{\prime}$-band filter from the Sloan Digital Sky Survey set \cite{SDSS_filters} 
    }
    \label{fig:transmission}
\end{figure}

\begin{figure}[h!]
    \centering
    \includegraphics[width=\columnwidth]{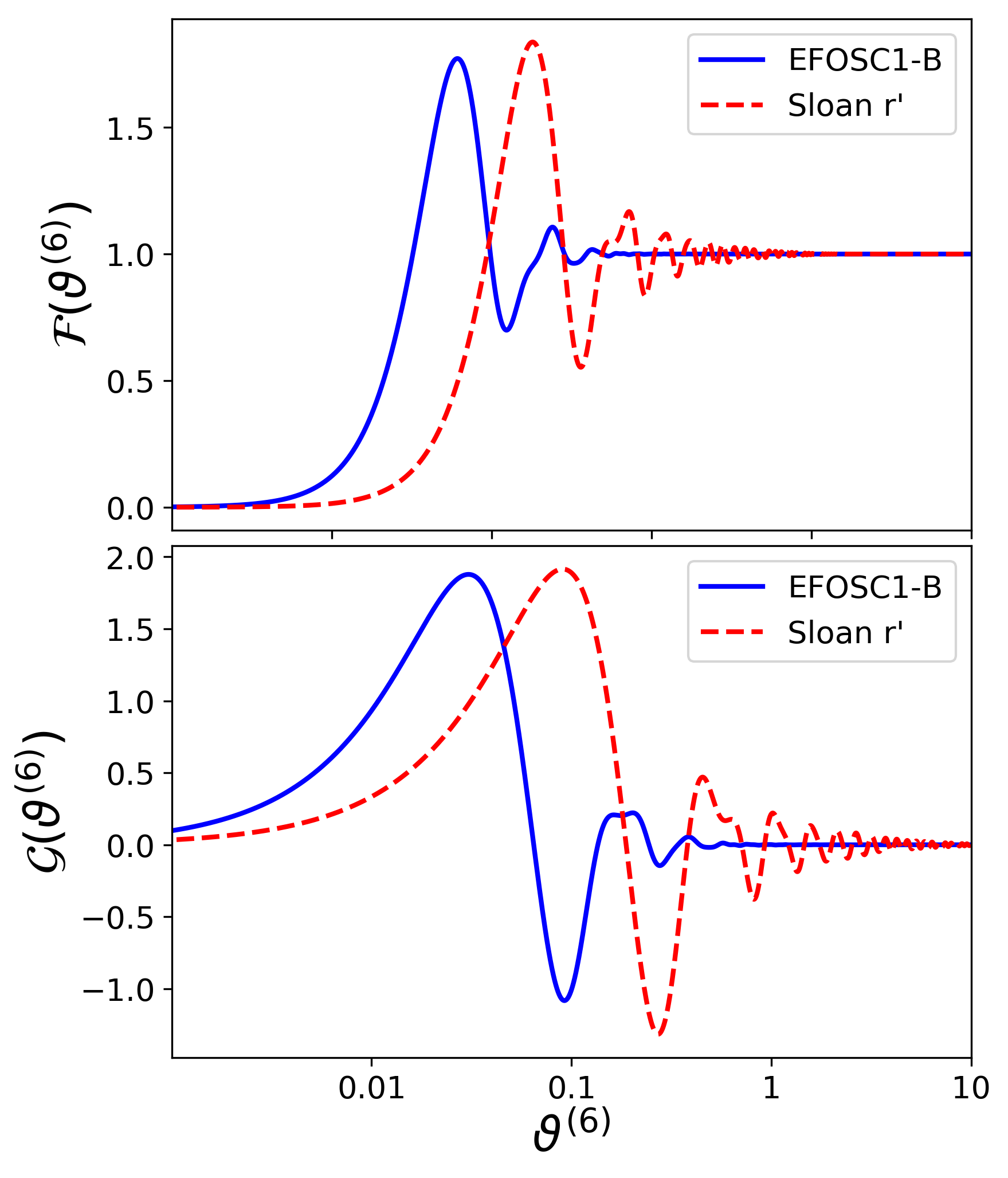}
    \vspace{\dskips}
    \caption{$\AAA$ and $\BB$ integrals defined in Eqs.~\eqref{eq:Avartheta} and \eqref{eq:Bvartheta} as functions of $\varthetad$ defined in Eq.~\eqref{eq:varthetad} for $d=6$ and the observation bands in \fig{}~\ref{fig:transmission}. The integrals encode the dependence of the maximum observable linear polarization, $\polmax$ on the band of observation, with a stronger effect for larger $d$.}
    \label{fig:F_G_integrals}
\end{figure}

In this scenario, the effective Stokes parameters observed through a given bandpass are given by 
\begin{alignat}{2}
    \mathpzc{Q}^{(d)\prime} & \equiv \NN \PqdprimeN  = \int T(E) \Pqdprime(E) dE  \nonumber \\
    & = \int T(E) \left( \Pqdzprime + \DPqdprime(E) \right) dE  \nonumber\\
    & = \Pqdzprime \int T(E) dE + \int T(E) \DPqdprime(E) dE\, ,  \\
    \mathpzc{U}^{(d)\prime} & \equiv \NN \PudprimeN  = \int T(E) \Pudprime(E) dE  \nonumber \\
    & = \int T(E) \left( \Pudzprime + \DPudprime(E) \right) dE  \nonumber \\
    & = \Pudzprime \int T(E) dE + \int T(E) \DPudprime(E) dE\, , 
    \label{eq:quprime}
\end{alignat}
where we define the instrument-dependent normalization constant 
\begin{equation}
    \NN = \int T(E) dE\, ,
    \label{eq:NN}
\end{equation}
and, following Ref.~\cite{kislat18}, we have conservatively assumed no Stokes parameter energy dependence at the source via $\Pqdzprime(E)= \Pqdzprime$ and $\Pudzprime(E)= \Pudzprime$. Substituting Eqs.~\eqref{eq:qzprime}-\eqref{eq:uzprime} for $\Pqdzprime$ and $\Pudzprime$ and Eqs.~\eqref{eq:deltaqprime}-\eqref{eq:deltauprime} for $\DPqdprime(E)$ and $\DPudprime(E)$ yields
\begin{alignat}{2}
\PqdprimeN 
& = 
\begin{cases}
         \Pqdzprime \Big[1 - \AAA\left(\varthetad(\nhat)\right)\Big]  \\
         \ \ \ \ - \frac{1}{2} \Pudzprime \GG\left(\varthetad(\nhat)\right)\, ,         
         & \text{ odd } d\, , \\
        \Pqdzprime\, ,  & \text{ even } d\, ,
\end{cases}
\label{eq:qdprimebroad}
\end{alignat}
and
\begin{alignat}{2}
\PudprimeN 
& = 
\begin{cases}
        \Pudzprime \Big[1-\FF\left(\varthetad(\nhat)\right)\Big]  \\
        \ \ \ \ + \frac{1}{2} \Pqdzprime \GG\left(\varthetad(\nhat)\right)  \, ,             & \text{ odd } d\, , \\
         \Pudzprime \Big[1 - \FF\left(\varthetad(\nhat)\right)\Big]\, ,  & \text{ even } d\, ,
\end{cases}
\label{eq:udprimebroad}
\end{alignat}
where we define the instrument-dependent integrals 
\begin{equation}
     \AAA(\varthetad) = \frac{2}{\NN}\int T(E) \sin^2\left(E^{d-3} \varthetad\right) dE\, , 
   \label{eq:Avartheta}
\end{equation}
\begin{equation}
     \BB(\varthetad) = \frac{2}{\NN}\int T(E) \sin\left(2 E^{d-3} \varthetad\right)  dE\, . 
    \label{eq:Bvartheta}
\end{equation}
For selected instruments with transmission profiles in \fig{}~\ref{fig:transmission}, sample plots of $\AAA$ and $\BB$ are available in \fig{}~\ref{fig:F_G_integrals}. 

Given a set of SME parameters for arbitrary mass dimension $d$, with the effective Stokes parameters $\PqdprimeN$ and $\PudprimeN$ given by Eqs.~\eqref{eq:qdprimebroad}-\eqref{eq:udprimebroad}, the maximum theoretically possible observed linear polarization fraction $\pmaxd$ is given by
\begin{alignat}{3}
\label{eq:pmax}
& \pmaxd = \sqrt{\Big(\PqdprimeN\Big)^2 + \Big(\PudprimeN\Big)^2}  \\
& = \pz
\begin{cases}
         \sqrt{ \Big[1 - \FF\left(\varthetad\right)\Big]^2 + \frac{1}{4}\GG\left(\varthetad\right)^2  }    \, , 
         & 
         \text{ odd } d\, , \nonumber \\
         \sqrt{ 1 - \uzoverpzprime \FF(\varthetad)\left(2 - \FF(\varthetad)\right) }  \, ,  
         & 
         \text{ even } d\, ,
\end{cases}
\end{alignat}
where we define the quantity
\begin{equation}
\uzoverpzprime \equiv \Bigg(\frac{\Pudzprime}{\pz}\Bigg)^2 \, ,
    \label{eq:uzoverpz}
\end{equation}
and we used the definition $\pz^2 = \left(\Pqdzprime\right)^2 + \left(\Pudzprime\right)^2$ to write Eq.~\eqref{eq:pmax} in terms of $\uzoverpzprime$.

\begin{figure*}[t]
    \centering
    \includegraphics[width=0.9\textwidth]{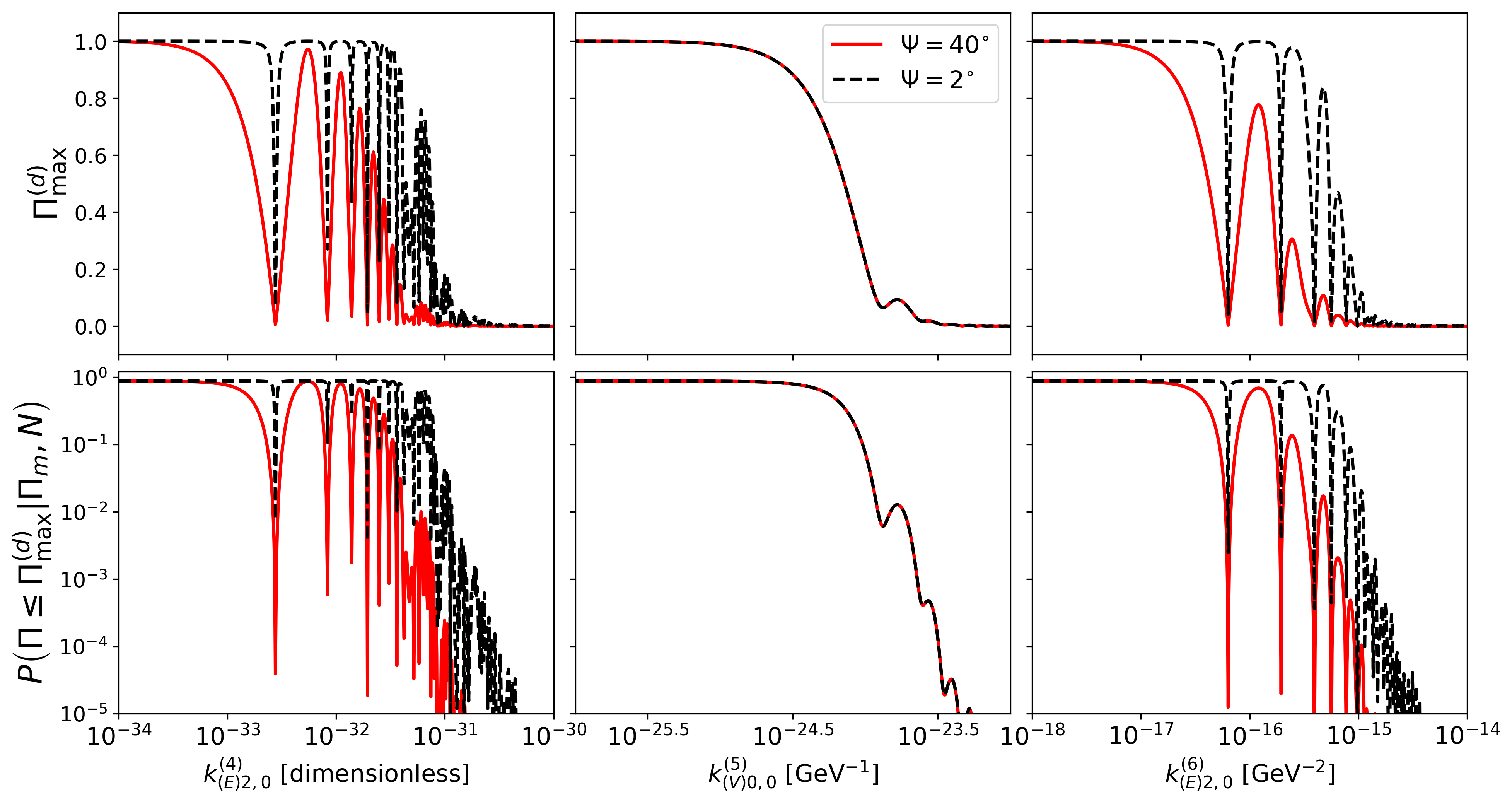}
    \vspace{\dskips}
    \caption{
    (\textit{Upper panels}) Maximum allowed linear polarization fraction from Eq.~\eqref{eq:pmax} through the Bessel V-band of the ESO Faint Object Spectrograph and Camera \cite{EFOSC1} as a function of one of the real SME coefficients with all other coefficients set to $0$. Plots are for mass dimensions $d=4,5,6$, left to right, as indicated by the x-axis labels. (\textit{Lower panels}) Probability of the same set of SME coefficients being compatible with a hypothetical observed linear polarization fraction of $\pol=0.5\pm 0.3$ given by Eq.~\eqref{eq:cumprob}. In all cases, the test source is positioned at RA $=2\mathrm{h}$, Dec $=-60\degree$, and $z=3$. All plots show a clear downward trend, as the depolarization effect of the SME-induced birefringence becomes more prominent for larger values of the chosen SME coefficient. The initial spectrum is assumed to be 100\% polarized, with $\pz=1$, with a fixed polarization angle at all wavelengths of either $40\degree$ (solid red line) or $2\degree$ (dashed black line). Due to the special alignment of the axis of birefringence in the CPT-odd case as shown in \fig{}~\ref{fig:stokes_rotation}, the middle column plots for $d=5$ do not depend on the initial polarization angle.
    The plots are symmetric about the origin, so only positive SME coefficients are shown.
     }
    \label{fig:Pi_max_VS_SME}
\end{figure*}

\begin{figure}[h!]
    \centering
    \includegraphics[width=0.9\columnwidth]{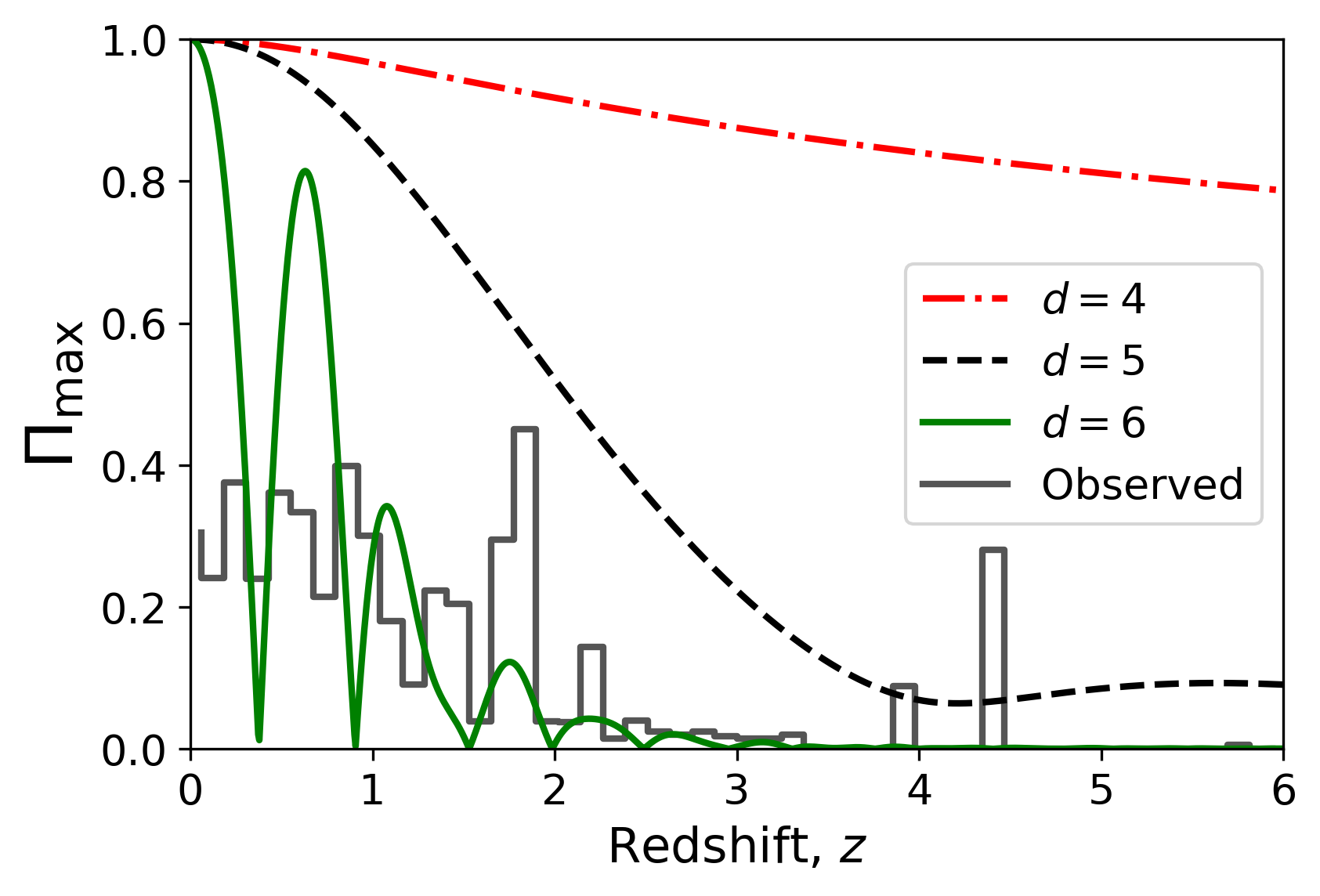}
    \vspace{\dskips}
    \caption{Maximum allowed linear polarization fraction from Eq.~\eqref{eq:pmax} through the Bessel V-band of ESO Faint Object Spectrograph and Camera \cite{EFOSC1} as a function of source redshift for different mass dimensions. In each case, all SME coefficients are set to $0$ except the ones in \fig{}~\ref{fig:Pi_max_VS_SME}, which are set to $10^{-33} \ev^{4-d}$. The source is positioned at RA $=2\mathrm{h}$, Dec $=-60\degree$. The initial spectrum is assumed to be 100\% polarized, with $\pz=1$, with a constant polarization angle of $30\degree$ at all wavelengths. The redshift dependence of $\pmaxd$ becomes stronger at increasing mass dimension, by lowering the upper envelope of the $\pmaxd(z)$ function, which asymptotes to a vanishing value, $\pmaxd \rightarrow 0$, at smaller redshifts as $d$ increases. For reference, the maximum measured linear polarization fractions from the compiled catalog of observational data are plotted in redshift bins of width $\Delta z \approx 0.122$.
    }
    \label{fig:SME_of_z}
\end{figure}

In the CPT-odd case, if we assume that $\pz$ is known, then we do not need to know the individual source frame Stokes parameters to compute $\pmaxd$, whereas in the CPT-even case, we do need to solve for the quantity $\uzoverpzprime$ defined in Eq.~\eqref{eq:uzoverpz} to compute $\pmaxd$ using Eq.~\eqref{eq:pmax}.
To do so, we use the fact that the observed polarization angle in the primed coordinate frame $\psioprime$ for the CPT-even case is given by
\begin{alignat}{3}
 \psioprime & = \psio - \varxi/2 = \frac{1}{2} \arctan\Bigg(\frac{\PudprimeN}{\PqdprimeN}\Bigg) \nonumber \\
  & = \frac{1}{2}\arctan\Bigg(\frac{1 - \FF(\varthetad)}{\sign\left(\Pqdzprime\right)\sqrt{\left(\uzoverpzprime\right)^{-1}-1}}\Bigg)\, .
\label{eq:psiprimearctan}
\end{alignat}
If we equate the theoretical and measured polarization angles in the unprimed frame, such that $\psio = \psim$, we can invert Eq.~\eqref{eq:psiprimearctan} to solve for $\uzoverpzprime$, which is given by
\begin{alignat}{3}
  \uzoverpzprime & = \Bigg[1 + \Bigg(\frac{1-\FF(\varthetad)}{\tan\left(2\psim -\varxi\right)}\Bigg)\Bigg]^{-1}\, , 
\label{eq:uzoverpzprime1}
\end{alignat}
which we then substitute back into Eq.~\eqref{eq:pmax} to solve for $\pmaxd$ in the CPT-odd case, which reveals that for both odd and even $d$, the intrinsic polarization fraction $\pz$ is indeed a simple multiplicative factor.

The rest of the broadband polarimetry analysis follows Ref.~\cite{kislat18}, where we model the probability to observe a measured polarization $\pol$ given a true polarization $\polhat$, following Refs.~\cite{weisskopf09,krawczynski11}, as given by
\begin{equation}
\begin{aligned} 
P(\pol | \hat{\pol}, N) 
&=\frac{N \pol}{2} \exp \left(-\frac{N(\pol-\hat{\pol})^{2}}{4}\right) i_{0}\left(\frac{N \pol \hat{\pol}}{2}\right)\, ,
\end{aligned}
\label{eq:probpol}
\end{equation}
where $I_0$ is the 0th order modified Bessel function, $i_0(x) = \exp(-|x|)I_0(x)$, and $N$ is related to the number of photons detected in a photon counting experiment. Following Refs.~\cite{weisskopf09,krawczynski11}, the expectation value $\polbar$ and standard deviation $\sigmapolbar$ of the observed polarization $\pol$ are given by
\begin{equation}
\begin{aligned} \polbar &=\sqrt{\frac{\pi}{16 N}} \exp \left(-\frac{N \polhat^{2}}{8}\right) \times \nonumber \\
  & \left[\left(4+N \polhat^{2}\right) I_{0}\left(\frac{N \polhat^{2}}{8}\right)+N \polhat^{2} I_{1}\left(\frac{N \polhat^{2}}{8}\right)\right]\, , \\ 
  \sigmapolbar &=\left(\polhat^{2}+\frac{4}{N}-\polbar^{2}\right)^{1 / 2}\, ,
\end{aligned}
\label{eq:expectation}
\end{equation}
where $I_1$ is the first order modified Bessel function. For a polarization measurement and error $\polmm \pm \sigmapolm$, $N$ can be computed numerically by solving $\sigmapolbar = \sigmapolm$ for $N$ assuming $\polhat=\polmm$. The cumulative probability distribution can then be found by numerically integrating Eq.~\eqref{eq:probpol} via
\begin{equation}
P\left(\pol \leq \pmaxd | \polmm, N\right)=\int_{0}^{\pmaxd} P\left(\pol | \polmm, N\right) d \Pi\, .
\label{eq:cumprob}
\end{equation}
Eq.~\eqref{eq:cumprob} thus specifies the probability that a specific set of SME coefficients for a mass dimension $d$ model, which allow a theoretical maximum polarization $\pmaxd$, is compatible with the broadband polarization measurement $\polmm \pm \sigmapolm$. 

\section{Constraining SME Coefficients}
\label{sec:constrain}

In this work, we wish to obtain constraints on the individual birefringent SME coefficients $\kdVjm$ for CPT-odd $d$, and for $\kdEjm$, and $\kdBjm$ for CPT-even $d$.  In each case, let us call these coefficients $\kdXjm$, where $X \in \{V, \{E,B\}\}$ for odd and even $d$, respectively. We can then combine broadband measurements from multiple sources, and multiple observations for each source, using the cumulative probability distribution in Eq.~\eqref{eq:cumprob}. By assuming $i$ independent measurements of individual astronomical sources, where observations of the same source at different times are also assumed to be independent, the combined probability distribution is given by
\begin{equation}
    P(\kdXjm) = \prod_{i} P_i(\kdXjm)\, .
    \label{eq:combinedprob}
\end{equation}

The multi-dimensional distribution in Eq.~\eqref{eq:combinedprob} is best probed using Markov-Chain Monte Carlo (MCMC) methods, for example, the Metropolis-Hastings algorithm used in Ref.~\cite{kislat18}. The likelihood space is sampled by placing one or more so-called \textit{walkers} at some initial positions (i.e. some values of the SME coefficients) and moving them in a chain of trials. On each trial, the direction and distance of the move are drawn randomly from some \textit{proposal distribution} for each walker. The ratio of the new likelihood to the old one is calculated and compared to a uniform-randomly chosen number between $0$ and $1$. The move is accepted if the former exceeds the latter. Otherwise, the walker remains at its current position. The random nature of each move allows the walkers to ``climb out" of possible local minima and explore the likelihood space more thoroughly. Once enough trials have been carried out, the posterior distribution of each SME coefficient at a given value is approximated as the fraction of the chain length that the walkers spent in its vicinity.
 
Mass dimension $d=4$, $5$ and $6$ SME universes span parameter spaces with $10$, $16$ and $42$ dimensions respectively, corresponding to the number of independent SME coefficients. Our MCMC chains explore those spaces with $400$, $640$ and $1680$ walkers, respectively ($40$ walkers for each dimension). The large number of walkers allowed us to efficiently distribute the computational demand among the nodes of a supercomputer.

Following Ref.~\cite{kislat18}, we chose an origin-centered scalar Gaussian proposal distribution. Since the desired posterior distributions are expected to fall close to the origin of the likelihood space, we draw the initial positions of the walkers from the proposal distribution as well. The standard deviation of the proposal distribution was individually tuned for each mass dimension to yield move acceptance rates close to 15\%-20\% for most walkers. Specifically, the standard deviations were set to $10^{-34}$, $0.4 \times 10^{-34}\ \mathrm{eV}^{-1}$ and $10^{-36} \times 10^{-36}\ \mathrm{eV}^{-2}$ for $d=4,5,6$, yielding the final average acceptance rates of $0.16$, $0.17$ and $0.17$ respectively. Each of the three chains was run for approximately $12500$ trials, corresponding to $0.5 \times 10^6$ moves (accepted or rejected) across all walkers per mass dimension. All calculations are performed using the Python \texttt{emcee} package\footnote{\url{https://pypi.org/project/emcee/} \cite{foreman-mackey13}}. Our results are described in \S\ref{sec:constraints}.

\section{Archival Catalog of Broadband Optical Polarimetry of Extragalactic Sources}
\label{sec:catalog}

Refs.~\cite{kislat17} (\cite{kislat18}) analyzed a preliminary set of  71 (70) AGN and GRB afterglows (including 44 (43) with only broadband polarimetry and 27 (27) with spectropolarimetry. For the catalog of broadband optical polarimetry displayed in \figs{}~\ref{fig:catalog}-\ref{fig:angles_catalog}, we compiled \nobs{} optical polarization measurements of \nsources{} extragalactic AGN and GRB afterglow sources from 23 references in the literature \cite{Steele_2017,Hovatta_2016,Pavlidou_2014,Heidt_2011,Angelakis_2018,Kumar_2018,Borguet_2008,Smith_2002,Tadhunter_2002,Jones_2012,Almeida_2016,Gorosabel_2014,Brindle_1986,Brindle_1990a,Brindle_1990b,Brindle_1991,Martin_1983,Cimatti_1993,Angelakis_2016,Itoh_2016,Sluse_2005,Wills_2011,Hutsemekers_2017}. All \nobs{} have measured linear polarization fractions and errors, and can be used to constrain the CPT-odd $d=5$ birefringent SME parameters, while only \nobsangle{} have measured polarization angles, which are required to constrain the CPT-even $d=4$ and $d=6$ SME coefficients analyzed here. Note that our conservative approach is remarkably insensitive to the \textit{uncertainty} in the measured polarization angle, so it is not used in the analysis, although we include it in our catalog where available.

Depending on the format, we extracted the data from machine-readable tables from VizieR or from journal websites for individual publications. Older data was parsed using optical character recognition (OCR) or manual input (checked twice to avoid typing errors) as needed. The complete selection criteria imposed on all extracted entries before analysis are described in Appendix~\ref{sec:catalog1}, while notes for individual references are detailed in Appendix~\ref{sec:catalog2}. 

To our knowledge, while far from exhaustive, this represents the most complete catalog of broadband polarization measurements of extragalactic sources to be compiled from the literature to date, in the spirit of the optical starlight polarimetry catalog compiled by Heiles in Ref.~\cite{heiles00}, which included polarization measurements of over 9000 Milky Way stars. A brief sample of the catalog is shown in Tables~\ref{tab:catalog}-\ref{tab:catalog1}. The complete catalog will be made available online in machine-readable format upon publication. 

Such a catalog may have many additional applications beyond Lorentz invariance and CPT violation tests, including tests for large scale alignment of quasar polarization vectors \cite{hutsemekers05,pelgrims14,hutsemekers14}, cold dark matter searches for axions based on polarization effects on extragalactic sources \cite{bassan10,day18}, and studies of the evolution of AGN optical polarization properties.

\begin{table*}[t!]
\begin{tabular}{|l|l|l|l|l|l|}
\hline
Observation \# & Reference & \texttt{Simbad} Source ID & $\pol$ [\%] & $\psi$ [deg]  & Filter\\ \hline
\hhline{|=|=|=|=|=|=|}
\multicolumn{6}{|l|}{...} \\ \hline
UB& Heidt+2011 \cite{heidt11}&\texttt{[MML2015] 5BZB J0925+5958}&$8.65\pm1.1$&$83.3\pm2.9$ &  Gunn-r\\ \hline
UC& Heidt+2011 \cite{heidt11}&\texttt{2MASS J09263881+5411270}&$7.02\pm0.93$&$24.9\pm3.0$ &  Gunn-r\\ \hline
UD& Heidt+2011 \cite{heidt11}&\texttt{2MASS J09291222+0300297}&$9.41\pm0.69$&$-88.6\pm2.1$ & EFOSC2-gunn-r\\ \hline
\multicolumn{6}{|l|}{...} \\ \hline
\end{tabular}
\caption{A portion of individual observations from our Broadband Optical Polarization Catalog of Extragalactic Sources described in \S\ref{sec:catalog} and Appendices~\ref{sec:catalog1}-\ref{sec:catalog2} is shown for format and guidance. A complete, machine-readable version of the catalog will be made available upon publication, including \nobs{} polarization fraction observations and \nobsangle{} polarization angle observations of \nsources{} unique sources from 23 unique references in the literature \cite{Steele_2017,Hovatta_2016,Pavlidou_2014,Heidt_2011,Angelakis_2018,Kumar_2018,Borguet_2008,Smith_2002,Tadhunter_2002,Jones_2012,Almeida_2016,Gorosabel_2014,Brindle_1986,Brindle_1990a,Brindle_1990b,Brindle_1991,Martin_1983,Cimatti_1993,Angelakis_2016,Itoh_2016,Sluse_2005,Wills_2011,Hutsemekers_2017}. The catalog columns include, left to right, a unique ID \# string for each observation (including repeated observations of the same source, where available), the \texttt{Simbad} Source ID, the observed polarization fraction $\Pi$ and error [in percent], the observed polarization angle $\psi$ and error [in degrees] (the polarization angle error may be missing in some cases since we did not use it in our analysis), and the name of the broadband optical filter (and/or the detector, where applicable) used to perform the polarization measurement. The transmission profiles of filters and (where necessary), response curves of detectors, are included 
in a machine-readable form with the catalog. Table~\ref{tab:catalog1} includes additional information for the \nsources{} individual sources, including the cosmological redshift $z$, the IRCS 2000 RA and Dec celestial coordinates.
}
\label{tab:catalog}
\end{table*}

\begin{table*}[t!]
\begin{tabular}{|l|l|l|l|l|}
\hline
\texttt{Simbad} Source ID & Redshift $z$ & RA J2000 & Dec J2000 & V magnitude \\ \hline
\hhline{|=|=|=|=|=|}
\multicolumn{5}{|l|}{...} \\ \hline
\texttt{[MML2015] 5BZB J0925+5958}&0.69 & 09h25m42.91s & 59d58m16.3s & 19.27\\ \hline
\texttt{2MASS J09263881+5411270} & 0.85 & 09h26m38.88s & 54d11m26.6s & 19.6 \\ \hline
\texttt{2MASS J09291222+0300297}& 2.21 & 09h29m12.26s & 03d00m29.9s & 20.87 \\ \hline
\multicolumn{5}{|l|}{...} \\ \hline
\end{tabular}
\caption{A portion of individual Extragalactic Sources from our Broadband Optical Polarization Catalog described in \S\ref{sec:catalog} and Appendices~\ref{sec:catalog1}-\ref{sec:catalog2} is shown for format and guidance. A complete, machine-readable version of the catalog will be made available upon publication. The catalog columns include, left to right, the \texttt{Simbad} Source ID, the redshift $z$, the IRCS 2000 RA and Dec celestial coordinates and the apparent magnitude of the source. Although not shown here, in the machine-readable version of the catalog, we also provide errors, bibliographic references and apparent magnitudes in other optical bands from \texttt{Simbad}. References for individual observations of each source, potentially from multiple publications, are included in Table~\ref{tab:catalog}. Tables~\ref{tab:catalog} and \ref{tab:catalog1} can be cross referenced via the common \texttt{Simbad} Source IDs.
}
\label{tab:catalog1}
\end{table*}

\section{Constraints on Lorentz Invariance Violation and CPT Violation}
\label{sec:constraints}

Our main results are presented in Tables~\ref{tab:constraints_4}-\ref{tab:constraints_6},
which present our upper limits on the $N(d)=10$, $16$, and $42$ anisotropic birefringent SME coefficients for mass dimensions $4$, $5$, and $6$, respectively, using our database of up to \nobs{} broadband optical polarization observations and \nsources{} unique lines of sight over the sky. These upper limits are computed as the maximum of the absolute value of the 5th and 95th percentiles from our MCMC posterior distributions, which are shown in Figs.~\ref{fig:histograms_4}-\ref{fig:histograms_6} in Appendix~\ref{sec:mcmc}, for $d=4$, $5$, and $6$.

\fig{}~\ref{fig:heatmap_6} shows heat maps of the Pearson correlation coefficients between various SME parameters for $d=4$, $5$, and $6$, which we choose to present instead of the 2D posterior distributions showing the correlation between various SME parameters. Selected pairs of SME coefficients show correlation coefficients as high as $\approx \pm 0.6$. This may perhaps be attributed to the uneven distribution of sources across the sky. An exceptionally well-sampled line of sight may be making a dominating contribution to the constraints on multiple SME coefficients, introducing a partial degeneracy between the two and, therefore, a statistically significant (anti)correlation. We however emphasize that our chosen probability distribution is only suitable for estimating the upper limits on the SME coefficients and is inadequate to make any more definitive statements about the specific values of the coefficients or the relationships between them.

\begin{table}[t!]
\begin{tabular}{rcl}
\hline
\hline
$|k^{(4)}_{{(E)2,0}}|$ $<$ & ${\bf 2.9} \times 10^{-34}$\\
$|k^{(4)}_{{(B)2,0}}|$ $<$ & ${\bf 3.0} \times 10^{-34}$\\
$|\mathrm{Re}\left[k^{(4)}_{{(E)2,1}}\right]|$ $<$ & ${\bf 2.9} \times 10^{-34}$\\
$|\mathrm{Re}\left[k^{(4)}_{{(B)2,1}}\right]|$ $<$ & ${\bf 2.8} \times 10^{-34}$\\
$|\mathrm{Im}\left[k^{(4)}_{{(E)2,1}}\right]|$ $<$ & ${\bf 2.1} \times 10^{-34}$\\
$|\mathrm{Im}\left[k^{(4)}_{{(B)2,1}}\right]|$ $<$ & ${\bf 2.1} \times 10^{-34}$\\
$|\mathrm{Re}\left[k^{(4)}_{{(E)2,2}}\right]|$ $<$ & ${\bf 4.0} \times 10^{-34}$\\
$|\mathrm{Re}\left[k^{(4)}_{{(B)2,2}}\right]|$ $<$ & ${\bf 3.5} \times 10^{-34}$\\
$|\mathrm{Im}\left[k^{(4)}_{{(E)2,2}}\right]|$ $<$ & ${\bf 3.3} \times 10^{-34}$\\
$|\mathrm{Im}\left[k^{(4)}_{{(B)2,2}}\right]|$ $<$ & ${\bf 3.4} \times 10^{-34}$\\
\hline
\hline
\end{tabular}
\caption{Mass dimension $d=4$ limits for all $N(4)=10$ independent anisotropic birefringent \textit{dimensionless} SME coefficients $|k^{(4)}_{(E)jm}|$ and $|k^{(4)}_{(B)jm}|$ constrained in this analysis. Upper limits are presented as the maximum of the absolute value of the 5th and 95th percentile constraints, as shown in \fig{}~\ref{fig:histograms_4}. For $d=4$, $j=2$ from Eq.~\eqref{eq:jm} for all values of $m \in [0,1,2]$. The dependent parameters $k^{(4)}_{(E)2(-m)}$ and $k^{(4)}_{(B)2(-m)}$ can be computed using Eq.~\eqref{eq:parityevenk}.
}
\label{tab:constraints_4}
\end{table}

\begin{table}[t!]
\begin{tabular}{rcl}
\hline
\hline
$|k^{(5)}_{{(V)0,0}}|$ $<$ & ${\bf 3.5} \times 10^{-25}$\\
$|k^{(5)}_{{(V)1,0}}|$ $<$ & ${\bf 4.0} \times 10^{-25}$\\
$|\mathrm{Re}\left[k^{(5)}_{{(V)1,1}}\right]|$ $<$ & ${\bf 2.3} \times 10^{-25}$\\
$|\mathrm{Im}\left[k^{(5)}_{{(V)1,1}}\right]|$ $<$ & ${\bf 2.2} \times 10^{-25}$\\
$|k^{(5)}_{{(V)2,0}}|$ $<$ & ${\bf 3.6} \times 10^{-25}$\\
$|\mathrm{Re}\left[k^{(5)}_{{(V)2,1}}\right]|$ $<$ & ${\bf 3.0} \times 10^{-25}$\\
$|\mathrm{Im}\left[k^{(5)}_{{(V)2,1}}\right]|$ $<$ & ${\bf 3.0} \times 10^{-25}$\\
$|\mathrm{Re}\left[k^{(5)}_{{(V)2,2}}\right]|$ $<$ & ${\bf 1.6} \times 10^{-25}$\\
$|\mathrm{Im}\left[k^{(5)}_{{(V)2,2}}\right]|$ $<$ & ${\bf 1.5} \times 10^{-25}$\\
$|k^{(5)}_{{(V)3,0}}|$ $<$ & ${\bf 2.7} \times 10^{-25}$\\
$|\mathrm{Re}\left[k^{(5)}_{{(V)3,1}}\right]|$ $<$ & ${\bf 2.8} \times 10^{-25}$\\
$|\mathrm{Im}\left[k^{(5)}_{{(V)3,1}}\right]|$ $<$ & ${\bf 2.7} \times 10^{-25}$\\
$|\mathrm{Re}\left[k^{(5)}_{{(V)3,2}}\right]|$ $<$ & ${\bf 2.5} \times 10^{-25}$\\
$|\mathrm{Im}\left[k^{(5)}_{{(V)3,2}}\right]|$ $<$ & ${\bf 2.0} \times 10^{-25}$\\
$|\mathrm{Re}\left[k^{(5)}_{{(V)3,3}}\right]|$ $<$ & ${\bf 1.8} \times 10^{-25}$\\
$|\mathrm{Im}\left[k^{(5)}_{{(V)3,3}}\right]|$ $<$ & ${\bf 1.6} \times 10^{-25}$\\
\hline
\hline
\end{tabular}
\caption{Mass dimension $d=5$ limits for all $N(5)=16$ independent anisotropic birefringent SME coefficients $k^{(5)}_{(V)jm}$ constrained in this analysis in GeV$^{-1}$. Upper limits are presented as the maximum of the absolute value of the 5th and 95th percentile constraints, as shown in \fig{}~\ref{fig:histograms_5}. The dependent parameters $k^{(5)}_{(V)j(-m)}$ can be computed using Eq.~\eqref{eq:parityoddk}.
}
\label{tab:constraints_5}
\end{table}

\begin{table}[t!]
\begin{tabular}{rl|rl}
\hline
\hline
$|k^{(6)}_{{(E)2,0}}|$ $<$ & ${\bf 8.5} \times 10^{-18}$ & $|k^{(6)}_{{(B)2,0}}|$ $<$ & ${\bf 8.2} \times 10^{-18}$\\
$|\mathrm{Re}\left[k^{(6)}_{{(E)2,1}}\right]|$ $<$ & ${\bf 7.8} \times 10^{-18}$ & $|\mathrm{Re}\left[k^{(6)}_{{(B)2,1}}\right]|$ $<$ & ${\bf 8.4} \times 10^{-18}$\\
$|\mathrm{Im}\left[k^{(6)}_{{(E)2,1}}\right]|$ $<$ & ${\bf 7.4} \times 10^{-18}$ & $|\mathrm{Im}\left[k^{(6)}_{{(B)2,1}}\right]|$ $<$ & ${\bf 7.6} \times 10^{-18}$\\
$|\mathrm{Re}\left[k^{(6)}_{{(E)2,2}}\right]|$ $<$ & ${\bf 7.7} \times 10^{-18}$ & $|\mathrm{Re}\left[k^{(6)}_{{(B)2,2}}\right]|$ $<$ & ${\bf 7.9} \times 10^{-18}$\\
$|\mathrm{Im}\left[k^{(6)}_{{(E)2,2}}\right]|$ $<$ & ${\bf 8.0} \times 10^{-18}$ & $|\mathrm{Im}\left[k^{(6)}_{{(B)2,2}}\right]|$ $<$ & ${\bf 8.1} \times 10^{-18}$\\
$|k^{(6)}_{{(E)3,0}}|$ $<$ & ${\bf 8.8} \times 10^{-18}$ & $|k^{(6)}_{{(B)3,0}}|$ $<$ & ${\bf 8.3} \times 10^{-18}$\\
$|\mathrm{Re}\left[k^{(6)}_{{(E)3,1}}\right]|$ $<$ & ${\bf 7.7} \times 10^{-18}$ & $|\mathrm{Re}\left[k^{(6)}_{{(B)3,1}}\right]|$ $<$ & ${\bf 7.5} \times 10^{-18}$\\
$|\mathrm{Im}\left[k^{(6)}_{{(E)3,1}}\right]|$ $<$ & ${\bf 8.0} \times 10^{-18}$ & $|\mathrm{Im}\left[k^{(6)}_{{(B)3,1}}\right]|$ $<$ & ${\bf 8.0} \times 10^{-18}$\\
$|\mathrm{Re}\left[k^{(6)}_{{(E)3,2}}\right]|$ $<$ & ${\bf 6.6} \times 10^{-18}$ & $|\mathrm{Re}\left[k^{(6)}_{{(B)3,2}}\right]|$ $<$ & ${\bf 6.8} \times 10^{-18}$\\
$|\mathrm{Im}\left[k^{(6)}_{{(E)3,2}}\right]|$ $<$ & ${\bf 7.1} \times 10^{-18}$ & $|\mathrm{Im}\left[k^{(6)}_{{(B)3,2}}\right]|$ $<$ & ${\bf 7.5} \times 10^{-18}$\\
$|\mathrm{Re}\left[k^{(6)}_{{(E)3,3}}\right]|$ $<$ & ${\bf 7.7} \times 10^{-18}$ & $|\mathrm{Re}\left[k^{(6)}_{{(B)3,3}}\right]|$ $<$ & ${\bf 8.1} \times 10^{-18}$\\
$|\mathrm{Im}\left[k^{(6)}_{{(E)3,3}}\right]|$ $<$ & ${\bf 8.2} \times 10^{-18}$ & $|\mathrm{Im}\left[k^{(6)}_{{(B)3,3}}\right]|$ $<$ & ${\bf 8.0} \times 10^{-18}$\\
$|k^{(6)}_{{(E)4,0}}|$ $<$ & ${\bf 8.4} \times 10^{-18}$ & $|k^{(6)}_{{(B)4,0}}|$ $<$ & ${\bf 8.6} \times 10^{-18}$\\
$|\mathrm{Re}\left[k^{(6)}_{{(E)4,1}}\right]|$ $<$ & ${\bf 7.8} \times 10^{-18}$ & $|\mathrm{Re}\left[k^{(6)}_{{(B)4,1}}\right]|$ $<$ & ${\bf 7.6} \times 10^{-18}$\\
$|\mathrm{Im}\left[k^{(6)}_{{(E)4,1}}\right]|$ $<$ & ${\bf 7.8} \times 10^{-18}$ & $|\mathrm{Im}\left[k^{(6)}_{{(B)4,1}}\right]|$ $<$ & ${\bf 7.7} \times 10^{-18}$\\
$|\mathrm{Re}\left[k^{(6)}_{{(E)4,2}}\right]|$ $<$ & ${\bf 7.1} \times 10^{-18}$ & $|\mathrm{Re}\left[k^{(6)}_{{(B)4,2}}\right]|$ $<$ & ${\bf 7.2} \times 10^{-18}$\\
$|\mathrm{Im}\left[k^{(6)}_{{(E)4,2}}\right]|$ $<$ & ${\bf 7.1} \times 10^{-18}$ & $|\mathrm{Im}\left[k^{(6)}_{{(B)4,2}}\right]|$ $<$ & ${\bf 7.5} \times 10^{-18}$\\
$|\mathrm{Re}\left[k^{(6)}_{{(E)4,3}}\right]|$ $<$ & ${\bf 7.2} \times 10^{-18}$ & $|\mathrm{Re}\left[k^{(6)}_{{(B)4,3}}\right]|$ $<$ & ${\bf 7.3} \times 10^{-18}$\\
$|\mathrm{Im}\left[k^{(6)}_{{(E)4,3}}\right]|$ $<$ & ${\bf 7.4} \times 10^{-18}$ & $|\mathrm{Im}\left[k^{(6)}_{{(B)4,3}}\right]|$ $<$ & ${\bf 7.4} \times 10^{-18}$\\
$|\mathrm{Re}\left[k^{(6)}_{{(E)4,4}}\right]|$ $<$ & ${\bf 7.2} \times 10^{-18}$ & $|\mathrm{Re}\left[k^{(6)}_{{(B)4,4}}\right]|$ $<$ & ${\bf 7.7} \times 10^{-18}$\\
$|\mathrm{Im}\left[k^{(6)}_{{(E)4,4}}\right]|$ $<$ & ${\bf 7.8} \times 10^{-18}$ & $|\mathrm{Im}\left[k^{(6)}_{{(B)4,4}}\right]|$ $<$ & ${\bf 7.6} \times 10^{-18}$\\
\hline
\hline
\end{tabular}
\caption{Mass dimension $d=6$ limits for all $N(6)=42$ independent anisotropic birefringent SME coefficients $|k^{(6)}_{(E)jm}|$ and $|k^{(6)}_{(E)jm}|$ constrained in this analysis in GeV$^{-2}$. Upper limits are presented as the maximum of the absolute value of the 5th and 95th percentile constraints, as shown in \fig{}~\ref{fig:histograms_6}. The dependent parameters $k^{(6)}_{(E)j(-m)}$ and $k^{(6)}_{(E)j(-m)}$ can be computed using Eq.~\eqref{eq:parityevenk}.
}
\label{tab:constraints_6}
\end{table}

\section{Addressing Systematic Errors}
\label{sec:sys}

In this section, we address systematic astrophysical effects which could mimic Lorentz invariance and CPT violation from cosmic birefringence, causing us to overestimate the tightness of our SME constraints and present smaller upper limits than are appropriate. Such effects would act in the same way as cosmic birefringence and depolarize light by rotating the plane of linear polarization, or by reducing the polarization via absorption, for example, by dust extinction along the line of sight. These effects could operate
either near the extragalactic source and/or as it travels to us over cosmological distances. 

We first note that, while Faraday rotation can theoretically rotate the plane of linear polarization for photons, it is negligible at optical wavelengths \cite{falomo14}. We are therefore most concerned with intrinsic source effects and astrophysical propagation effects on polarized light incident on our galaxy, which can either be further polarized or depolarized depending on the dust column it traverses.

When attempting to upper bound any LIV/CPTV effects, larger broadband polarization measurements lead to tighter SME constraints because non-zero SME effects observed in a broad bandpass would tend to depolarize the light as it travels from the source to the observer. As such, our conservative approach, which assumes the source is 70\% polarized at all energies, has the advantage of being insensitive to additional astrophysical line-of-sight effects which could further depolarize light beyond any cosmic birefringence, e.g.~dilution by unpolarized host galaxy light \cite{marin18}, or passage through multiple dust clouds in the Milky Way interstellar medium \cite{bagnulo17,siebenmorgen18}, since modeling these effects would only tighten our constraints.

Outside our galaxy, intergalactic dust in damped Lyman-$\alpha$ absorbers along the line of sight toward the extragalactic source (e.g. \cite{heintz18}) could theoretically depolarize optical light from the source of interest, but such dust is rarely seen, and unlikely to be significant along lines of sight where optical polarization was observed for objects in our catalog. Future work could exclude sources that additionally showed a depletion in Ultraviolet flux, which could indicate such intergalactic dust.

Ultimately, the most important astrophysical effect which could sometimes \textit{increase} polarization along lines-of-sight to extragalactic sources --- causing us to overestimate the tightness of our birefringence constraints --- is due to interstellar polarization from Milky Way dust \cite{bagnulo17,siebenmorgen18}. Therefore, such tests ideally require subtracting a conservative upper bound for the estimated interstellar polarization, e.g.~using field star polarimetry as in Ref.~\cite{friedman19b}, or some other method, in addition to accounting for any systematic polarization inside the instrument. 

Nevertheless, we argue that our overall constraints are insensitive to this particular systematic for the following reasons. First of all, linearly polarized light incident on a Milky Way dust cloud will either emerge from it with greater or smaller linear polarization depending on the local magnetic field orientation in the cloud. When averaging over sufficiently many lines of sight, this type of systematic error will behave like a random error that averages out. Future work will test this using realistic simulations of the interstellar medium, following \cite{kritsuk18}.

In this work, we simply present the polarization data set in our catalog as it was published. Additional analysis could require optical starlight polarimetry of $\gtrsim 2$-$3$ stars along lines of sight within a few arcminutes of each extragalactic source, under the assumption that the interstellar polarization through the entire column of the galaxy was constant over that sky area \cite{friedman19b}. In addition, the existing stellar optical polarimetry catalogs, e.g. \cite{heiles00,weitenbeck08,meade12,bagnulo17,siebenmorgen18}, do not have sufficient sky density to suffice for this purpose, and data from the RoboPol survey \cite{king14,panopoulou15,angelakis16,angelakis18,skalidis18} primarily focused on linear polarization measurements of AGN in the centers of their fields, rather than nearby stars, so we defer such an analysis to future work using simulations or when sufficient observations become available. Future optical polarization surveys like PASIPHAE \cite{tassis18}, for example, will also significantly improve optical stellar polarimetry sky coverage out to $R < 16.5$ mag at high and low galactic latitudes $|b| \gtrsim +55^{\circ}$, while also obtaining polarimetry of all point sources, including AGN, in their fields.

However, even in the worst case scenario, where every line of sight had its polarization overestimated, neglecting this potential systematic error does not significantly affect our results. First of all, typical stellar polarization values of 0.5\%-1\% are often comparable to, or smaller than, the errors of the polarization measurements in our catalog.  Furthermore, even if we conservatively subtracted a typical optical stellar linear polarization fraction of 0.5\%-1\% \cite{weitenbeck08,meade12,siebenmorgen18} from every measurement in our catalog as an estimate of the added interstellar polarization, it would increase the numerical values of our $d=4$ upper limits in Table~\ref{tab:constraints_4}, for example, by no more than $\sim30\%$. This conservative systematic upper limit was derived from artificially subtracting 1\% linear polarization from each of the 45 sources in Ref.~\cite{kislat18}, and repeating their analysis using our MCMC simulations. In our actual sample of \nobs{} sources, since our constraints are dominated by the most highly polarized sources, with $p > 2\%$, any such effects would be significantly smaller. Future work could also test this with additional MCMC simulations on our entire catalog, which are beyond the scope of this work. 

\section{Discussion and Conclusions}
\label{sec:disc}

Using \nobs{} linear broadband optical polarization measurements and \nobsangle{} polarization angle measurements of \nagn{} extragalactic sources from the literature --- which comprises the most comprehensive such optical polarization database in the literature to date --- we constrained anisotropic Lorentz invariance and CPT violation in the context of the Standard Model Extension. We derived conservative upper limits on each of the $N(d)=10$, $16$, and $42$ anisotropic birefringent SME coefficients with mass dimensions $d=4$, $5$, and $6$, respectively.

Useful metrics to quantify birefringent SME constraints for arbitrary $d$ include the mean $K(d)$ of the $N(d)$ SME coefficient upper bounds, e.g., from Tables~\ref{tab:constraints_4}-\ref{tab:constraints_6}, or the \textit{product} of all upper bounds $V(d) \approx K(d)^{N(d)}$, which represents the $d$-dimensional parameter space volume. Both $K(d)$ and $V(d)$ decrease as constraints improve. The predicted improvement ratios 
\begin{equation}
K'(d) \equiv \frac{K_{\rm before}(d)}{K_{\rm after}(d)}
\label{eq:Kprime}\, ,
\end{equation}
and
\begin{equation}
V'(d) \equiv \log_{10}\Bigg(\frac{V_{\rm before}(d)}{V_{\rm after}(d)}\Bigg) \approx N(d) \log_{10}(K'(d))\, ,
\label{eq:Vprime}
\end{equation}
before and after analyzing more archival data represent powerful ways to quantify improved anisotropic LIV/CPTV constraints.

The results summarized in Table~\ref{tab:constraints_4} show that using a database of broadband optical polarimetry with more than an order of magnitude as many lines of sight and over two orders of magnitude as many individual observations as studied in Ref.~\cite{kislat18}, we constrain the minimal SME $d=4$ dimensionless coefficients at the level of $10^{-34}$. This yields average constraints that are $K^{\prime}(4)= \dfourbetterfactor{}$ times better than the broadband-only constraints from Ref.~\cite{kislat18}, with a reduction in the allowed $N(4)=10$-dimensional parameter space volume of $V^{\prime}(4)= \dfourbettervol{}$ orders of magnitude. Remarkably, our average $d=4$ constraints are actually comparable to the constraints in Ref.~\cite{kislat18}, which also analyzed 27 sources with optical spectropolarimetry, to within a factor of two. This holds despite the fact that spectropolarimetry can provide significantly improved $d=4$ constraints along each line of sight that are each $\sim 1$-$2$ orders of magnitude better than from broadband polarimetry. At least for $d=4$, compared to Ref.~\cite{kislat18} , the additional lines of sight analyzed here compensate for the improved constraining power of spectropolarimetry along individual lines of sight, which stems from the $E^{d-3}=E$ energy dependence in Eq.~\eqref{eq:phiz1} for $d=4$.

In addition, our average $d=5$ constraints are $K^{\prime}(5)= \dfivebetterfactor{}$ times better than the broadband-only constraints from Ref.~\cite{kislat17} --- which we re-computed using the linear least squares analysis method in that work --- yielding a reduction in the allowed $N(5)=16$-dimensional parameter space volume of $V^{\prime}(5)= \dfivebettervol{}$ orders of magnitude. This improvement stems, in part, from the fact that Ref.~\cite{kislat17} assumed an intrinsic polarization fraction of $\pz=1$, whereas this work assumes $\pz=\pzup$. Due to the $E^{d-3}=E^2$ energy dependence in Eq.~\eqref{eq:deltaphiz1} at $d=5$, spectropolarimetry can yield line of sight constraints that are $\sim 2$-$3$ times better than broadband polarimetry \cite{kislat17}. Despite these advantages of spectropolarimetry at increasing mass dimension, our $d=5$ constraints at the level of $10^{-25}$ GeV$^{-1}$ in Table~\ref{tab:constraints_5} are only \dfiveworsefactorspec{} times worse than the constraints using the 27 sources with optical spectropolarimetry analyzed in Ref.~\cite{kislat17}, while using a completely independent broadband data set and analysis method.

Finally, Table~\ref{tab:constraints_6} presents $d=6$ constraints at the $10^{-18}$ GeV$^{-2}$ level for all $N(6)=42$ anisotropic birefringent SME coefficients, which are the first constraints of their kind in the literature. This work is also the first to constrain all anisotropic birefringent coefficients for a CPT-even case at a higher mass dimension beyond the minimal SME $d=4$ case analyzed in Ref.~\cite{kislat18}.

To derive these constraints, we modeled the theoretically predicted effects due to cosmic birefringence and generalized the analysis to arbitrary mass dimension for the first time. We developed a method to upper bound the strength of the relevant anisotropic birefringent SME coefficients that are consistent with the observed broadband polarization data, and we computed the posterior probability distributions for the relevant SME parameters using MCMC simulations.

While this paper focused on broadband optical polarimetry, multi-wavelength observations can yield significantly stronger constraints \cite{kislat17,kislat18,friedman19b,friedman19c}. We note that the methods in this work can be easily generalized to analyze spectropolarimetry or multi-band polarimetry from any wavelength range, building upon Ref.~\cite{kislat18}. Increasingly tighter constraints on anisotropic cosmic birefringence from spectropolarimetry and simultaneous multi-band broadband polarimetry will be presented in future work.

In addition, birefringence effects in the SME are predicted to increase towards higher redshifts and energies. While significantly stronger constraints along individual lines-of-sight are also possible using higher energy broadband x-ray/$\gamma$-ray polarization measurements of GRBs (e.g. \cite{kostelecky13}), such measurements --- which require space or balloon instruments --- do not yet exist in sufficient number and quality \cite{hunter14,weisskopf16,moiseev17,mcconnell17,yang18,pearce19} to fully constrain the SME parameters for the most natural SME models at increasing mass dimension $d=4,5,6,\ldots$ \cite{kislat17,kislat18,friedman19b,friedman19c}. In addition, the statistical and systematic errors of existing x-ray/$\gamma$-ray polarization measurements --- many of which were derived from earlier instruments that were not primarily designed to directly measure linear polarization --- are larger and much less well understood than those at optical wavelengths \cite{kislat17,kislat18}, so we defer inclusion of such data to future work. However, all of the analysis methods presented here will be directly applicable to existing and future x-ray/$\gamma$-ray polarization data.

It would also be interesting to repeat the analysis performed here on larger samples, which can be divided into different redshift bins, and for different AGN sub-classes, to test for redshift-dependent effects in the polarization signatures used to constrain Lorentz invariance and CPT violation or to search for redshift dependence in the best fit values of the SME coefficients themselves. To perform such tests for redshift dependence in individual redshift bins, $\MM >> N(d)$ sources are required \cite{friedman19c}. Such data are already available using archival optical polarimetry, but it will be years to decades before x-ray/$\gamma$-ray data have comparable statistics \cite{hunter14,weisskopf16,moiseev17,mcconnell17,yang18,pearce19}. For example, the IXPE X-ray polarimetry spacecraft \cite{weisskopf16} will likely target only $\sim 10$ AGN during its baseline 2021-2023 mission (Alan Marscher and Roger Romani --- private communication).

Future work could also include potential tests for circular polarization, which could be incorporated into our analysis, should sufficient extragalactic Stokes $V$ data become available, or future methods be developed to simulate circular polarization even in the absence of astrophysical observations comparable in number and quality to the existing Stokes $Q$ and $U$ measurements. 

Finally, it will be useful to investigate new astrophysical approaches which go beyond merely constraining or ruling out various sectors of SME parameter space, in order to search directly for positive evidence of cosmic birefringence and Lorentz invariance and CPT violation in nature. Such searches will require increasing numbers of sources over a wider range of sky positions and energies, as well as detailed theoretical modeling of systematic uncertainties, to account for confounding intrinsic source effects and line-of-sight astrophysical effects, including polarization or depolarization of extragalactic light due to passage through the turbulent interstellar medium. Overall, the growing polarimetric database of extragalactic sources analyzed here represents the largest existing catalog that could also be used for future astroparticle physics tests, which will continue to complement traditional particle physics searches using accelerators and other laboratory tests on Earth.

\acknowledgments

The authors would like to thank David I. Kaiser, Gary M. Cole, Brandon Hensley, Jason Gallicchio, Calvin Leung, Jack Steiner, and Dave Mattingly for helpful conversations. We would also like to thank Gina Panopoulou for help understanding the available broadband polarimetry data products from the RoboPol survey. This research has made use of the Simbad and VizieR databases, both operated at CDS, Strasbourg, France, along with NASA's Astrophysics Data System Bibliographic Services. We performed computations for this work using the Triton Shared Computing Cluster at the San Diego Supercomputing Center at the University of California, San Diego. A.S.F. acknowledges support from NSF Award PHYS 1541160 and NASA Hubble Space Telescope Award HST GO-15889. A.S.F, R.G., D.L., W.S. and B.G.K. gratefully acknowledge support from UCSD's Ax Center for Experimental Cosmology.

\appendix

\section{MCMC Posterior Distributions and Correlations Between SME Coefficients}
\label{sec:mcmc}

\begin{figure*}[h!]
    \centering
    \includegraphics[width=0.82\textwidth]{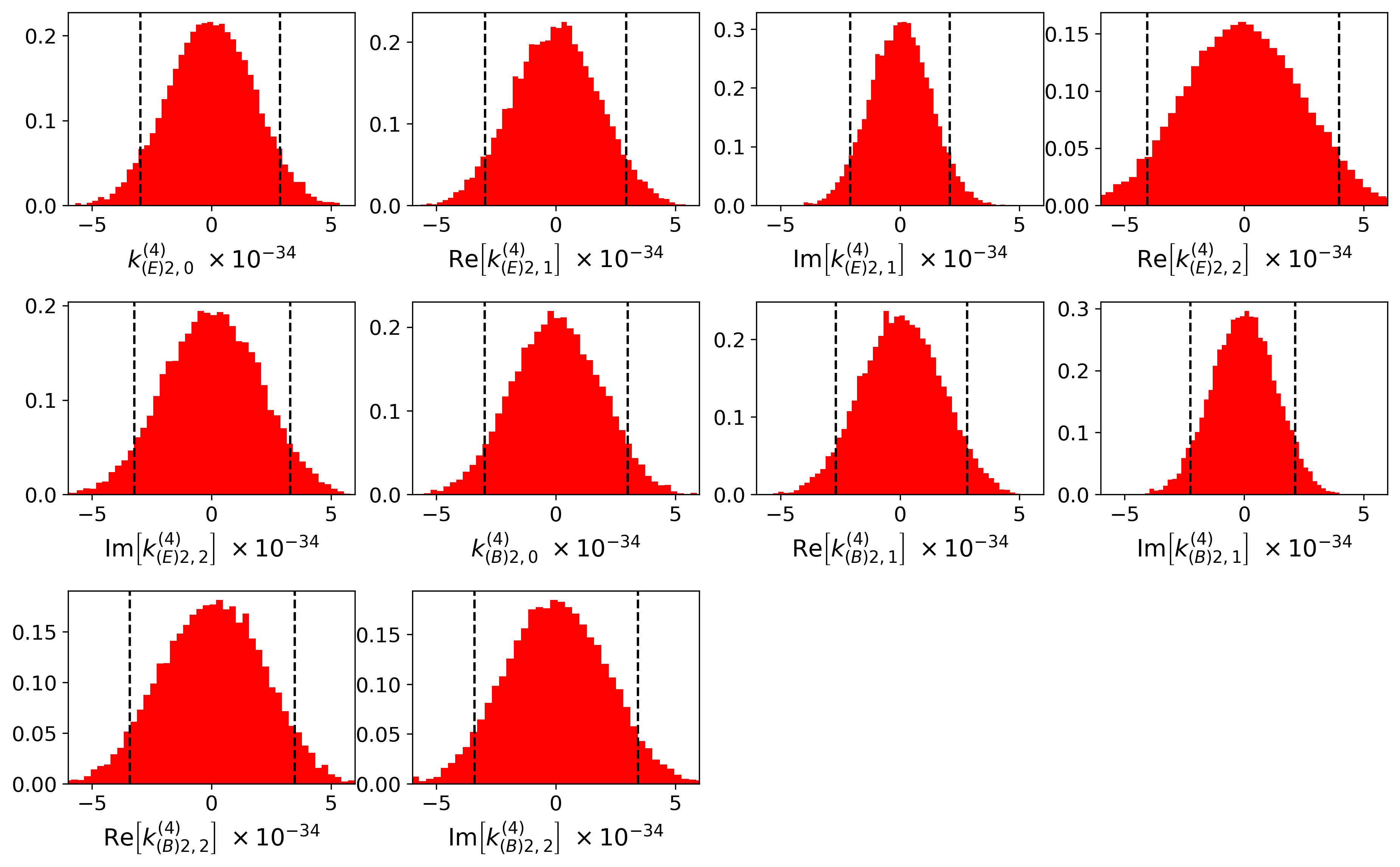}
    \vspace{\dskips}
    \caption{Posterior probability distributions of the $N(4)=10$ dimensionless $d=4$ anisotropic birefringent SME coefficients from our MCMC simulations, each marginalized over the remaining coefficients. For each coefficient, we show the 5th and 95th percentile constraints (vertical dashed lines). 
    }
    \label{fig:histograms_4}
\end{figure*}

\begin{figure*}[h!]
    \centering
    \includegraphics[width=0.82\textwidth]{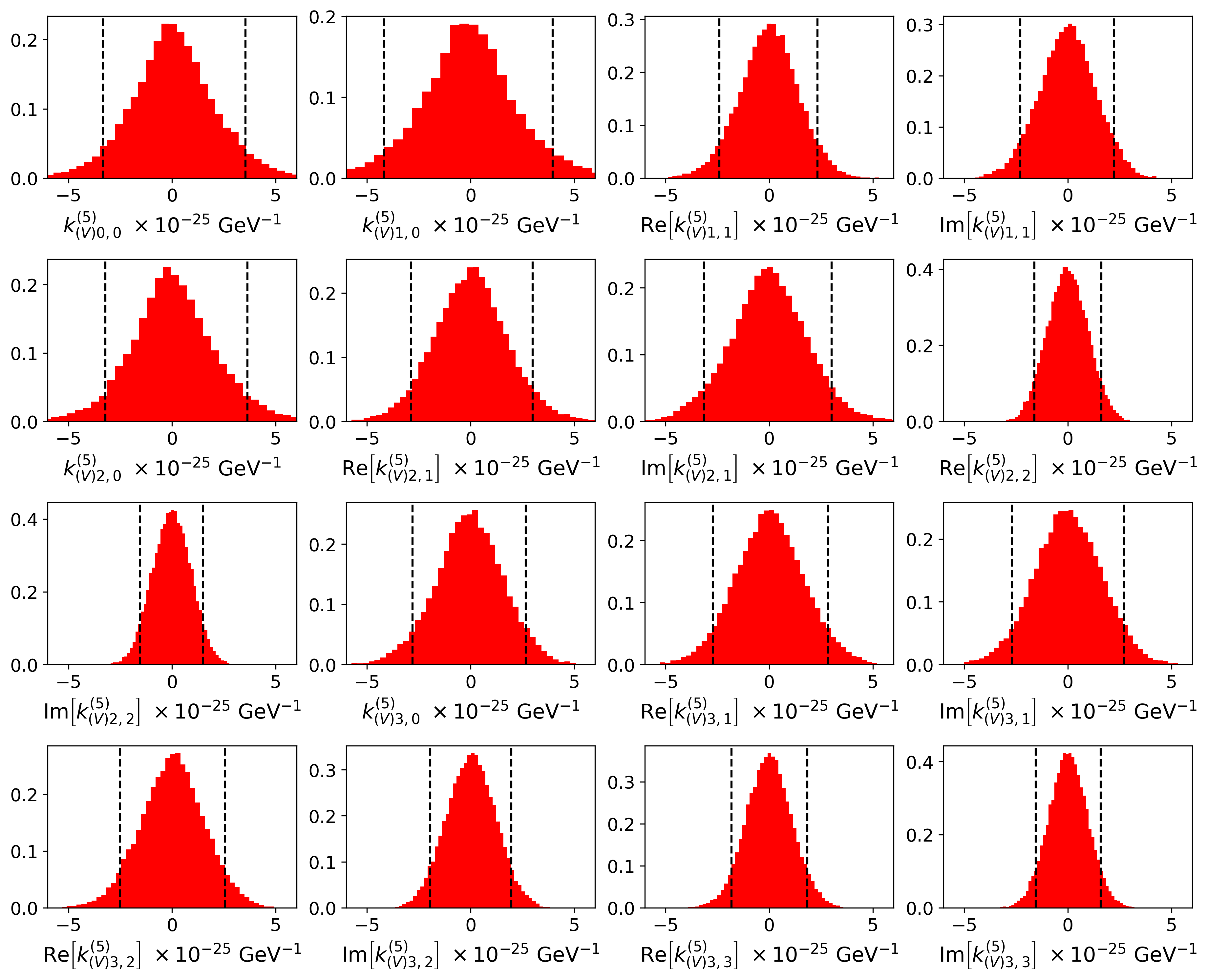}
    \vspace{\dskips}
    \caption{Same as \fig{}~\ref{fig:histograms_4}, but for the $N(5)=16$ anisotropic birefringent SME coefficients at $d=5$.
    }
    \label{fig:histograms_5}
\end{figure*}

\begin{figure*}[h!]
    \centering
    \includegraphics[width=0.99\textwidth]{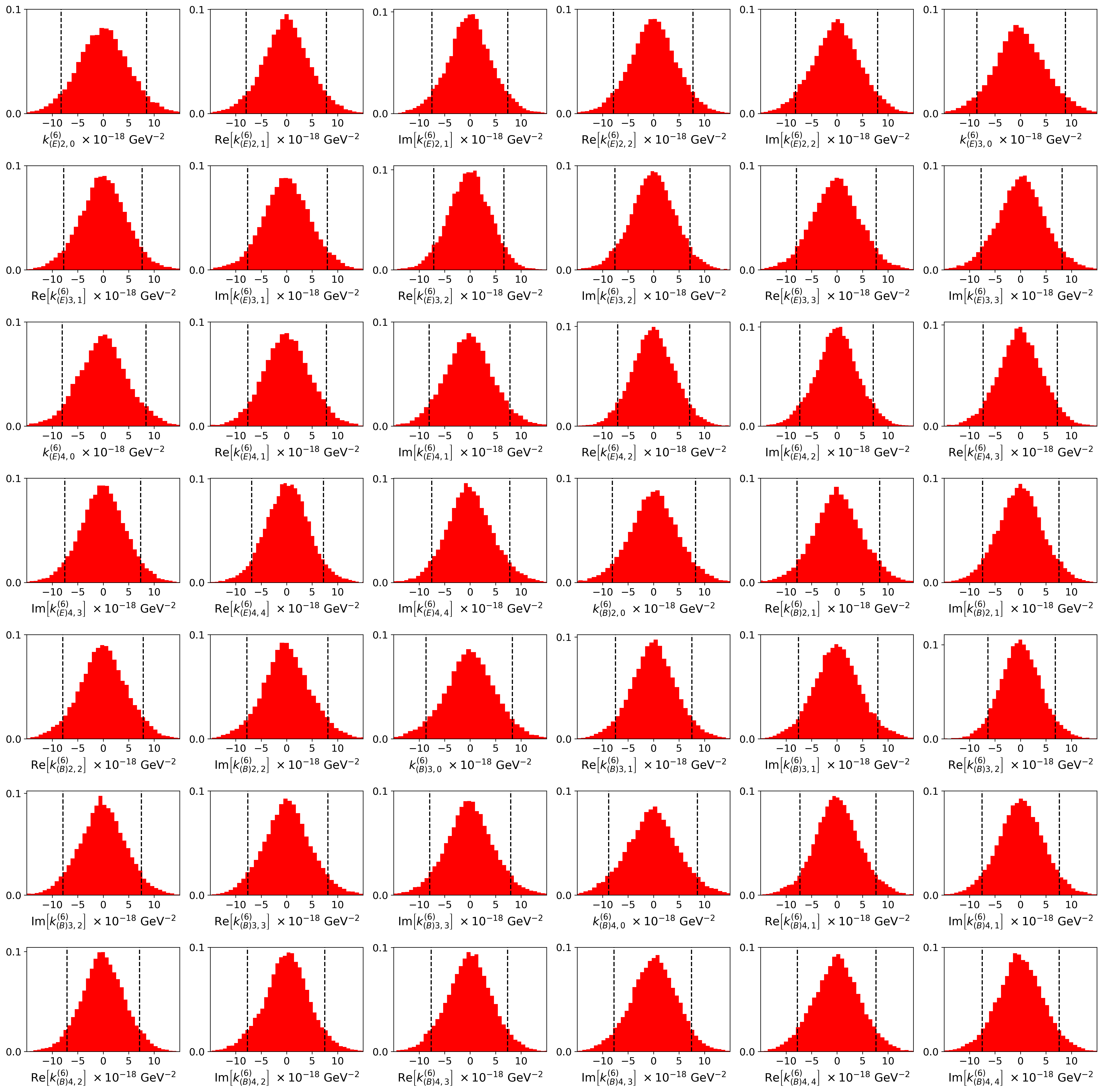}
    \vspace{\dskips}
    \caption{Same as \figs{}~\ref{fig:histograms_4}-\ref{fig:histograms_5}, but for the $N(6)=42$ anisotropic birefringent SME coefficients at $d=6$.
    }
    \label{fig:histograms_6}
\end{figure*}

\begin{figure*}[h!]
    \centering
    \includegraphics[width=0.99\textwidth]{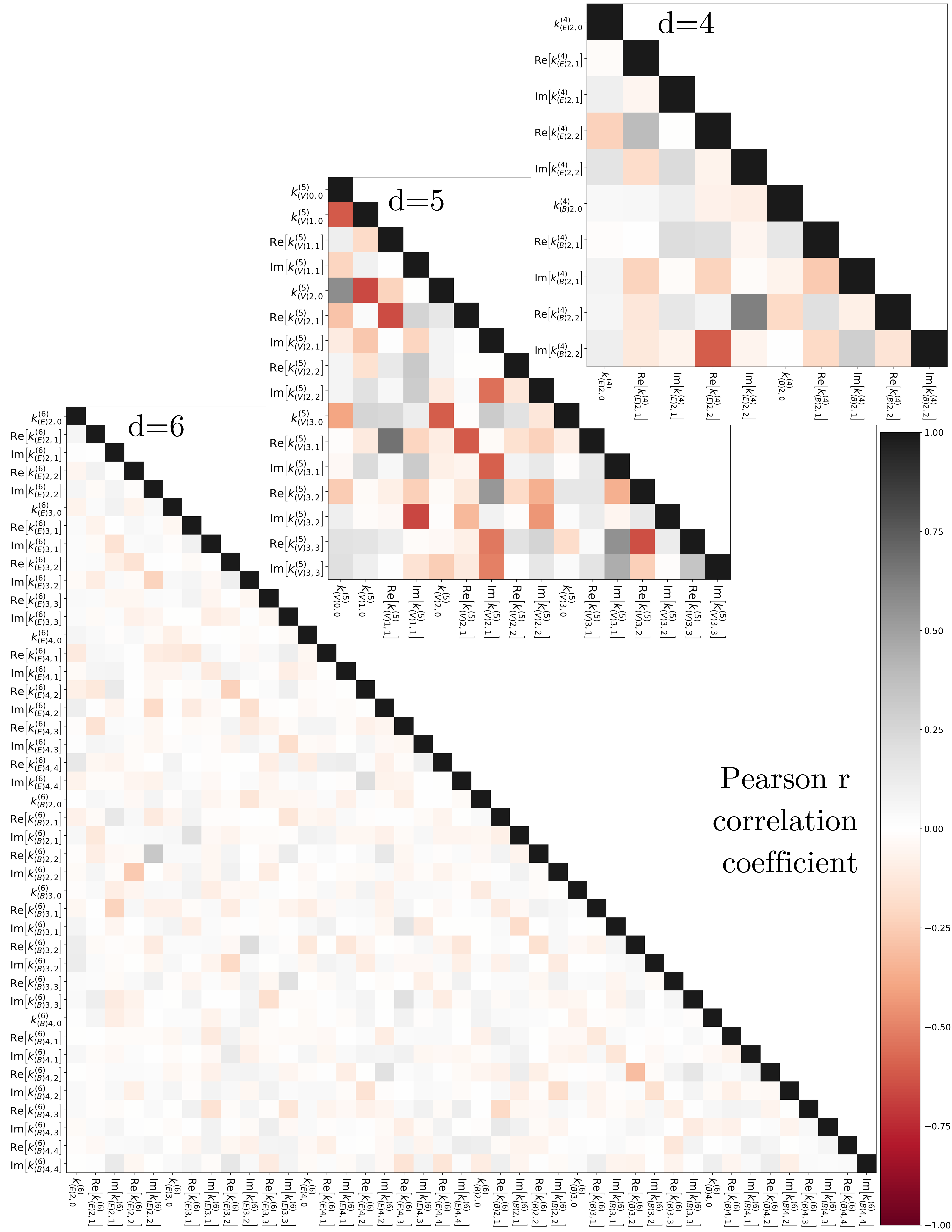}
    \vspace{\dskips}
    \caption{Pearson correlation coefficients extracted from our MCMC simulations between pairs of anisotropic birefringent $d=4$ SME parameters $k^{(4)}_{(E,B)jm}$,  $d=5$ SME parameters $k^{(5)}_{(V)jm}$, and $d=6$ SME parameters $k^{(6)}_{(E,B)jm}$. The same colorbar applies for each mass dimension.
    }
    \label{fig:heatmap_6}
\end{figure*}

As described in \S\ref{sec:constraints},  Figs.~\ref{fig:histograms_4}-\ref{fig:histograms_6} show the MCMC posterior distributions of the 10, 16, and 42 anisotropic birefringent SME coefficients for mass dimensions $d=4$, $5$, and $6$, respectively, while \andyinadd{\fig{}-\ref{fig:heatmap_6}} show heat maps of the Pearson correlation coefficients between these SME parameters.

\clearpage

\section{Catalog of extragalactic polarization: General requirements}
\label{sec:catalog1}

\renewcommand{\UrlFont}{\footnotesize}

The following criteria were applied to all data included in our catalog of broadband extragalactic polarization measurements.

\begin{enumerate}
\setlength{\itemsep}{0pt}%
\setlength{\parskip}{0pt}%
	\item \label{itm:i1} The measured source can be unambiguously linked to an entry on the \href{http://simbad.u-strasbg.fr/simbad/}{CDS \texttt{Simbad} database} 
	\cite{wenger00}.\footnote{\url{http://simbad.u-strasbg.fr/simbad/}}
	\item \texttt{Simbad} lists some measure of redshift that is non-negative.
	\item The parent publication lists the measured linear polarization fraction of the source with its uncertainty and the latter is non-zero.
	\item For the CPT-even case, we also require the measured polarization angle, 
	but its uncertainty is not strictly required in our approach, since our conservative CPT-even constraints are essentially insensitive to it, and completely insensitive to both the polarization angle and its uncertainty in the CPT-odd case.
	\item \label{itm:i5} If the observation is fully filtered, we require enough information to straightforwardly determine the transmission profile of the band.
	\item \label{itm:i6} If the observation is unfiltered or cut-on/cut-off filtered, we require both the transmission profile of the band (if applicable) and the spectral sensitivity of the detector.
\end{enumerate}

We will refer to the cases of observations that do not satisfy items \ref{itm:i5} or \ref{itm:i6} as \textit{instrumental ambiguity}. Once imported, our catalog is further processed as follows:

\begin{enumerate}
\setlength{\itemsep}{0pt}%
\setlength{\parskip}{0pt}%
	\item All sources resolved by \texttt{Simbad} as stellar are checked for available proper motion and parallax measurements. If any are present and are statistically significant, the source is excluded.
	\item All duplicated measurements from different publications are removed.
	\item All polarization angles are wrapped such that the values fall between $-\pi/2$ and $\pi/2$. We assume all extracted polarization angles to be provided in the standard IAU convention, i.e. measured East from North.
\end{enumerate}

\section{Catalog of extragalactic polarization: References and Notes}
\label{sec:catalog2}

\subsection{Steele+2017 \cite{Steele_2017}}

Early-time photometry and polarimetry of optical gamma-ray burst afterglows. The data of interest are available in table III of the publication as well as through VizieR in \href{http://cdsarc.u-strasbg.fr/viz-bin/Cat?J/ApJ/843/143}{\texttt{J/ApJ/843/143}}. The instrument used for all observations is RINGO2 (Liverpool Telescope), which uses a V+R filter whose transmission profile is available on the instrument's website.\footnote{\url{https://telescope.livjm.ac.uk/TelInst/Inst/RINGO2/}}

\subsection{Hovatta+2016 \cite{Hovatta_2016}}

Comparative study of the optical properties of TeV-loud versus TeV-undetected BL Lac objects. Polarization data were acquired in the R band with RoboPol (Skinakas Observatory) and ALFOSC (Nordic Optical Telescope). The former employs a standard Johnson-Cousins R filter \cite{robopol}. For the latter, two different R band transmission profiles are available in the online documentation\footnote{\url{http://www.not.iac.es/instruments/alfosc/stdfilt/stdfilt.html}} corresponding to two generations of detectors denoted as CCD8 and CCD14. We assume that CCD8 was used in this publication given the observation dates (03/2014-11/2014) and the CCD14 commissioning date (2016/03/30).
All data are available through VizieR in \href{http://cdsarc.u-strasbg.fr/viz-bin/Cat?J/A+A/596/A78}{\texttt{J/A+A/596/A78}}.

\subsection{Pavlidou+2014 \cite{Pavlidou_2014}}

Polarization survey of a statistically unbiased sample of blazars. All data were taken with RoboPol (Johnson-Cousins R) and published through VizieR in \href{http://cdsarc.u-strasbg.fr/viz-bin/Cat?J/MNRAS/442/1693}{\texttt{J/MNRAS/442/1693}}.

\subsection{Heidt+2011 \cite{Heidt_2011}}

Polarimetric analysis of optically selected BL Lac candidates on three instruments:  EFOSC2 on ESO's New Technology Telescope, CAFOS at Calar Alto observatory and ALFOSC on Nordic Optical Telescope. The filters are identified in the publication as ESO \#786, Gunn-r and SDSS-r respectively. The transmission profiles of ESO filters are available online\footnote{\url{https://www.eso.org/sci/facilities/lasilla/instruments/efosc/inst/Efosc2Filters.html}} (note that \#786 and \#784 are almost identical). For CAFOS, we used a standard Gunn profile, while ALFOSC filters are described in the instrument's online documentation\footnote{\url{http://www.not.iac.es/instruments/alfosc/stdfilt/stdfilt.html}}, where we again assumed CCD8 based on the observation dates. All data are available through VizieR in \href{http://cdsarc.u-strasbg.fr/viz-bin/Cat?J/A+A/529/A162}{\texttt{J/A+A/529/A162}}. 

\subsection{Angelakis+2018 \cite{Angelakis_2018}}

Search for time-dependent behaviour of polarization in a sample of Seyfert 1 galaxies. The measurements in the publication were obtained with RoboPol (Skinakas Observatory), PRISM (Lowell Observatory) and HOWPol (Higashi-Hiroshima Observatory). Furthermore, a small fraction of data were retrieved from the Steward observatory archive, which we had to reject from our catalog due to instrumental ambiguity.

As before, the standard Johnson-Cousins R profile was assumed for all RoboPol measurements. The same profile was adopted for all PRISM measurements, as suggested in the publication. Finally, the R-band profile of HOWPol is given in the instrument's online documentation.\footnote{\url{http://hasc.hiroshima-u.ac.jp/instruments/howpol/specification-e.html}} All measurements are accessible through VizieR in \href{http://cdsarc.u-strasbg.fr/viz-bin/Cat?J/A+A/618/A92}{\texttt{J/A+A/618/A92}}.

\subsection{Kumar+2018 \cite{Kumar_2018}}

Test for misclassification of BL Lac sources as radio-quiet quasars through optical polarimetry. All observations were obtained with EFOSC2 (ESO's New Technology Telescope). The filter in the optical path can be identified as \#642 (Bessel R) by cross-referencing the observation dates listed in the publication (04/25/2006-04/28/2006) with the publicly available ESO observing logs\footnote{\url{http://archive.eso.org/eso/eso\_archive\_main.html}}. The transmission profiles of all ESO filters are available online.\footnote{\url{https://www.eso.org/sci/facilities/lasilla/instruments/efosc/inst/Efosc2Filters.html}}

\subsection{Borguet+2008 \cite{Borguet_2008}}

Study of the correlation between the optical polarization of quasars and their morphology. All polarization data employed in the paper were chosen from 20 other references based on their reliability and absence of significant temporal variations. The corresponding VizieR repository (\href{http://cdsarc.u-strasbg.fr/viz-bin/Cat?J/A+A/478/321}{\texttt{J/A+A/478/321}}) contains all measurements as well as identifies the designations used for each of the secondary references. The data from a number of said references were rejected either due to instrumental ambiguity or because we were able to include them in our catalog as a primary reference. Overall, this covers approximately $1/3$ of the measurements. The other $2/3$ were incorporated in our catalog, including the following references listed here by their designations: \textit{Ta92}, \textit{Wi80}, \textit{We93}, \textit{Sc99}, \textit{Wi92}, \textit{Mo84}, \textit{Vi98}, \textit{Im90}, \textit{Im91}, \textit{St84}, \textit{Be90}, \textit{Za06}.

The measurements in \textit{Be90}, \textit{St84}, \textit{Mo84}, \textit{Im91} and \textit{Im90} were taken with an unfiltered Ga-As photomultiplier. For all of those, we adopt a typical Ga-As profile from \cite{olympus}. The measurements in \textit{Wi80}, \textit{Wi92} and \textit{Sc99} were obtained with EMI-9658 -- a borosilicate-filtered Na-K-Cs-Sb photomultiplier -- whose transmission profile is available in \cite{emi}. \textit{Za06} observations were conducted with the Hubble Space Telescope and use the F550M filter on ASC with a detailed manual available online.\footnote{\url{http://www.stsci.edu/hst/acs/documents/handbooks/current/c05\_imaging2.html}} \textit{Ta92} include measurements on the Isaac Newton Telescope with the filter identified as \textit{broad Johnson V}. Unfortunately, the telescope underwent a major refurbishment after the data were acquired, leaving little available information on the old setup. For our purposes, we took the standard Johnson-Cousins filter and scaled/translated its transmission to the central wavelength and FWHM quoted in the paper. \textit{We93} use standard filters from the Johnson set. \textit{Vi98} employ another Na-K-Cs-Sb photomultiplier, but do not specify the exact flavour. Hence, we adopt a typical characteristic profile from \cite{ebdon_evans_1998}.

\subsection{Smith+2002 \cite{Smith_2002}}

Follow-up polarimetry of photometrically identified quasars. All measurements are available through VizieR in \href{http://cdsarc.u-strasbg.fr/viz-bin/Cat?J/ApJ/569/23}{\texttt{J/ApJ/569/23}}. Specifically, the \textit{Comm} column of the table indicates the instrument used for each observation. About $1/3$ of the measurements were taken with the Two-Holer Polarimeter (2H), which uses a Ga-As photomultiplier \cite{2H}. As before, we use the profile from \cite{olympus} for such measurements.

For observations in this publication, 2H was installed on two different telescopes: Mt. Lemmon $1.5\ \mathrm{m}$ and Bok $2.3\ \mathrm{m}$. In the former case, the observations were taken unfiltered, implying that the nominal Ga-As profile can be used. In the latter case, a UV-blocking glass was installed in the optical path. To account for this difference, we multiplied the Ga-As response profile by the transmission profile of Edmund Optics N-SF10 glass.\footnote{\url{https://www.edmundoptics.com/knowledge-center/application-notes/optics/optical-glass/}} which has a blue cut-off similar to that quoted in the paper

Other measurements in this publication were obtained using a CCD with the KPNO (Kitt Peak National Observatory) nearly-Mould R filter, which we recognize as those corresponding to empty \textit{Comm} values. Most KPNO filters have published transmission profiles online.\footnote{\url{https://www.noao.edu/kpno/filters/2Inch\_List.html}} Additionally, two  measurements have been obtained with spectropolarimetry, which we exclude from our catalog due to instrumental ambiguity.

\subsection{Tadhunter+2002 \cite{Tadhunter_2002}}

Optical polarimetry of galaxies to differentiate different potential origins of UV emission. All measurements were obtained on ESO's EFOSC1 with the Bessel B filter installed in the optical path. The exact transmission profile of the filter is available in the instrument manual.\footnote{\url{http://www.eso.org/sci/libraries/historicaldocuments/Operating\_Manuals/Operating\_Manual\_No.4\_A1b.pdf}}

Note that the paper offers ``measured" and ``intrinsic" linear polarization fractions for each object, of which the former was included in our catalog for consistency. Intrinsic polarization is estimated via model fitting.

\subsection{Jones+2012 \cite{Jones_2012}}

A study into the relationship between polarization and other properties of a sample of nearby galaxies. All data were collected with the Imaging Grism Polarimeter at McDonald Observatory. In each case, the standard Johnson-Cousins B filter was placed in the optical path.

\subsection{Almeida+2016 \cite{Almeida_2016}}

Spectropolarimetry of selected Seyfert 2 galaxies to differentiate hidden and non-hidden broad-line regions. Synthetic broadband polarization through a standard Johnson-Cousins B filter is offered in table III of the publication. We ignore all narrow-band polarimetry for consistency with the rest of the catalog.

\subsection{Gorosabel+2014 \cite{Gorosabel_2014}}

Polarimetric time series of the optical afterglow of GRB 020813. All measurements were obtained on ESO's FORS1 through the Bessel V filter. The relevant transmission profile is listed in the instrument's operation manual.\footnote{\url{http://www.eso.org/sci/facilities/paranal/instruments/fors/doc/VLT-MAN-ESO-13100-1543\_v82.pdf}}

\subsection{Brindle+1986 \cite{Brindle_1986}, Brindle+1990a \cite{Brindle_1990a}, Brindle+1990b \cite{Brindle_1990b}, Brindle+1991 \cite{Brindle_1991}}

All four publications share a similar format, presenting simultaneous optical and infrared polarimetry of galaxies. While no specific references to the filters used can be found in the papers, most have listed central and half-power wavelengths. This allows us to vaguely match some of the filters to either the standard Johnson-Cousins system (UBVRI) or Glass system (JHK). The data appears to have been taken through two different K-band filters (denoted with \textit{K1} and \textit{K2}), of which we match the latter to the standard Glass K filter and reject the former due to instrumental ambiguity.

All measurements marked with \textit{RI} are assumed to have been taken with a superposition of R and I standard filters. All other filters mentioned in the publications (e.g. \textit{BY}, \textit{WB} and more) could not be linked to known transmission profiles and had to be similarly discarded. Those measurements, however, comprise a small minority of the available data.

\subsection{Martin+1983 \cite{Martin_1983}}

A study of polarization properties of Seyfert galaxies. The survey was mostly conducted using a two-channel photoelectric Pockels cell polarimeter described in \cite{pockels} with the Corning 4-96 filter in the optical path. We assume that the transmission profile can be approximation by that of Grayglass 9782\footnote{\url{http://www.grayglass.net/glass.cfm/Filters/Kopp-Standard-Filters/catid/45/conid/102}}, as they have the same color specification number.

All measurements are listed in table I. We exclude all values, for which the \textit{Remarks} column indicates that some setup other than the one described above was used.

\subsection{Cimatti+1993 \cite{Cimatti_1993}}

An investigation of the polarimetric properties of $z>0.1$ galaxies. This publication uses archival data from 10 other references, denoted with various designations in the \textit{Ref} column of table I. All measurements from \textit{A84}, \textit{GC92} and \textit{FM88} were excluded due to instrumental ambiguity and \textit{Ta92} measurements were ignored as they have already been imported from \cite{Borguet_2008}.

Of those measurements that have been kept, \textit{R83} and \textit{I91} appear to have mostly been taken with unfiltered Ga-As photomultipliers apart from a minority of data obtained through non-standard filters that had to be excluded. As before, the Ga-As response profile from \cite{olympus} was used. \textit{dSA93} measurements were taken on ESO's EFOSC1 through the Bessel filter set. All relevant transmission profiles can be obtained from the instrument's operation manual.\footnote{\url{http://www.eso.org/sci/libraries/historicaldocuments/Operating\_Manuals/Operating\_Manual\_No.4\_A1b.pdf}} \textit{C93} measurements are assumed to have been taken through standard Johnson-Cousins filters. Finally, \textit{JE91} measurements were obtained through nearly Mould R and B filters, whose transmission profiles can be retrieved from the KPNO website.\footnote{\url{https://www.noao.edu/kpno/filters/2Inch\_List.html}}

\subsection{Angelakis+2016 \cite{Angelakis_2016}}

Polarimetric survey to study the differences between gamma-ray loud and quiet quasars. The survey was conducted on RoboPol (Skinakas Observatory), which uses the standard Johnson-Cousins R filter \cite{robopol}. All data are available in table II of the publication, distributed as supplementary material. The table lists minimum, maximum and mean linear polarization fractions, of which only the latter have listed polarization angles. For this reason, only the mean values were included in our catalog.

\subsection{Itoh+2016 \cite{Itoh_2016}}

Observational program to study the temporal variability in polarization of core-dominated quasars. All data are available on VizieR in \href{http://cdsarc.u-strasbg.fr/viz-bin/Cat?J/ApJ/833/77}{\texttt{J/ApJ/833/77}}. The acquisition instrument is HOWPol (Higashi-Hiroshima Observatory). The transmission profiles of the available filters can be found on the specification website.\footnote{\url{http://hasc.hiroshima-u.ac.jp/instruments/howpol/specification-e.html}}

\subsection{Sluse+2005 \cite{Sluse_2005}}

Polarization survey of quasars in both hemispheres. The data were mostly obtained on ESO's EFOSC2 and are fully available through VizieR in \href{http://cdsarc.u-strasbg.fr/viz-bin/Cat?J/A+A/433/757}{\texttt{J/A+A/433/757}}. The observation band is Bessel V for most entries, except a handful of measurements that were taken in i or R bands and can be identified by the \textit{Remarks} column of the table. Furthermore, a small fraction of measurements were taken on ESO's FORS1 in the V band and can be identified by the observation date (02/25/2003). A few measurements are marked as contaminated or potentially contaminated, which have been excluded from our catalog.

The transmission profiles of all relevant filters are available in the operation manuals of the corresponding instruments.

\subsection{Wills+2011 \cite{Wills_2011}}

30 years of previously unpublished data from McDonald observatory. All measurements are listed on VizieR in \texttt{J/ApJS/194/19}. Most of them are unfiltered with the detector identifiable by date as either the EMI-9658 Na-K-Cs-Sb photomultiplier (before 1987) or the R943-02 Ga-As photomultiplier (after 1987). A few measurements acquired in 1987 had to be excluded, as it is unclear which of the photomultipliers was in use at the time. The response curves of both devices can be found in \cite{emi} and \cite{R943_02}. The minority of filtered measurements were taken in one of the UBVRI bands. Of those, UBV are suspected to refer to standard Johnson-Cousins filters, while the nature of R and I filters is less clear. Due to the inherently small number of such measurements, both bands are conservatively discarded due to instrumental ambiguity. A few measurements were obtained through cut-on/cut-off filters including \textit{GG395}, \textit{RG630}, \textit{OG570}, \textit{OG580} and the $\mathrm{CuSO}_4$ filter. In those cases, the transmission profiles of the filters were multiplied by the response profile of the underlying detector. In most cases, the filter profiles could be found on the manufacturer's website.\footnote{\url{https://www.sydor.com/}} Otherwise, the corresponding measurements were excluded from the catalog. When importing the VizieR table, we paid particular attention to the \textit{Notes} and \textit{n\_} columns to exclude all calibration measurements as well as values that may have been affected by other factors such as failed pointing and contamination.

\subsection{Hutsemekers+2017 \cite{Hutsemekers_2017}}

192 previously unpublished polarization measurements of quasars from ESO's EFOSC2. All values are tabulated on VizieR in \href{http://cdsarc.u-strasbg.fr/viz-bin/Cat?J/A+A/606/A101}{\texttt{J/A+A/606/A101}}. All EFOSC2 filter transmission profiles are available online.\footnote{\url{https://www.eso.org/sci/facilities/lasilla/instruments/efosc/inst/Efosc2Filters.html}} The measurements that are marked as potentially contaminated have been excluded from the catalog. A single measurement was taken unfiltered, which we exclude as well.

\renewcommand{\UrlFont}{\scriptsize}



\end{document}